\documentclass[11pt]{article}
\pdfoutput=1
\usepackage{jheppub}

\usepackage[svgnames,table]{xcolor}
\usepackage{graphicx}
\usepackage{amsfonts,amsmath,amssymb,amsthm}
\usepackage{mathrsfs,bm,bbm,esint}
\usepackage[mathscr]{eucal}
\usepackage{physics}
\usepackage{slashed,soul}
\usepackage{rotating,pdflscape}
\usepackage{enumitem}
\usepackage{multirow,array}
\usepackage[numbers,sort&compress]{natbib}

\usepackage{tikz}
\usepackage{pgfplots}
\usetikzlibrary{cd}
\usetikzlibrary{decorations.pathmorphing}
\usetikzlibrary{decorations.pathreplacing}
\usetikzlibrary{decorations.markings}
\usetikzlibrary{snakes}
\usetikzlibrary{shapes.misc}
\tikzset{->-/.style={decoration={
  markings,
  mark=at position .5 with {\arrow{>}}},postaction={decorate}}}
\tikzset{-<-/.style={decoration={
  markings,
  mark=at position .5 with {\arrow{<}}},postaction={decorate}}}
\usepackage{subcaption}
\usepackage{float}
\usepackage{afterpage}
\usepackage{cleveref}
\crefname{section}{\S\!\!}{\S\S\!\!}
\Crefname{section}{\S}{\S\S}
\crefname{appendix}{Appendix}{Appendices\!}
\crefname{figure}{Fig.\!}{Figs.\!}

\definecolor{rust}{rgb}{0.8,0.2,0.2}

\newcommand{\be}{\begin{equation}}
\newcommand{\ee}{\end{equation}}
\newcommand{\bi}{\begin{itemize}}
\newcommand{\ei}{\end{itemize}}

\newcommand{\AdS}[1]{AdS$_{#1}$}
\newcommand{\lads}{\ell_\text{AdS}}

\newcommand{\bulk}{{\cal M}}
\newcommand{\bdy}{{\cal B}}

\newcommand{\bdyk}{{\sf B}^\text{k}}
\newcommand{\bdyb}{{\sf B}^\text{b}}
\newcommand{\bulkket}{{\sf M}^\text{k}}
\newcommand{\bulkbra}{{\sf M}^\text{b}}
\newcommand{\tx}{\tilde{x}}
\newcommand{\yt}{y}
\newcommand{\xs}{\chi_{_S}}
\newcommand{\cx}{x}
\newcommand{\xf}{\mathscr{T}}
\newcommand{\xfh}{\mathfrak{t}}
\newcommand{\bv}{\bar{v}}

\newcommand{\regA}{\mathcal{A}}
\newcommand{\regAc}{\mathcal{A}^c}
\newcommand{\rhoA}{\rho_{_\regA}}

\newcommand{\entsurf}{\partial\regA}
\newcommand{\Cbdy}{\Sigma_{_t}}
\newcommand{\Cbulk}{{\tilde \Sigma}_{_t}}

\newcommand{\fixM}{\bm{\gamma}}

\newcommand{\sn}{\text{sn}}
\newcommand{\cn}{\text{cn}}
\newcommand{\dn}{\text{dn}}

\title{Real-time gravitational replicas: Low dimensional examples}

\author[a]{Sean Colin-Ellerin,}
\author[b]{Xi Dong,}
\author[b]{Donald Marolf,}
\author[a]{Mukund Rangamani,}
\author[b]{Zhencheng Wang}
\affiliation[a]{Center for Quantum Mathematics and Physics (QMAP)\\
Department of Physics \& Astronomy, University of California, Davis, CA 95616, USA}
\affiliation[b]{
Department of Physics, University of California, Santa Barbara, CA 93106, USA}
\emailAdd{scolinellerin@ucdavis.edu}
\emailAdd{xidong@ucsb.edu}
\emailAdd{marolf@ucsb.edu}
\emailAdd{mukund@physics.ucdavis.edu}
\emailAdd{zhencheng@ucsb.edu}

\abstract{
We continue the study of  real-time replica wormholes initiated in \cite{Colin-Ellerin:2020mva}. Previously, we had discussed the general principles and had outlined a variational principle for obtaining stationary points of the real-time gravitational path integral. In the current work we present several explicit examples in low-dimensional gravitational theories where the dynamics is amenable to analytic computation. We demonstrate the computation of R\'enyi entropies in the cases of JT gravity and for holographic two-dimensional CFTs (using the dual gravitational dynamics). In  particular, we explain how to obtain the large central charge result for subregions  comprising of disjoint intervals directly from the real-time path integral.
}

\begin{document}
\maketitle


\section{Introduction}
\label{sec:intro}

Real-time computation of correlation functions, both time-ordered and out-of-time-order, as well as density operator matrix elements and their moments,  in any quantum system either with or without dynamical gravity, requires the use of a suitable timefolded contour, with segments of forward and backward evolution. One often however eschews the use of such contours, relying instead on computations in the Euclidean domain, and then analytically continuing the answers thus obtained into the real-time domain (see e.g., \cite{Calabrese:2005in,Calabrese:2007rg} for non-gravitational theories as well as the more recent analysis in gravitational context in \cite{Goto:2020wnk}), a strategy that works well when  the quantum evolution is not subject to non-analytic sources. While this is strategy  is efficient in extracting information about the non-perturbative aspects  of the theory, it does not lend insight into the physical dynamical evolution directly.

These issues have been well appreciated in the context of quantum field theory for many decades, but have come to fore with recent analyses  of new semiclassical configurations that address the black hole information problem. Inspired by the Euclidean path integral arguments  \cite{Lewkowycz:2013nqa,Faulkner:2013ana,Dong:2016fnf,Dong:2017xht} that helped derive the static holographic entanglement entropy formula \cite{Ryu:2006ef} and its quantum generalization  \cite{Engelhardt:2014gca}, recent investigations in low-dimensional gravity theories have argued for the contribution of replica wormhole saddles  \cite{Penington:2019kki,Almheiri:2019qdq}  in the gravitational path integral. For a review of these developments in the context of the black hole information problem, see \cite{Almheiri:2020cfm}. Furthermore, as argued for in \cite{Marolf:2020xie} such replica wormhole configurations are quite generic in the Euclidean formalism.

Motivated by  these developments, and by  earlier efforts \cite{Dong:2016hjy} to derive the covariant holographic entanglement entropy prescription of \cite{Hubeny:2007xt}, in a companion paper \cite{Colin-Ellerin:2020mva} we outlined the general formalism for understanding the stationary phase approximation of the real-time gravitational functional integral.  In addition, connections to  the black hole information problem and baby universes have also been discussed recently in \cite{Marolf:2020rpm}. Our goal in this current paper  is to exemplify the formal discussion in \cite{Colin-Ellerin:2020mva}  with some concrete examples. For technical reasons our examples will rely on gravitational dynamics  in low dimensions, especially in 2 and 3 spacetime dimensions, where  one can write  down explicit geometries that provide the appropriate stationary points. It should however be clear from our discussion that the construction can in principle be carried out, at least numerically,  in higher-dimensions with dynamical gravitational degrees of freedom.

\begin{figure}
\centering
\begin{tikzpicture}
[scale=1.0,
tzinstate/.style={rectangle,draw=orange!90,fill=red!20,thick, inner sep=0pt,minimum size=2pt},
]
\foreach \x in {0, 4,6,8}
{
	\draw[tzinstate]  (\x,0)   -- ++(1,0) -- ++(-0.,0.1)  -- ++(-1,0) -- cycle;
	\draw[color=black,thick, ->-] (\x,0)  -- ++ (0,4);
	\draw[color=black,thick, -<-] (\x+1,0) --  ++ (0,4);
	\node at (\x+0.5,0) [below=0.5pt]{$\rho_0$};
}
\node at (0,2)  [left]{$\mathcal{U}(t;t_0)$};
\node at (1,2) [right=2pt]{$\mathcal{U}(t;t_0)^\dagger$};
\foreach\x in {5,7}
{
	\draw[color=black,thick] (\x,4) -- ++ (0.5,0.5) -- ++ (0.5,-0.5);
}
	\draw[color=black,thick] (4,4) -- ++ (2.5,2.5) -- ++(2.5,-2.5);
\end{tikzpicture}
\caption{ An illustration of the real-time contours for the computation of the density matrix $\rho(t)$ (left) and traces of its powers
 ($\Tr(\rho(t)^3)$ on right) . The past boundary condition is supplied by the prescribed initial state $\rho_0$ and the direction of time evolution is explicitly indicated by the arrows.  }
\label{fig:skdmat}
\end{figure}
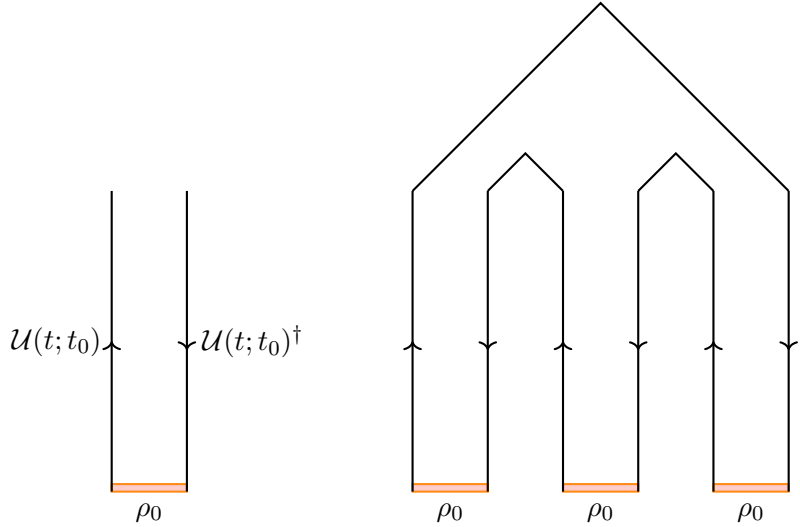
The specific class of problems we study  herein are those that correspond to computation of R\'enyi entropies  in holographic field theories in low dimensions, specifically \AdS{2} and \AdS{3}. We recall that in the field theory one is instructed to consider path integral contours of the form illustrated  in \cref{fig:skdmat}.
Reduced density matrices $\rhoA(t)$ associated with spatial subregions $\regA$ on a Cauchy slice $\Cbdy$ are obtained by sewing together the  ket and bra parts along the complementary domain $\regAc$, leaving open the parts along $\regA$. Traces of powers of $\rhoA(t)$ are computed by taking $n$-copies of the geometry and cyclically  gluing the parts associated with $\regA$ across the copies in a replica $\mathbb{Z}_n$ symmetric manner. This boundary geometry provides the asymptotic boundary conditions for our AdS gravity dual, which we seek to determine. In what  follows we will adhere to the terminology of \cite{Colin-Ellerin:2020mva} referring to the $n$-fold replica boundary geometry as  the branched cover spacetime $\bdy_n$, its dual bulk gravity stationary phase solution as the covering space geometry, $\bulk_n$, and the quotient of the bulk geometry by the $\mathbb{Z}_n$ replica symmetry as the fundamental domain, $\widehat{\bulk}_n = \bulk_n/\mathbb{Z}_n$.

The boundary and bulk spacetimes are composed of elementary building blocks which are the ket ($\bdyk$ and $\bulkket$) and bra components ($\bdyb$ and $\bulkbra$), which we indicate with $k$ and $b$ superscripts, respectively. We will be interested in computing the R\'enyi (or swap) entropy, which will be obtained from the stationary phase evaluation of the gravitational path integral. The $n^\text{th}$ R\'enyi entropy will be given by
\begin{equation}\label{eq:osRenyi}
\begin{split}
S^{(n)}
&=
	\frac{1}{1-n} \, \log \left( \frac{\mathcal{Z}[\bdy_n]}{\mathcal{Z}[\bdy]^n}\right) = \frac{1}{n-1} (I_n - n\, I_1) \,,\\
I_n &:=
	-\log \mathcal{Z}[\bdy_n] =
	\begin{cases}
	  S^E_\text{gr}[\bulk_n]  \,, \qquad \quad \; \text{Eulidean} \\
	  -i\,S_\text{gr}[\bulk_n] \,, \qquad  \text{Lorentzian}
	 \end{cases}
\end{split}
\end{equation}	
where $\bdy=\bdy_1$. The Lorentzian action with the general time-ordering necessary to compute replica path integrals takes a Schwinger-Keldysh form:
\begin{equation}\label{eq:SLorentz}
S_\text{gr}[\bulk_n] = S^k_\text{gr}[\bulk_n] -  S^b_\text{gr}[\bulk_n]  \,,
\end{equation}	
where we have forward evolution for the `kets' ($k$) and backward evolution for the `bras' ($b$), resulting in the relative sign above. As argued in \cite{Colin-Ellerin:2020mva}  (and earlier in \cite{Dong:2016hjy,Marolf:2020rpm}), the on-shell action $I_n$ in the Lorentzian context is real, and is given by
\begin{equation}\label{eq:InLorentz}
I_n = 2 \, \Im(S^k_\text{gr}[\bulk_n]) \;\; \Longrightarrow \;\; S^{(n)} = \frac{2}{n-1} \left[ \Im(S^k_\text{gr}[\bulk_n]) - n \, \Im(S^k_\text{gr}[\bulk]) \right],
\end{equation}
where $\bulk=\bulk_1$.
While the general arguments for these statements were presented in our companion paper \cite{Colin-Ellerin:2020mva}, we will verify these statements explicitly in some specific contents herein.

The examples we discuss in the bulk of the paper are the following. In \cref{sec:2dgrav} we examine the computation of R\'enyi entropy in an excited state with a localized dilaton excitation in Jackiw-Teitelboim  (JT) gravity \cite{Jackiw:1984je,Teitelboim:1983ux}. This provides a concrete context to contextualize the general discussion of \cite{Colin-Ellerin:2020mva} and understand the geometry in some detail. To orient the reader we present both the Euclidean approach  as well as the real-time computation, for the state we consider will be time-reversal symmetric, thereby providing a further check on the results we obtain.  In \cref{sec:ads31} and \cref{sec:ads32} we then turn to examples in 2d CFTs starting first with the case of a single-interval in \cref{sec:ads31}. This example has been well studied both  in field theory and gravity  and we  again use it to provide an illustration of the geometry of the real-time gravitational solution. In \cref{sec:ads32} we then turn to a more interesting case,  that of two disjoint intervals in a CFT on $\mathbb{R}^{1,1}$. We first begin by illustrating the geometry and the computation of the second R\'enyi entropy when the two intervals lie on a fixed time slice, and subsequently generalize to the case when the intervals are relatively boosted with respect to each other. We conclude with a brief discussion of other interesting avenues to explore in \cref{sec:discuss}.

We include in the appendices various technical details that enter into our calculations.  \cref{sec:rindreg} computes the Lorentzian on-shell action for a semi-infinite interval in a 2d CFT using a Rindler regulator to contrast with the discussion in the main text. In \cref{sec:lrenNnApp} we give further details for the evaluation of the Lorentzian on-shell action for disjoint intervals supplementing the discussion in \cref{sec:lrennN}.  \cref{sec:22ren} is a quick overview of the Schottky construction of the covering space geometry (both on the boundary and in the bulk) for the computation of second R\'enyi entropy for 2 disjoint intervals.  For this case we present an explicit evaluation of the Euclidean action from the bulk solution in \cref{sec:direct2ren}  (as far as we are aware this computation has not  hitherto been reported in the literature). Finally, \cref{sec:actions} summarizes some familiar sign conventions and useful identities that we employ in the course of our calculation.

\section{A toy model in 2d gravity}
\label{sec:2dgrav}

As our first example, we will consider a two dimensional scenario and examine the real-time contours for computing moments of the density matrix.  The particular example we pick is the ground state of  JT gravity. In Euclidean signature one may prepare this state by considering the  thermal \AdS{2}  geometry with the Euclidean time identified with period $\beta$ and taking $\beta \rightarrow \infty$.  For finite $\beta$ we may also slice open this geometry to expose the thermofield double (or Hartle-Hawking) state $\ket{\text{TFD}(\beta)} $  at temperature $T =\beta^{-1}$   at time $t=0$ (which we can think of as a pure entangled state of two quantum systems, one on each asymptotic boundary of  the Lorentzian geometry). If we focus on one of the boundaries we end up with a thermal density matrix $\rho_{_\beta}(t=0)$ at  temperature $\beta$ by the usual thermofield double construction. The entropy we compute may be viewed as the thermal entropy of this density matrix in the limit $\beta \rightarrow \infty$ or equivalently as the entanglement entropy between the two boundaries  \cite{Azeyanagi:2007bj,Sen:2008yk,Maldacena:2016hyu}. For earlier investigations of entanglement entropy in JT gravity see \cite{Lin:2018xkj,Mertens:2019tcm} and  \cite{Jafferis:2019wkd} which computes the subleading corrections and discusses a Lorentzian interpretation of the Euclidean replica trick.

We will focus on computing the  moments $\Tr(\rho_{_\beta}^n(t=0))$ at $\beta =\infty$.  The geometry computing this is obtained by stringing together $n$-copies of that preparing $\rho_{_\beta}(t=0)$ cyclically  and gluing them together. Once again in Euclidean signature we know the resulting spacetime: the $n$-fold replica geometry is thermal \AdS{2}, albeit now with a thermal circle that is $n$ times larger \cite{Lewkowycz:2013nqa}.

As described in \cite{Colin-Ellerin:2020mva} once one has the ansatz for the geometry $\bulk_n$ which is dual to the $n$-fold replica, we can either work in the covering space, or take a replica $\mathbb{Z}_n$ quotient and work in a single fundamental domain $\widehat{\bulk}_n = \bulk_n/\mathbb{Z}_ n$. In the present example the covering spacetime $\bulk_n$ is simply \AdS{2}. When we take the $\mathbb{Z}_n$ quotient we will obtain the fundamental  domain $\widehat{\bulk}_n$ which has a fixed point of the  $\mathbb{Z}_n$ action at the locus $\fixM = \{x =  t=0\}$. We will describe below the real-time geometry, delineating the various domains of interest, and then proceed to compute the on-shell action. To help orient the reader given that the configuration is time-reversal symmetric about $t=0$ (in fact it is globally static), we will describe both the Euclidean and the Lorentzian constructions and computations therein.

Before proceeding, it is worth recording the actual answer for the moments of the ground state density operator are not all that illuminating.  The ground state entropy in JT gravity is set by the value of the dilaton, and since there is a finite large $\beta$ limit it gives $\Tr(\rho(t)^n) = \Tr(\rho(t))$. Nevertheless, the example is instructive to consider, as it provides for useful illustration of the general issues encountered in real-time replica geometries which are easy to discern and intuit.
\subsection{The Hartle-Hawking state in JT gravity}
\label{secjtgrav:}

The two-dimensional JT gravity is a dilaton-gravity theory with the following action in Lorentz signature:\footnote{We will only quote explicitly the Lorentz signature action for the gravitational dynamics. The Euclidean action is given by $S^E_\text{gr} = - S_\text{gr}$ with the Lagrangian density evaluated on the appropriate signature metric in both cases; see \cref{sec:EHsigns}. The overall negative sign is consistent with the general intuition the Euclidean action is the Hamiltonian for imaginary time evolution. }
\begin{equation}\label{eq:JTact}
\begin{split}
 S_\text{gr}^{JT}
&=
	\frac{\phi_{0}}{16\pi G_N}	\left[\int_{\bulk}\, d^2x\,  \sqrt{-g} \, R+2 \int_{\bdy} dx\, \sqrt{-\gamma} K\right] \\
&\qquad 	
		+ \frac{1}{16\pi G_N}\left[\int_{\bulk} d^2x\, \sqrt{-g}\,  \phi\,  (R+2)+2 \int_{\bdy} \, \sqrt{-\gamma}\, \phi \, (K-1)\right] ,\\
&\equiv
	S_0 + S_\phi\,,
\end{split}
\end{equation}
where $S_0$ is the topological 2d gravity action and $S_\phi$ the dilatonic contribution. The classical equations of motion obtained by varying the dilaton and metric demand
\begin{equation}\label{eq:JTeom}
\begin{split}
& R +2 =0 \,, \qquad   \left( \nabla_{\mu} \nabla_\nu -g_{\mu\nu} \right) \phi=0 \,,
\end{split}
\end{equation}
respectively. We now proceed to solve these in Euclidean signature where the geometries are familiar and thence explain the Lorentz counterparts.

\subsubsection{Replicas in Euclidean signature}
\label{sec:Erep2}

The thermofield double state where the Euclidean time coordinate $t_{_\text{E}}$ has period $\beta$ is simply thermal \AdS{2} by virtue of the first equation in \eqref{eq:JTeom}.  The $n$-fold replica is likewise the same geometry albeit now with the thermal circle being $n$-times larger.

\begin{figure}
\centering
\begin{tikzpicture}
\draw[thick, fill=orange!10] (-2,0) circle [radius=1];
\draw[thick, fill=orange!10] (4,0) circle [radius =3];
\draw[thick,black] (4,0) -- ({4+3*cos(240)}, {3*sin(240)});
\filldraw[draw=black, thick,fill=red!20] ({4 +3*cos(120)}, {3*sin(120)}) -- (4,0) -- (7, 0) arc (0:120:3);
\draw[thick,red, fill=red] (4,0) circle [radius=2pt];
\draw[thick, blue] (4.4,0) arc(0:120:0.4);
\node at (4.25,0.35) [above] {$\color{blue}{\scriptstyle{r=\epsilon}}$};
\node at ({4+1.5*cos(45)}, {1.5*sin(45)}) [above] {$\widehat{\bulk}_3$};
\end{tikzpicture}
\caption{The Poincar\'e disc geometry dual to the thermofield double (or Hartle-Hawking) state of JT gravity and its $n$-fold replica depicted here for $n=3$. We have shaded the single fundamental domain obtained by taking the replica quotient and indicated the interior boundary at $r=\epsilon$ one introduces while computing the on-shell Euclidean action contribution from a single fundamental domain.
}
\label{fig:Eads2}
\end{figure}
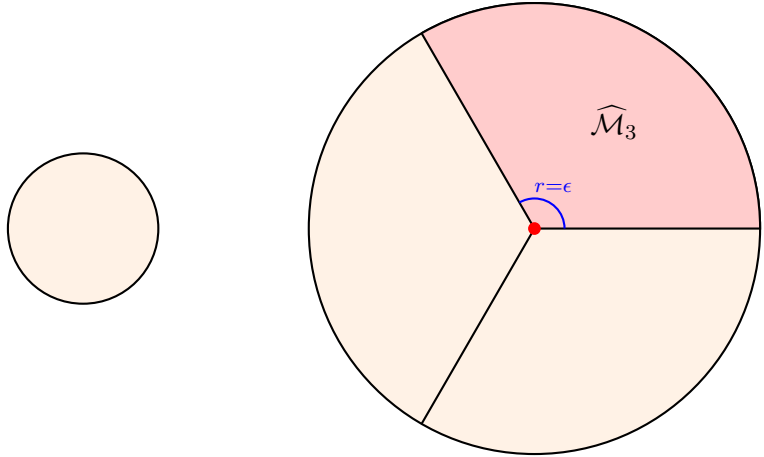

\paragraph{Covering space:} It is useful to write down the geometry using the  Poincar\'e disc model, and parameterize the Euclidean covering space $\bulk_n$ by complex coordinates $z,\bar{z}$ as
\begin{equation}\label{eq:eads2}
ds^2 = \frac{4\,dz\,  d\bar{z}}{(1 - z\, \bar{z})^2}  = 4\, \frac{dr^2 + \frac{1}{n^2} \, r^2\, d\tau^2}{(1-r^2)^2}  \,, \qquad z = r\, e^{i\, \tau/n} \,,
\end{equation}	
with the identification $\tau \sim \tau + 2 \pi \, n $ on the Poincar\'e disc to account for the $n$-fold cover.

A general solution for the dilaton can be easily written down:\footnote{ The easiest way to obtain the solution is to view Euclidean-\AdS{2} as a hyperboloid embedded in $\mathbb{R}^{2,1}$. The embedding coordinates are $\{X_0, X_{\pm 1}\} $ with the mapping
\[
	 X_0 = -i\, \frac{z-\bar{z}}{1-z\bar{z}} \,, \qquad  X_{-1} = \frac{1+z\bar{z}}{1-z\bar{z}}\,, \qquad X_1 = \frac{z+\bar{z}}{1-z\bar{z}}
\]
to the Poincar\'e model. It is easy to see that \eqref{eq:JTeom} requires $\phi = \alpha_{-}\, X_{-1} + \alpha_0\, X_0 + \alpha_{+}\, X_{1}$. }
\begin{equation}\label{eq:dilsolgen}
\phi=\frac{1}{1-z \bar{z}} \big[\alpha_- (1+z \bar{z})-i \, \alpha_0\,  (z-\bar{z})+ \alpha_+\,  (z+\bar{z}) \big] .
\end{equation}	
The covering space is an $n$-fold branched cover over a single Euclidean-\AdS{2} geometry;  we will require that the fields respect the replica $\mathbb{Z}_n$ symmetry which acts by $\tau\to \tau+2\pi$. The dilaton solution \eqref{eq:dilsolgen} will be admissible only it is invariant under $z \to z\, e^{2\pi i/n}$. This forces $\alpha_+ = \alpha_0 =0$ and thus the solution for the dilaton in covering space \AdS{2} is simply
\begin{equation}
\phi=\alpha \, \frac{1+z \bar{z}}{1-z \bar{z}} \,,
\end{equation}
where we have renamed $\alpha_- \to \alpha$ for simplicity.

\paragraph{A single fundamental domain:}  The  $\mathbb{Z}_n$ replica symmetry acts  on this geometry  by $\tau \to \tau  + 2\pi$, or equivalently as $z \to z\, e^{2\pi i  /n}$. Consequently, we can let $ v = z^n$  be coordinates on a single fundamental domain \AdS{2}$/\mathbb{Z}_n$. On the quotient space the metric and dilaton are then given by
\begin{subequations} \label{eq:eads2fd}
\begin{align}
ds^2
&
	= \frac{4\left(v\bv\right)^{\frac{1-n}{n}}}{n^2\left(1-(v\bv)^{\frac{1}{n}}\right)^2}dvd\bv
	=4\, \frac{n^2\, dr^2 +r^2\, d\tau^2}{n^2(1-r^2)^2} \,,
\label{eq:eads2fdmet}\\
\phi& =
	\alpha\frac{1+(v \bv)^{\frac{1}{n}}}{1-(v \bv)^{\frac{1}{n}}} \,.
\label{eq:eads2fddil}	
\end{align}
\end{subequations}
We have depicted the replica geometries of interest in Euclidean signature in \cref{fig:Eads2}. In what follows we will find it more convenient to use a Cartesian chart for the fundamental domain, so will let
$\{v,\bv\} \equiv \{x +i\, t_{_\text{E}}, x-i\, t_{_\text{E}}\}$.

\subsubsection{Lorentz signature replicas}
\label{sec:Lrep2}

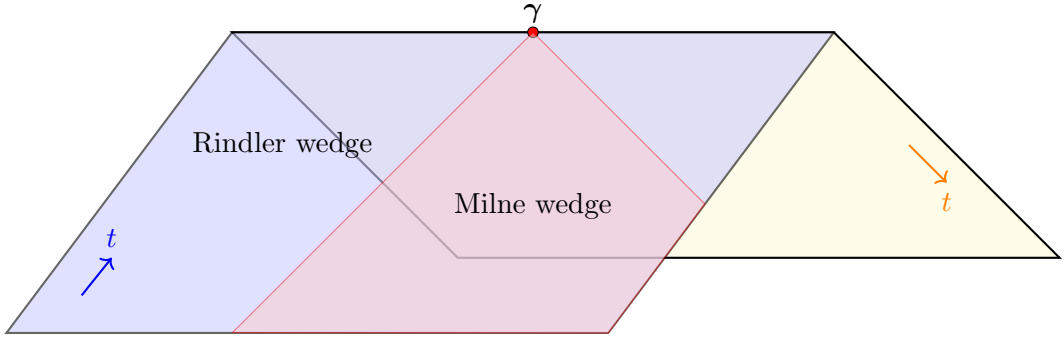
\begin{figure}[h]
\begin{center}
\begin{tikzpicture}[scale=1]
\draw[thick,black, fill=yellow!10] (-4,0) -- ++ (8,0) -- ++(3,-3) -- ++(-8,0) --cycle;
\draw[thick,black, fill=blue!20,opacity=0.6] (-4,0) -- ++ (8,0) -- ++ (-3,-4) -- ++(-8,0) -- cycle;
\draw[fill=red] (0,0) circle [radius=2pt] node [above] {$\fixM$};
\draw[thin,red,fill=red!20,opacity=0.5]  ({-5.65*cos(45)}, {-5.65*sin(45)}) --  (0,0) --  ({3.23*cos(45)}, {-3.23*sin(45)}) -- (1,-4) -- cycle;
\node at (0,-2) [below] {Milne wedge};
\node at (-2,-1.5) [left] {Rindler wedge};
\draw[thick,blue,->] (-6,-3.5) -- (-5.6,-3) node [above] {$t$};
\draw[thick,orange,->] (5,-1.5) -- (5.5,-2) node [below] {$t$};
\end{tikzpicture}
\caption{The domains in the Lorentzian geometry dual to a single fundamental domain $\widehat{\bulk}_n$. We have indicated both the `ket' and `bra' components of the spacetime ${\sf M}^k$ and ${\sf M}^b$ which are each a copy of the \AdS{2} geometry past of the Cauchy slice at $t=0$. The geometry $\widehat{\bulk}_n$ has a fixed point locus of the replica $\mathbb{Z}_n$ action at the splitting surface $\fixM$. The ket and bra geometries are real in the Rindler wedges, regions spacelike separated from $\fixM$, but are complex in the Milne wedge, the causal past of $\fixM$. }
\label{fig:Lads2}
\end{center}
\end{figure}

In Lorentz signature we work with coordinates $\{t,x\}$ with light-cone like combinations $\tx^\pm = x\pm t$ which are adapted to be positive in the spacelike domain as they will be natural analytic continuations of Euclidean variables. The metric in the covering space is that of \AdS{2} itself, with no identifications. It is more interesting to examine the geometry in a single fundamental domain. Owing to the time translational symmetry of the background, we may analytically continue and obtain the metric and dilaton profiles on $\widehat{\bulk}_n$ to be:
\begin{subequations} \label{eq:lads2fd}
\begin{align}
ds^2
&=
	\frac{4\left(\tx^+\tx^-\right)^{\frac{1-n}{n}}}{n^2\left(1-(\tx^+\tx^-)^{\frac{1}{n}}\right)^2} \ d\tx^+ d\tx^- \,,
\label{eq:lads2fdmet}		\\
\phi
&=
	\alpha\, \frac{1+(\tx^+ \tx^-)^{\frac{1}{n}}}{1-(\tx^+ \tx^-)^{\frac{1}{n}}} \,.
\label{eq:lads2fddil}	
\end{align}
\end{subequations}

The metric and dilaton profile in \eqref{eq:lads2fd} clearly solves \eqref{eq:JTeom}. However, it remains to fix the value of $\alpha$.  Since we wish to model the ground state, we should impose a positive frequency condition as described in \cite{Colin-Ellerin:2020mva}.  But \eqref{eq:lads2fd} is not positive frequency, so the only allowed solution is $\alpha =0$ for which $\phi=0$ everywhere.  This is somewhat degenerate in our description, but we can certainly study the limit $\alpha \rightarrow 0$ for all replica numbers $n$.  Note that this is in fact precisely the way in which our Euclidean analysis was performed.

One can add excitations to this state by allowing for time-dependent sources to be turned on in the real-time evolution. In this example one can give a clear picture of the  positive-frequency boundary conditions necessary to define the initial state $\rho_0$.  Let us consider this for finite $\alpha$, after which we can again take the limit $\alpha \rightarrow 0$. A  massless scalar field $\Phi$, by virtue of its conformal invariance satisfies the standard wave equation $ (-\partial^2_t+\partial^2_x)\Phi=\delta (t-t_0,x)$ in the \AdS{2} geometry. The solution in the presence of the source term will
be given by
\begin{equation}
\Phi(t,x)=\frac{1}{2\pi} \, \int \frac{-e^{-i\omega (t-t_0)-ik x}}{-\omega^2+k^2} d\omega dk\,.
\end{equation}
The positive frequency mode here can be isolated by an  $i\varepsilon$ prescription; we pick the $\omega=-|k|$ pole when integrating over $\omega$. The result is the familiar retarded solution for the scalar field ($\gamma$ is the  Euler-Mascheroni constant)
\begin{equation}
\begin{aligned}
\Phi(t,x) = i\operatorname{sgn}(t-t_0) \left(\frac{1}{2}\log \left( (t-t_0)^2-x^2\right)+\gamma\right) .
\end{aligned}
\end{equation}
We will not be considering excitations of the thermofield double state for simplicity, but the above analysis makes clear that we can easily add additional fields coupled gravitationally and study their effects.

Let us examine  the Lorentzian geometry: the metric \eqref{eq:lads2fdmet} describes the metric on the `ket' part of a single fundamental domain which we denote as ${\sf M}^k$, see \cref{fig:Lads2}. As described in \cite{Colin-Ellerin:2020mva} the cyclic $\mathbb{Z}_n$ replica symmetry together with the ${\sf CPT}$ symmetry that exchanges the bra and ket ${\sf M}^k \leftrightarrow {\sf M}^b$ requires that the geometric configurations be real in the homology wedge which is the region of spacetime spacelike separated from the fixed point locus $\fixM$, also referred to as the \emph{splitting surface} \cite{Colin-Ellerin:2020mva}.  Since the fixed point locus $\fixM$ in the present case is at $x = t = 0$, the homology wedges are the past Rindler wedges $\abs{x} > \abs{t}$ with $t \leq 0$. This is ensured in \eqref{eq:lads2fd} by the choice of analytic continuation: $\tx^\pm$ are both positive in the right Rindler wedge, and both negative in the left Rindler wedge. However, the solution is complex in the Milne wedge, the causal past of $\fixM$ where  $\tx^->0$ and $\tx^+<0$. Additionally, we need to choose $\alpha$ to be real owing to the $\mathbb{Z}_2 $ symmetry at $t=0$. This may be achieved by our choice of the initial state.

We can exhibit a manifestly real form of the configuration in the right Rindler wedge by the following coordinate transformation:
\begin{equation}\label{eq:RRindc}
t=(n\, \rho)^n \sinh {\sf t}_{_R}, \quad x=(n\,\rho)^n \cosh {\sf t}_{_R} \,, \qquad \rho \in \mathbb{R}_{\geq 0} \; \text{and} \; {\sf t}_{_R} < 0\,.
\end{equation}
which maps \eqref{eq:lads2fd} into
\begin{equation}\label{eq:RRind2fd}
ds^2=4\, \frac{n^2\, d\rho^2-\rho^2\, d {\sf t}_{_R}^2}{(1-n^2\rho^2)^2} \,, \qquad \phi=\alpha \frac{1+n^2\,\rho^2}{1-n^2\,\rho^2}.
\end{equation}
One can pass to the other wedges by effectively rotating ${\sf t}_{_R}$ by a phase as we cross the past horizon of $\fixM$,  with the result,
\begin{equation}
\begin{aligned}
\text{left Rindler wedge}:\quad
 t &= (n\rho)^n\, \sinh {\sf t}_{_L} \,, && x = -(n\rho)^n \,	 \cosh {\sf t}_{_L} \,, && \rho \in \mathbb{R}_{\geq 0} \; \text{and} \; {\sf t}_{_L} < 0\,,\\
\text{lower Milne wedge}:\quad
 t &= - (i\, n\rho)^n\, \cosh {\sf t}_{_M} \,, && x = (i\, n\rho)^n \,	 \sinh {\sf t}_{_M} \,, && \rho \in \mathbb{R}_{\geq 0} \; \text{and} \; {\sf t}_{_M} \in \mathbb{R}\,.\\
\end{aligned}
\end{equation}

\subsection{The R\'enyi entropy computation}
\label{sec:renyi2d}

Now that we have our replica spacetime we need to evaluate the on-shell action. We will first do so in the Euclidean setting just to remind ourselves of the expected answer, and then proceed with the real-time computation.

\subsubsection{Euclidean action calculation}
\label{sec:Eren2d}

The on-shell Euclidean action we need to evaluate is
\begin{equation}
\mathcal{Z} = e^{-I} =  e^{-S^E_\text{gr}} \big|_{\text{on-shell}} = e^{-S_0 -S_\phi}\big|_{\text{on-shell}} \,.
\end{equation}	
Recall that the counter-terms are designed to make the action finite, and recall also that our limit $\alpha \rightarrow 0$ sends $\phi \rightarrow 0$ everywhere.  Thus $\lim_{\alpha \rightarrow 0} S_\phi =0$.  It thus remains only to evaluate the contribution from $S_0$.  

The boundary conditions we need are that the radial coordinate is cut-off at $r=r_c$ and the proper length of the boundary thermal circle is $\beta/\epsilon$ with the boundary value of the dilaton being
$\phi_b = \phi_c/\epsilon$.

In this example it is simplest to work in the covering space, where $S_0$ can be trivially evaluated. One simply notes that the Gauss-Bonnet theorem gives us the gravitational contribution to be the Euler character of a disc, and hence
\begin{equation}
S_0\big|_\text{on-shell} =  - \frac{\phi_0}{16\pi G_N} \times 4\pi =- \frac{\phi_0}{4\, G_N}\,.
\end{equation}	
One can also directly verify this result by computing the Einstein-Hilbert and Gibbons-Hawking terms in $S_0$ separately with a radial cut-off at $r_c$ and the thermal periodicity as required.  One has the extrinsic curvature $K = \frac{1+r_c^2}{2\,r_c}$ for the constant $r=r_c$ slice and thus
\begin{equation}
\begin{aligned}
S_0\big|_\text{on-shell}
&=
	-\frac{\phi_{0}}{16\pi G_N} \left[\int_{\bulk_n} \, d^2x\, \sqrt{g}\,  R+2 \int_{\bdy_n}\, dx \,  \sqrt{\gamma}\,  K\right] \\
&=
		-\frac{\phi_{0}}{16\pi G_N} \times \left[\, \int_0^{r_c} \frac{4r\, dr}{n\, (1-r^2)^2} \times (-2)
		+2\, \frac{2r_c}{1-r_c^2}  \frac{1+r_c^2}{2\,n\,r_c}\right] \times \int_0^{2\pi n} \,d\tau \\
&= -\frac{\phi_{0}}{4\, G_N}\,.
\end{aligned}
\end{equation}

In principle there is a further contribution from the dilaton action (the Schwarzian term). For the thermofield double state at $\beta \to \infty$ however this can be checked to vanish at tree level (Schwarzian fluctuations will give the near-extremal result \cite{Maldacena:2016hyu}). 

Let us also check the result directly by working in a single fundamental domain. We will again use the Gauss-Bonnet theorem, but we will be careful to excise the contribution from the cosmic brane, the singular codimension-2 locus of the replica $\mathbb{Z}_n$ symmetry fixed point at $r=0$. The fastest way to proceed is to excise a  disc $\mathscr{D}_\epsilon$ of radius $r= \epsilon  $ around the origin. One then computes $S_0$  in terms of the Euler character of the resulting annulus and the contribution from the inner boundary term at $r= \epsilon$ which is another copy of the Gibbons-Hawking term now on a circle of radius $\epsilon$.   To wit,
\begin{equation}
\begin{split}
S_0\big|_\text{on-shell}
&=
	-n\, (S_0)_{\text{fund}} \\
&=
	-n\, \frac{\phi_0}{16\pi G_N} \left( \int_{\widehat{\bulk}_n} \, d^2x \, \sqrt{g}\, R + 2\,\int_{\bdy} \, dx \, \sqrt{\gamma} \, K -2 \, \int_{r=\epsilon} dx\, \sqrt{h}\, K \right)	\\
&
	= -\frac{n\, \phi_0}{16\pi G_N} \left(  0-  2  \left(\frac{2\epsilon}{1-n^2\epsilon^2} \right) \left( -\frac{1+n^2\epsilon^2}{2\, n\, \epsilon}\right)\int_0^{2\pi}\, d\tau \right) \\
&=-\frac{\phi_0}{4\,G_N} \,.
\end{split}
\end{equation}
where we used the fact that the Euler characteristic of the annulus vanishes and  $K =-\frac{1+n^2\epsilon^2}{2\, n\, \epsilon} $ on the regulating surface at $r=\epsilon$ (note the change in orientation of the normal gives us an extra negative sign).  

With the on-shell action at hand we can compute the $n^\text{th}$  R\'enyi entropy for the thermofield double (Hartle-Hawking) state. Since $I_n = I_1$ it immediately follows from \eqref{eq:osRenyi} that
\begin{equation}\label{eq:renHH2d}
S^{(n)}  = \frac{\phi_0}{4\,G_N}\,.
\end{equation}	
which is the promised temperature independent answer.

\subsubsection{Lorentzian action calculation}
\label{sec:Lren2d}

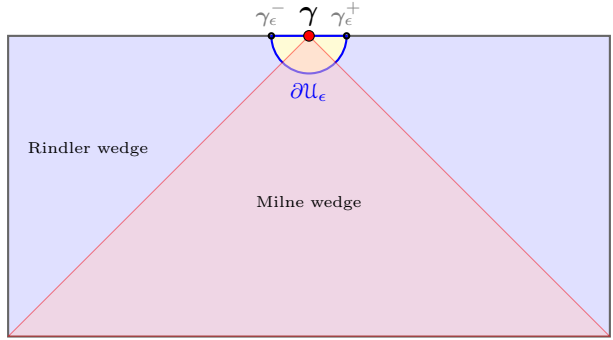
\begin{figure}[h]
\begin{center}
\begin{tikzpicture}[scale=1]
\draw[thick,black, fill=blue!20,opacity=0.6] (-4,0) -- ++ (8,0) -- ++ (0,-4) -- ++(-8,0) -- cycle;
\draw[thick,blue, fill=yellow!20] (-0.5,0) arc(-180:0:0.5) -- cycle;
\draw[thin,red,fill=red!20,opacity=0.5]  ({-5.65*cos(45)}, {-5.65*sin(45)}) --  (0,0) --  ({5.65*cos(45)}, {-5.65*sin(45)}) -- cycle;
\node at (0,-2) [below] {\tiny{Milne wedge}};
\node at (-2,-1.5) [left] {\tiny{Rindler wedge}};
\draw[thick,black] (0.5,0) circle[radius=1pt];
\draw[thick,black] (-0.5,0) circle[radius=1pt];
\draw[fill=red] (0,0) circle [radius=2pt] node [above] {$\fixM$};
\node at (0,-0.5) [below] {$\color{blue}{\scriptstyle{\partial \mathscr{U}_\epsilon}}$};
\node at (-0.5,0) [above]{$\color{gray}{\scriptstyle{\fixM_\epsilon^-}}$};
\node at (0.5,0) [above]{$\color{gray}{\scriptstyle{\fixM_\epsilon^+}}$};
\end{tikzpicture}
\caption{The  geometry  in the vicinity of the splitting surface $\fixM$  in the Lorentzian geometry dual to a single fundamental domain $\widehat{\bulk}_n$. We have excised a neighbourhood $\mathscr{U}_\epsilon$  of $\fixM$ with boundary $\partial \mathscr{U}_\epsilon$ to regulate the contribution from the fixed point locus. We take $\partial \mathscr{U}_\epsilon$ to be parametrized by an arbitrary curve $\tx^+ = U(\tx^-)$ in the $\tx^\pm$ plane. }
\label{fig:Lads2wedges}
\end{center}
\end{figure}

Let us now compute the result for the on-shell action in Lorentz signature.  Again, the limit $\alpha \rightarrow 0$ sends $\phi \rightarrow 0$ at all points, so we should understand $S_\phi$ as vanishing in the limit.  To compute the gravitational contributions, we will work in a single fundamental domain. Recall that the metric on $\widehat{\bulk}_n$ is given by \eqref{eq:lads2fd}. We will organize the computation as follows: $\widehat{\bulk}_n$ has two components ${\sf M}^k$ corresponding to the forward evolution of the ket and ${\sf M}^b$ corresponding to the backward evolution of the bra. The direction of time evolution being reversed in the two, one needs to compute as described in \eqref{eq:InLorentz}
\begin{equation}\label{eq:Lfundomos}
S^\text{fund}_n(\widehat{\bulk}_n) = \frac{1}{n} \left[S_{\text{gr},n}^k - S_{\text{gr},n}^b\right] \;\; \Longrightarrow \;\;
I_n = -i\,n\, S^\text{fund}_n = 2\, n\, \Im(S_\text{gr,fund}^k)\,.
\end{equation}	

We can thus focus on computing the imaginary part of $S^k_\text{gr,fund}$ from the ket.  In implementing this computation, we will organize the pieces in the following manner: we first excise a region $\mathscr{U}_\epsilon$ around $\tx^\pm = 0$, the fixed point locus $\fixM$ with boundary $\partial \mathscr{U}_\epsilon$. This cut-off region with the topology of a disc, intersects the Cauchy slice at $t=0$ on two corners $\fixM_\epsilon^\pm$, respectively, as depicted in \cref{fig:Lads2wedges}. We will take $\partial \mathscr{U}_\epsilon$ to be parameterized by a function $\tilde{x }^+ = U(\tx^-)$. We can implement the Gauss-Bonnet theorem on the lower-half plane after excising $\mathscr{U}_\epsilon$, provided we include a boundary term at the excision surface $\partial \mathscr{U}_\epsilon$ and the corner terms where this cut-off region meets the Cauchy surface at $t=0$. Specifically, focusing on the gravitational contribution of the JT action \eqref{eq:JTact} we have
\begin{equation}\label{eq:JTactgr}
\begin{split}
S_0
&=
	\frac{\phi_{0}}{16\pi G_N}	\left[\int_{\bulk}\, d^2x\,  \sqrt{-g} \, R+2 \int_{\bdy} dx\, \sqrt{-\gamma} K\right] \\
&=
	\frac{\phi_{0}}{16\pi G_N}	\left[4\pi\, \chi - 2 \int_{\partial \mathscr{U}_\epsilon} dx\, \sqrt{h} \, K - S_\text{corner}\right] ,
\end{split}
\end{equation}	
with $\chi$ being the Euler character. The bulk term encoded in $\chi$ does not give any imaginary contribution -- these are completely subsumed into the Gibbons-Hawking term on the Cauchy slice and the corner term. We will evaluate these in turn.

To facilitate the computation for the metric \eqref{eq:lads2fdmet} let us write the prefactor as $\sigma(\tx^+, \tx^-)$ and compute the extrinsic curvature of the surface $\partial \mathscr{U}_\epsilon$. Given the normal vector
\begin{equation}
n^\mu \partial_\mu = \sqrt{\frac{U'(\tx^-)}{\sigma}}\, \pdv{\tx^+} - \sqrt{\frac{1}{\sigma \, U'(\tx^-)} } \, \pdv{\tx^-} \,,
\end{equation}	
one finds:
\begin{equation}
K =
	\frac{1}{2\, \left(U'\, \sigma\right)^\frac{3}{2}} 	
	\left[ \sigma\, U'' - U' \left( \pdv{\sigma}{\tx^-} - U'\, \pdv{\sigma}{\tx^+}\right)
\right] .
\end{equation}	
Factoring in the induced measure $\sqrt{h} = \sqrt{\sigma\, U'}$ we end up with the Gibbons-Hawking contribution evaluating to
\begin{equation}
2 \int_{\mathscr{U}_\epsilon} dx\, \sqrt{h} \, K = \int\, d\tx^- \, \left(\frac{U''}{U'} - \pdv{\log \sigma}{\tx^-} + U'\, \pdv{\log \sigma}{\tx^+}
\right) \equiv T_0 + T_- +T_+\,,
\end{equation}	
where we have chosen to split the integrand and label the three integrals as $T_{0,\pm}$ for convenience.
We now note the following $T_0$, which is an integral of our cut-off function $U(\tx^-)$ alone, can  be seen to be purely real. We can pick for instance a smooth function and realize that the integral is over some domain of the form: $\tx^- \in [-\delta, x_* + \delta]$ with $x_*$ being a zero locus of $U(x)$ and $\delta >0$.  Important to this argument is the fact that the integrand can be made a regular function of $\tx^-$. Furthermore,
\begin{equation}
T_+ = \int \, dx^+\, \pdv{\log \sigma}{\tx^+} \,,
\end{equation}	
which is obtained by a $\tx^- \leftrightarrow \tx^+$ swap from $T_-$ and we record that $\sigma(\tx^-, \tx^+))$ is a symmetric function. We will see below that $\Im(T_- + T_+) = 2\Im(T_-)$, so we will simply focus on its evaluation for now.

Plugging in the conformal factor $\sigma$ from \eqref{eq:lads2fdmet} we have
\begin{equation}
T_-  = -\int\, d\tx^- \, \pdv{\log \sigma}{\tx^-}  = \int \, \frac{d\tx^-}{n\, \tx^-}\,  \left( 1+n - \frac{2}{ 1-\left(\tx^+ \tx^-\right)^\frac{1}{n} }  \right).
\end{equation}
We see that the integral over $\tx^-$ has a pole at the origin which needs to be accounted for. We will do so using an $i\epsilon$ regulator and defining the integrand by a principal value prescription. Recall,
\begin{equation}\label{eq:prinvalue}
\frac{1}{x\pm i\epsilon} = \mathcal{P}\frac{1}{x}  \mp i\, \pi \delta (x)\,.
\end{equation}	
The natural choice of the contours is such that $\tx^- \to \tx^- + i\epsilon$ \cite{Colin-Ellerin:2020mva}. We then have
\begin{equation}
\begin{split}
&T_- =
	\frac{1}{n} \, \int d\tx^- \, \left[ \mathcal{P}\frac{1}{\tx^-}  \mp i\, \pi \delta (\tx^-)\right]
	\left( 1+ n -  \frac{2}{ 1-\left(U(\tx^-) \,\tx^-\right)^\frac{1}{n} } \right) ,  \\
& \Longrightarrow \;\;
\Im(T_-) = -\frac{1}{n} (n-1) \,\pi \,.
\end{split}
\end{equation}	
In evaluating the integral we have finally restricted to the cut-off surface and used the smoothness of $U(x)$ to obtain the final result.
The evaluation of $T_+$ proceeds similarly with the $i\epsilon$ prescription reading now $\tx^+ \to \tx^+ -i\epsilon$. The relative sign of the $i\epsilon$ implies that the imaginary part from $T_-$ is doubled, so that
\begin{equation}\label{eq:LGB2}
\Im\left[2 \int_{\partial \mathscr{U}_\epsilon} dx\, \sqrt{h} \, K \right]  = 2 \pi \left(\frac{1}{n} -1 \right) .
\end{equation}	

The final piece we need is the corner term where the spacelike Cauchy surface $\Cbulk$ intersects with the chosen cut-off $\partial \mathscr{U}_\epsilon$. As explained in \cite{Colin-Ellerin:2020mva} this contribution arises when the regulator surface  $\partial \mathscr{U}_\epsilon$ does not intersect the Cauchy surface orthogonally.\footnote{We pause to note here that these contributions have been discussed earlier in \cite{Neiman:2013ap}  (in the context of applications to black hole entropy computations) and were treated in full generality quite elegantly in \cite{Jubb:2016qzt}. We also note its use in the holographic entanglement entropy computations in  \cite{Dong:2016hjy}.}  For our purposes we simply need to know that the  integral of the extrinsic curvature along the boundary in two dimensions is the same as adding up the infinitesimal rotation angles of the normal $n^\mu$. At the corner the boost angle associated  with the normal vector jumps by a factor $i\,\frac{\pi}{2}$ as originally computed in \cite{Farhi:1989yr}.
Specifically, at each corner $\fixM^\pm$ we get a contribution from the relative boost that arises in going from the ket  to the bra component  $\bulkket$ of $\widehat{\bulk}_n$\footnote{ There is a useful heuristic for this calculation which underlies the complex Gauss-Bonnet theorem employed in \cite{Louko:1995jw} -- the cut-off surface has to pass from the timelike Milne region to the spacelike Rindler region and each crossing involves a $i\frac{\pi}{2}$ jump in the normal (see also \cite{Neiman:2013ap}). This is the piece we pick up in the corner contribution if we have a non-orthogonal intersection at the Cauchy slice; see Appendix A of \cite{Colin-Ellerin:2020mva} for a brief discussion. }
\begin{equation}
\int \sqrt{-h} K = \cosh^{-1} \left(n^{\sf{k}}_\epsilon \cdot n^{\sf{b}}_\epsilon\right) =  i\, \frac{\pi}{2} \,.
\end{equation}
We have two corners $\fixM_\epsilon^\pm$ with opposing orientations and hence
\begin{equation}\label{eq:Lcorner2}
\Im(S_\text{corner}) = \Im\left(2\int_{\fixM_\epsilon^+}\, \sqrt{-h}\, K  + 2\int_{\fixM_\epsilon^-}\, \sqrt{-h}\, K\right)  = 2\pi\,.
\end{equation}	
Adding all the contributions from \eqref{eq:LGB2} and \eqref{eq:Lcorner2}, we get the full Lorentzian action,
\begin{equation}
I_n = -2\, n\, \frac{\phi_0}{16\pi\,G_N} \Im \left[2 \int_{\partial \mathscr{U}_\epsilon}\, \sqrt{h}\, K  + S_\text{corner} \right]=
-\frac{\phi_0}{4\,G_N } \,.
\end{equation}
This indeed is the expected answer for one immediately recovers from the above the result for the $n^\text{th}$ R\'enyi entropy obtained from the Euclidean computation \eqref{eq:renHH2d}.

\section{R\'enyi entropies in 2d CFTs: A single interval}
\label{sec:ads31}

As our next example we will examine the much studied example of a single-interval R\'enyi entropy in the vacuum state of a two dimensional conformal field theory on the plane. This computation which was first carried out in \cite{Holzhey:1994we} and re-examined in \cite{Calabrese:2004eu} exploits the fact that the computation of the R\'enyi entropies can either be viewed as the computation of the partition function on a $n$-folded branch cover, or equivalently as the correlation function of $\mathbb{Z}_n$ twist operators. The key point is that the $n$-fold branched cover of the complex plane is a genus-zero Riemann surface which can be uniformized by a simple map.

To be concrete let us consider the CFT$_2$ on $\mathbb{R}^{1,1}$ and let $\regA$ be a codimension-1 spacelike region on some Cauchy surface with  $\entsurf$ comprising of two-points $a_1 = (0,0)$ and $a_2 = (t_0,x_0)$ with $t_0< x_0$. The CFT computation gives ($\delta$ is a UV regulator)
\begin{equation}\label{eq:ren2d1int}
\begin{split}
S_\regA^{(n)}
&=
	\frac{1}{1-n}\, \log\,\Tr(\rhoA^n)
	= \frac{1}{1-n}\, \log\,\expval{\mathcal{T}_n(a_1)\, \mathcal{T}_{-n}(a_2)}  \\
&
	=  \frac{c}{12} \left(1+\frac{1}{n}\right) \log\left(\frac{\abs{a_2-a_1}^2}{\delta^2}\right)
	= \frac{c}{12}\left(1+\frac{1}{n}\right)   \log\left(\frac{x_0^2 - t_0^2}{\delta^2}\right) .
\end{split}
\end{equation}	
Here $\mathcal{T}_n$, $\mathcal{T}_{-n}$ are the $\mathbb{Z}_n$ twist operators and we have exploited the fact that the partition function on the $n$-fold cover $\bdy_n$ can be mapped to a two-point function of these twist operators.

We would like to reproduce this answer from a gravity computation. We will take the bulk theory to be Einstein-Hilbert gravity in \AdS{3} which has by the classic analysis of \cite{Brown:1986nw} an asymptotic Virasoro symmetry with central charge $c= \frac{3\lads}{2\, G_N}$. We will use this relation explicitly and rewrite the strength of the gravitation interaction $\frac{\lads}{16\pi\, G_N} = \frac{1}{24\pi} \,c$.\footnote{ We will set $\lads=1$ in most of our analysis below, but will  quote the result in terms of the dimensionless CFT central charge.}

\subsection{The boundary replica geometry}
\label{sec:repbdy31}

Let us first examine the boundary replica geometry in Euclidean signature obtained by Wick rotating $t \to -i\,t_{_\text{E}}$.\footnote{For any $t_0<x_0$ we can pick a Cauchy surface of  $\mathbb{R}^{1,1}$ to be defined by $\frac{x}{x_0} = \frac{t}{t_0}$ -- its normal is a timelike vector: $x_0\, \pdv{t} + t_0 \, \pdv{x}$. We can Wick rotate this vector  and obtain the Euclidean spacetime of interest. It is simpler to visualize the case when $t_0 = 0$. However, for the $\text{SL}(2)$ invariant CFT$_2$ vacuum, all foliations by slices of constant $-\frac{x}{x_0} + \frac{t}{t_0}$ are equivalent by the underlying boost invariance. } The original geometry $\bdy$ is the complex plane with coordinates $\{v = x + i\, t_{_\text{E}},\bv = x-i\,t_{_\text{E}}\}$, and hence the branched cover replica space $\bdy_n$ is topologically a sphere, with branch points at $a_1$ and $a_2$ where it has  a conical excess  given by $2\pi(n-1)$. Let $z$ be the complex coordinate on the covering space.  The complex structure on $\bdy_n$, $\frac{v-a_1}{v-a_2}$, defines a uniformization map to the smooth covering space, which itself is a complex plane with coordinate $z$ defined by
\begin{equation}\label{eq:zuniform}
z = \left(\frac{v-a_1}{v-a_2} \right)^\frac{1}{n}
\end{equation}	
In the $z$-plane the $n$-sheets of the branched cover are mapped to $n$ wedges with opening angle $\frac{2\pi}{n}$ as depicted in \cref{fig:Eads2}.  The uniformization map can be viewed as a conformal transformation since
\begin{equation}\label{eq:zmet1}
dz  d\bar{z} = \Omega^2\, dv\,d\bv \,, \qquad   \Omega^2  \equiv
	 \frac{1}{n^2} \, \frac{\abs{a_2-a_1}^2}{\abs{(v-a_1)^{1-\frac{1}{n}} \, (v-a_2)^{1+\frac{1}{n}} }^2} \,.
\end{equation}	
%

\begin{figure}[h]
\begin{center}
\begin{tikzpicture}[scale=0.7]
\draw[thick,black] (-6,0) -- (6,2) node[right]{$\Cbdy$};
\draw[thin,red,fill=red!20,opacity=0.5]  (-6,-4) -- (-2,2/3) -- (2,-4);
\draw[thin,red,fill=red!20,opacity=0.5]  (-2,-4) -- (2,4/3) -- (6,-4);
\draw[fill=red] (-2,2/3) circle [radius=2pt]  node[above]{$a_1$} ;
\draw[fill=red] (2,4/3) circle [radius=2pt]  node[above]{$a_2$};
\node at (-2.5,-2) [below]{\small{complex}};
\node at (2.5,-2) [below]{\small{complex}};
\node at (0,-3) [below]{\small{real}};
\node at (0,0.2) [below]{\small{real}};
\node at (-5,-1) [below]{\small{real}};
\node at (5,0.5) [below]{\small{real}};
\end{tikzpicture}
\caption{Causal domains on the boundary ket spacetime ${\sf B}^k$ for a two dimensional field theory with the region $\regA$ taken to be a spacelike segment of a boosted Cauchy slice. We indicate the regions where the resulting metric is real and complex, respectively. In general the metric is not guaranteed to be real in regions that are in the causal past of the entangling surface $\entsurf$ which here comprises of the two points $a_1$ and $a_2$.}
\label{fig:Lads31}
\end{center}
\end{figure}
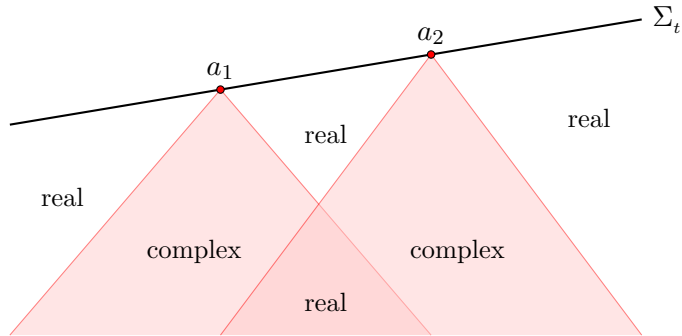

The passage to Lorentz signature can be achieved by the inverse Wick rotation and in terms of our light-cone coordinates $\tx^\pm = x \pm t$, the metric is
\begin{equation}\label{eq:lmet31}
ds^2 = \frac{\abs{a_2-a_1}^2}{n^2}\, \frac{d\tx^+\, d\tx^-}{(\tx^- -a_1)^{\frac{1}{2}-\frac{1}{2n} } (\tx^- - a_2)^{\frac{1}{2}+\frac{1}{2n} } \, (\tx^+ -a_1)^{\frac{1}{2}-\frac{1}{2n} } (\tx^+ - a_2)^{\frac{1}{2}+\frac{1}{2n} } } .
\end{equation}	
Note that the Wick rotation is carried out with respect to the time-coordinate on the base space $\bdy$ where the physical quantum fields reside. The Lorentzian metric on $\bdy_n/\mathbb{Z}_n$ is not real everywhere: it is complex in regions that lie in the causal past of $\entsurf$. For the present example this is the domain that  is timelike separated from one endpoint, but spacelike separated from the other as depicted in \cref{fig:Lads31}.\footnote{
As noted in \cite{Colin-Ellerin:2020mva} the boundary conditions at the asymptotic AdS boundary are specified by a real boundary metric (with conical singularities at the entangling surface).  The reason for the complex metric in \eqref{eq:lmet31} is because we have made a specific choice for the boundary conformal frame which is related to the real boundary metric by a complex Weyl factor.  We have analytically continued the Euclidean boundary geometry  \eqref{eq:zmet1} obtained via the uniformization and it is this choice that is responsible for the complex Weyl factor. }

In arriving at this answer we have used the Euclidean construction of the branched cover as a crutch, but one can verify this directly by taking $n$-copies of the ket and bra spacetimes with the replica gluing conditions. A simple way to see this is to consider a conformal transformation which makes $\regA$ a semi-infinite interval, mapping in the process its past domain of dependence to a Rindler wedge of the resulting Minkowski spacetime (on $\bdy$) \cite{Casini:2011kv}. The $n$-fold cover is obtained by gluing the Rindler wedges of $\regA$ cyclically across the replica bras and kets (while those of $\regAc$ are glued together within the bra-ket combination of each replica copy). The combination of $\mathbb{Z}_n$ replica symmetry and the $\mathbb{Z}_2$ ${\sf CPT}$-conjugation swapping bras and kets, ensures that the resulting spacetime has a real Lorentz signature geometry in the Rindler wedges, but not necessarily so in the Milne wedges \cite{Colin-Ellerin:2020mva}. The example above makes this manifest.

\subsection{The bulk R\'enyi geometries}
\label{sec:repbulk31}

Given the boundary geometry $\bdy_n$ we are tasked with constructing the bulk dual $\bulk_n$. We will first describe the geometry in Euclidean signature and then outline the Lorentzian description. The covering space geometry $\bulk_n$ is simply \AdS{3}, since the $z$-plane is a copy of $\mathbb{C}$.  It is more interesting to examine the geometry of the fundamental domain $\widehat{\bulk}_n$ where the boundary has the conical singularities associated with the branch points.

We will proceed by exploiting the fact that the Fefferman-Graham expansion converges in \AdS{3} (since all geometries are locally diffeomorphic to \AdS{3}). Using the general results of \cite{Skenderis:1999nb} one can write the metric dual to the state of interest in terms of the boundary stress tensor data (this was used by \cite{Hung:2011nu}  to compute holographic  R\'enyi entropies in \AdS{3}). The physical state we are considering on one fundamental domain of the CFT is the state obtained by acting on the vacuum with the twist operators (which thence create the appropriate monodromy around the branch points).

The standard Fefferman-Graham expansion in \AdS{3} with boundary metric $\gamma_{ij}$ and $\rho$ being the Fefferman-Graham radial coordinate, is given as  \cite{Skenderis:1999nb}
\begin{equation}
ds^2=
	\frac{d\rho^2}{4\,\rho^2} + \frac{1}{\rho}
	\left[ \left(1 - \frac{\rho}{4}\, \Tr(T) \right) \delta_i^{\ k} +  \frac{\rho}{4}\, T_i^{\ k}\right]
	\gamma_{kl}\left[\left(1- \frac{\rho}{4}\, \Tr(T) \right) \delta^l_{\ j} +  \frac{\rho}{4}\, T^l_{\ j}\right] \,dx^i \, dx^j \,.
\end{equation}	

Working in the complex coordinates $v,\bv$  the geometry takes the form:
\begin{equation}\label{eq:efgmetads3}
\begin{split}
ds^2
&=	
	\frac{d\rho^2}{4\,\rho^2} + \frac{dv\,d\bv}{\rho}
	- \frac{1}{2} \left[ T_{v\bv}\, dv  d\bv  - T_{vv}\, dv^2 - T_{\bv\bv} \, d\bv^2\right]\\
&\qquad
	+ 	\frac{\rho}{8} \left[ \left( T_{vv}\,T_{\bv\bv} +T_{v\bv}^2\right) dv  d\bv
	- 2\, T_{v\bv}  \left( T_{vv}\,dv^2+T_{\bv\bv} \, d\bv^2\right)\right] .
\end{split}
\end{equation}	

For the case of interest we need to know the boundary stress tensor, which is easily obtained by the conformal map \eqref{eq:zuniform}. One has the result given in terms of the Schwarzian map for the diagonal components, viz.,
\begin{equation}\label{eq:eTd31}
\begin{split}
T_{vv}
&=
	\text{Sch}(z,v) = \frac{1}{2}\, \left( 1-\frac{1}{n^2}\right)\, \frac{\abs{a_2-a_1}^2}{(v-a_1)^2(v-a_2)^2} ,\\
T_{\bv \bv}
&=
	\text{Sch}(\bar{z},\bv) = \frac{1}{2}\, \left( 1-\frac{1}{n^2}\right)\, \frac{\abs{a_2-a_1}^2}{(\bv-a_1)^2(\bv-a_2)^2}\,.
\end{split}
\end{equation}	
The off-diagonal term is instead given by the conformal anomaly term:
\begin{equation}\label{eq:eTvvb31}
T_{v\bv}
=
	 -2\, \partial_v \partial{\bv}\, \log \Omega
=
	\pi \left[ \left(1-\frac{1}{n}\right) \, \delta(\abs{v-a_1})
		+ \left(1+\frac{1}{n}\right) \, \delta(\abs{v-a_2})
	\right].
\end{equation}	
Plugging in these expressions into \eqref{eq:efgmetads3} we obtain the metric on a single fundamental domain $\widehat{M}_n$ in Euclidean signature.

One can exhibit the fact that the Euclidean geometry on $\bulk_n$ is smooth by constructing an explicit diffeomorphism (see \cite{Krasnov:2001cu}) from the
$(\rho, v, \bv)$ coordinates above to a new set of coordinates $(\xi, y, \bar{y})$. All we need is for this diffeomorphism to act as the desired conformal transformation implementing the uniformization.  Explicitly, we have
\begin{equation}\label{eq:FGglMap}
\xi  = \frac{\sqrt{\rho} \, \Omega}{1+ \rho\, \Omega^2 \, \abs{\partial_z \log \Omega}^2} \,, \qquad
y = z  +\frac{\rho \, \Omega^2 \, \partial_{\bar{z}} \log \Omega}{1+ \rho\, \Omega^2 \, \abs{\partial_z \log \Omega}^2} \,,
\end{equation}	
which maps the metric on the covering space to the standard Poincar\'e metric:
\begin{equation}
ds^2 = \frac{d\xi^2+ dy d\bar{y}}{\xi^2} \,.
\end{equation}	
On this covering space the replica $\mathbb{Z}_n$ symmetry acts as $z \to e^{\frac{2\pi i }{n}} \,z$ or equivalently  $y \to e^{\frac{2\pi i }{n}} \,y$. The fixed points of the symmetry are the branch points $v=a_1$ and $v = a_2$ on the boundary, and a bulk locus $\fixM$ which in this particular case is a geodesic that connects the two boundary points. In the regular $(\xi,y,\bar{y})$ coordinates this is the geodesic that connects the north and south poles of the boundary Riemann sphere.

The Lorentzian geometry on the ket part, ${\sf M}^k$,  of a single fundamental domain $\widehat{M}_n$ can be obtained from the
above. One might naively think this is simply an analytic continuation of  the $(v,\bv)$ coordinates. However, we should exercise some care since the naive analytic continuation of the $T_{v\bv}$ component of the stress tensor which has delta function sources would indicate that we have shockwaves propagating along the past-light cones of the branch points. This is incorrect and inconsistent with the boundary conditions of the variational problem described in \cite{Colin-Ellerin:2020mva}. The single fundamental domain has a fixed point locus from the replica $\mathbb{Z}_n$ action, and a complex metric in the causal past of $\entsurf$, but no singularities along the light-cone. Instead the correct metric in real-time is one where we Wick rotate $T_{vv} \to T_{\tx^-\, \tx^-}$ and $T_{\bv\bv} \to T_{\tx^+ \tx^+} $ but  define the analytic continuation of $T_{v\bv} \to T_{\tx^-\tx^+}$ to only have delta function singularities at the fixed point locus. To wit, (with $T_{-+} \equiv  T_{\tx^-\tx^+} $ etc)
\begin{equation}\label{eq:lT31}
\begin{split}
T_{-+}
&=
	2\pi i  \left[ \left(1-\frac{1}{n}\right) \, \delta(\tx^--a_1) \,\delta(\tx^+-a_1)
		+ \left(1+\frac{1}{n}\right) \, \delta(\tx^--a_2) \, \delta(\tx^+-a_2) \right]  , \\
T_{--} \
&=
	\frac{1}{2} \left(1-\frac{1}{n^2}\right) \, \frac{\abs{a_2-a_1}^2}{(\tx^- -a_1)^2\, (\tx^- -a_2)^2} \,,\\
T_{++}
&=
	\frac{1}{2} \left(1-\frac{1}{n^2}\right) \, \frac{\abs{a_2-a_1}^2}{(\tx^+ -a_1)^2\, (\tx^+ - a_2)^2} \,,
\end{split}
\end{equation}	
in terms of which we can parameterize the bulk real-time metric  on ${\sf M}^k$ as
\begin{equation}\label{eq:lfgmetads3}
\begin{split}
ds^2
&=
	\frac{d\rho^2}{4\rho^2} + \frac{d\tx^+\, d\tx^-}{\rho}
	+ \frac{1}{2} \left(- T_{-+}\, d\tx^-\, d\tx^+ + T_{--}\, (d\tx^-)^2 + T_{++}\, (d\tx^+)^2\right) \\
&\qquad
	+\frac{\rho}{8}\left[ \left(T_{--} T_{++} + T_{-+}^2\right)\, d\tx^+ d\tx^-
	- 2\, T_{-+} \left( T_{--} \, (d\tx^-)^2 + T_{++} \, (d\tx^+)^2\right)
	\right]	.
\end{split}
\end{equation}	

The choice of analytic continuation made in \eqref{eq:lT31} is really a question of correctly interpreting the codimension-2 delta functions therein. One can justify this by an integral representation in momentum space.  We recall that the $T_{-+} $ component is determined by the conformal factor $\Omega$ since
\begin{equation}\label{eq:Tpm31justify}
\begin{split}
T_{-+}
&=
	2\, \partial_{-} \partial_{+} \, \log \Omega(\tx^+, \tx^-) \\
&=
 	2\left(1-\frac{1}{n}\right) \, 	\partial_{-} \partial_{+} \log  \sqrt{(\tx^- - a_1) (\tx^+-a_1)}  \\
& \qquad \qquad
		+ 2\left(1+\frac{1}{n}\right) \, 	\partial_{-} \partial_{+} \log  \sqrt{(\tx^- -a_2) (\tx^+-a_2)} .
\end{split}
\end{equation}	
We need to define the argument of the logarithm by analytic continuation, which we do by using a Fourier transform trick. Consider the following  regulated integral which  in Euclidean space, $\vb{x}   \equiv (x, t_{_\text{E}})$,	 provides the standard integral representation of the modified Bessel function of the second kind $K_0( \vb{x} ) = -\log(\abs{\vb{x}}) +\text{constant}$:\footnote{The Pauli-Villars mass term here is introduced to remove the IR divergence. We are also allowing for a constant shift which will not affect the analysis.}
\begin{equation}
\begin{split}
\log(\abs{\vb{x} })
&
	= -\lim_{m \to 0}\, \frac{1}{2\pi} \,
		\int d^{2}\vb{p} \; \frac{e^{i \vb{p} \cdot \vb{x} }}{\abs{\vb{p}}^2+m^2}  \\	
&
	\to -\lim_{m \to 0} \, \frac{i}{2\pi} \, \int d^{2}p \; \frac{e^{ip \cdot x}}{p^{2}+m^{2}-i\tilde{\epsilon}}  \\
&=
	-\lim_{m \to 0} \, \frac{i}{4\pi}\,
		\int dp^{+}dp^{-} \; \frac{e^{-\frac{i}{2}(p^{+}\tx^{-}+p^{-}\tx^{+})}}{p^{+}p^{-}-m^{2}-i\tilde{\epsilon}}\,.
\end{split}
\end{equation}
Using the last line of the expression above it can be checked that one does recover \eqref{eq:lT31} from \eqref{eq:Tpm31justify}.
\subsection{R\'enyi entropies from gravity}
\label{sec:ren31}

We will now outline the computation of the R\'enyi entropies from the bulk geometries constructed in \cref{sec:repbulk31}. We will first revisit the computation in Euclidean signature as before just to set the stage and then proceed to describe how the Lorentzian computation works. The logic we follow will roughly parallel the discussion in \cref{sec:renyi2d} though we now have to deal with the fact that the geometry in a single fundamental domain is more complicated.

\subsubsection{Euclidean on-shell action in a fundamental domain}
\label{sec:eren31}

We will compute the R\'enyi entropies using \eqref{eq:osRenyi}.  As remarked above, we carry out the computation of $I_n$ in a single fundamental domain and  then scale it up to the covering space. In evaluating the fundamental domain action, as explained in \cite{Lewkowycz:2013nqa}, we need to ensure that we do not include the contribution from the cosmic-brane, i.e., from the delta-function singularities arising as a result of taking the quotient. We thus want to evaluate
\begin{equation}\label{eq:eInhIn}
I_n =  S^E_\text{gr}[\bulk_n] = n \,\hat{I}_n \equiv n\,  S^E_\text{gr}[\widehat{\bulk}_n] \bigg|_\text{cosmic brane excised} \,.
\end{equation}	
We will start by outlining the contributions to $S^E_\text{gr}[\widehat{\bulk}_n] $ and then note the pieces that we need to remove to excise the cosmic brane contribution.

The on-shell action in gravity has three distinct contributions: a bulk term from the Einstein-Hilbert action, a boundary Gibbons-Hawking term, and finally boundary counterterms necessary to regulate the divergences. For definiteness we will regulate the spacetime by cutting-off the radial coordinate at $\rho = \rho_c$ and thence take the limit $\rho_c \to 0 $ at the end of the computation. Denoting the induced metric on the cut-off timelike boundary $\bdy_c$ by $\gamma_{\mu\nu}$ we have the action as the sum of the aforementioned three contributions:
\begin{equation}\label{eq:er31exp}
S^E_\text{gr}[\widehat{\bulk}_n]= -\frac{1}{16\pi G_N} \left[ \int_{\widehat{\bulk}_n} \, d^3x\, \sqrt{g}\, (R+2)  + 2\, \int_{\bdy_c} \sqrt{\gamma}\, K - \int_{\bdy_c} \, \sqrt{\gamma} \left(2+{}^\gamma R \, \log\rho_c\right)\right] .
\end{equation}	

We can evaluate each of these in turn. Firstly, since $R = -6$ it follows that the bulk contribution can be evaluated explicitly to be
\begin{equation}\label{eq:er31eh}
\begin{split}
\int_{\widehat{\bulk}_n} \, d^3x\, \sqrt{g}\, (R+2)
&=
	- 4  \int_{\widehat{\bulk}_n}  \sqrt{g}  \\
&=
	-\int\, dv d\bv \int_{\rho_c}^{\rho_*}\, d\rho \left[\frac{1}{\rho^2} + \frac{\Tr(T)}{4\rho} + \frac{\det(T)}{16}\right] \\
&=
	-  \,\int\, dv d\bv \left[\frac{1}{\rho_c} -\frac{\Tr(T)}{4} \, \log\frac{\rho_*}{\rho_c}-   \sqrt{\abs{T_{vv}}^2} \right] .
\end{split}
\end{equation}	
In this expression $\rho_*$ is the value of $\rho$ at the origin of \AdS{3}. In the Fefferman-Graham chart this is the point where the  determinant of the metric vanishes. Explicitly one finds
\begin{equation}\label{eq:rhostar}
\rho_* = \frac{8}{\Tr(T) + \sqrt{\Tr(T)^2 -4\, \det(T)}}\,.
\end{equation}	

The boundary terms follow easily once we note that $K = -\frac{2}{\sqrt{g}}\, \rho \, \pdv{\sqrt{g}}{\rho} +2 $ evaluated at $\rho=\rho_c$ and that the curvatures of the induced metric on the cut-off boundary are related to the stress tensor. One has
\begin{equation}\label{eq:er31gh}
\begin{split}
 2\, \int_{\bdy_c} \sqrt{\gamma}\, K
 &=
 	- 4\, \int_{\bdy_c} \, d^2x \, \sqrt{\gamma}\, \left(\frac{1}{\sqrt{g}}\, \rho \, \pdv{\sqrt{g}}{\rho} -1 \right) \\
 &=
 		2\, \int_{\bdy_c} \, dvd\bv \, \frac{1}{\rho_c}  \,.
 \end{split}
\end{equation}	
The counterterm piece evaluates to
\begin{equation}\label{eq:er31ct}
\int_{\bdy_c} \, d^2x\, \sqrt{\gamma} \left(2+{}^\gamma R \, \log\rho_c\right)
=
  \int dv  d\bv \left( \frac{1}{\rho_c} - \frac{\Tr(T)}{4} \left( 1+ \log\rho_c\right)\right) \,.
\end{equation}	
Putting the pieces together we find
\begin{equation}\label{eq:er31all}
S^E_\text{gr}[\widehat{\bulk}_n]
=
 -\frac{c}{24\pi}\, \int\,dv d\bv\, \left[ \frac{\Tr(T)}{4} \, \left(1+ \log \rho_*\right)  +  \sqrt{\abs{T_{vv}}^2} \right] .
\end{equation}	

Now as remarked we need to exclude the contribution from the cosmic brane. In the form written above in \eqref{eq:er31all} this term is completely isolated in the contribution to $\Tr(T)$. Dropping these terms will in fact suffice to extract for us the part that is the cosmic-brane excised action. As a result:
\begin{equation}\label{eq:eren31fin}
\begin{split}
\hat{I}_n = -\frac{c}{24\pi}\, \int\,dv d\bv\,\sqrt{\abs{T_{vv}}^2} \,.
\end{split}
\end{equation}
We can evaluate this integral using the explicit form of the stress tensor quoted in \eqref{eq:eTd31}. One has
\begin{equation}\label{eq:eren31intbyparts}
\begin{split}
\hat{I}_n
&=
	-\frac{c}{48\pi} \left(1-\frac{1}{n^2}\right)   \int\,dv d\bv\, \frac{(a_2-a_1)^2}{\abs{v-a_1}^2\, \abs{v-a_2}^2} \\
&=
	-\frac{c}{48\pi}  \left(1-\frac{1}{n^2}\right)   \int\,dv d\bv\, \partial_v\, \mathcal{Q} \, \partial_{\bv}\, \mathcal{Q} \,, \qquad \quad
	\mathcal{Q}(v,\bv) \equiv \log \abs{\frac{v-a_1}{v-a_2}}^2 \\
&=
	-\frac{c\, \delta}{96\pi} \left(1-\frac{1}{n^2}\right)
	\bigg[\oint_{a_1} \, \mathcal{Q} \partial_{\abs{v}} \mathcal{Q} +
	 \oint_{a_2} \, \mathcal{Q} \partial_{\abs{v}} \mathcal{Q}  \bigg] \\
&=
	\frac{c}{6} \left(1-\frac{1}{n^2}\right) \log \frac{\abs{a_2-a_1}}{\delta} \,.
\end{split}
\end{equation}
This integral has been evaluated by using the fact that $\mathcal{Q}(v,\bv)$ is a Green's function on the plane with sources at $a_1$ and $a_2$. Massaging the integral and integrating by parts, we find source $\delta$-function contributions and the above boundary terms. We  discard the former since the conical singularities on $\bdy_n/\mathbb{Z}_n$ also ought not  be included in the cosmic-brane excised action. This leaves us with a contour integral around each branch point which we have evaluated with a UV regulator $\delta$.  Finally, from \eqref{eq:osRenyi} and \eqref{eq:eInhIn} we obtain on using $I_1 =0$, the expected answer \eqref{eq:ren2d1int} of the $n^{\rm th}$ R\'enyi entropy, viz.,
\begin{equation}\label{eq:1nren}
S_\regA^{(n)} = \frac{n}{n-1}\, \left[ \, \hat{I}_n - I_1\right] = \frac{c}{6} \left(1+\frac{1}{n}\right) \log \frac{\abs{a_2-a_1}}{\delta} \,.
\end{equation}	
%

\subsubsection{Lorentzian on-shell action in a fundamental domain}
\label{sec:lren31}

Let us now turn to the computation of the on-shell action for the real-time geometry \eqref{eq:lfgmetads3}. The on-shell Lorentzian action we need is given by \eqref{eq:Lfundomos} which we rewrite here for convenience as
\begin{equation}\label{eq:InLorentz3d}
\begin{split}
I_n
&=
	- i\, S_\text{gr}^L[\bulk_n]
		= -i\,n \, \left[S^k_\text{gr}[\widehat{\bulk}_n]-S^b_\text{gr}[\widehat{\bulk}_n]  \right]_\text{cosmic-brane excised} \\
&=
	2\, n \, \Im\left(S^k_{\text{gr,fund}}\right) .
\end{split}	
\end{equation}	
We will as before focus on the ket part of the geometry and try to directly isolate the imaginary part of the on-shell Lorentzian action. In fact, we have already computed the various pieces hitherto in the Euclidean context and we can simply take the contributions from \eqref{eq:er31eh}, \eqref{eq:er31gh}, and \eqref{eq:er31ct} and continue $\{v,\bv\} \to \{\tx^-, \tx^+\}$. We would now find prior to excising the cosmic-brane contribution the following integral to evaluate:
\begin{equation}\label{eq:Lket31}
S^k_\text{gr}[\widehat{\bulk}_n] = \frac{c}{24\pi}\, \int\, d\tx^-\, d\tx^+\,\left[ -\frac{\Tr(T)}{4} \left(1+\log \rho_*\right) +   \sqrt{T_{++}\, T_{--}} \right] .
\end{equation}	
where the stress tensor components are given in \eqref{eq:lT31}. In writing this expression we have performed the radial integral and converted the computation of the on-shell action into an integral over the boundary directions alone. This is somewhat different from the basic philosophy outlined in \cite{Colin-Ellerin:2020mva}, so let us pause a moment to record them.

The evaluation of the on-shell action with a neighbourhood of the cosmic brane excised is easiest to implement in coordinates which are adapted to the brane. In the present case the locus is a curve in three dimensions. We pick coordinates  $y^i$ tangent to the brane and a Gaussian normal chart in the normal plane (which is locally $\mathbb{R}^{1,1}$). The regulator around the brane then is a simple matter of excising a disc shaped domain in the normal plane.

However, this coordinate chart which is adapted to $\fixM$ is not the Fefferman-Graham chart used in \eqref{eq:lfgmetads3}. This may a-priori seem surprising since the normal plane for each fixed $\rho$ is parameterized by $\tx^\pm$ and the cosmic brane is located at the same coordinate positions in this Minkowski plane. This is misleading, since the range of $\rho$ is constrained to lie within the interval $\rho \in [\rho_c, \rho_*]$ and the right-end point $\rho_*$ is a non-trivial function of $\tx^\pm$ from \eqref{eq:rhostar}.  In our coordinates, the radial direction in the normal $\mathbb{R}^{1,1}$ plane is an admixture of the Fefferman-Graham radial coordinate $\rho$ and a timelike combination made up from $\tx^\pm$. Adapting coordinates to the cosmic brane locus is in principle possible, but quite messy, since the stress tensor is a non-trivial function of $\tx^\pm$.\footnote{ If the boundary stress tensor is constant, then the transformation is straightforward, and can be inferred from the BTZ solution. }

 Rather than attempt to convert this to the Gaussian normal chart in the neighbourhood of $\fixM$, we will instead demonstrate a direct way to compute the on-shell action in Lorentz signature. Our starting point is the integral in \eqref{eq:Lket31} and we first excise a neighbourhood of the cosmic brane.  This removes the piece $\Tr(T)$ which only has delta function support on $\fixM$ owing to \eqref{eq:lT31}. Dropping this piece in the excised geometry we have
\begin{equation}\label{eq: LSkn31}
\begin{split}
S^k_\text{gr,fund}
& =
	\frac{c}{24\pi}\, \int_{t<0}\, d\tx^-\, d\tx^+\,  \sqrt{T_{++}\, T_{--}} \\
&=
	\frac{c}{48\pi} \left(1-\frac{1}{n^2}\right) \, (a_2-a_1)^2\;\mathfrak{I}(a_1,a_2) \,,\\
\mathfrak{I}(a_1,a_2)
&\equiv
	\int_{t<0}\,\frac{ d\tx^-\, d\tx^+}{(\tx^--a_1)(\tx^+-a_1)(\tx^--a_2)(\tx^+ - a_2)} \,.
\end{split}
\end{equation}	
We need to evaluate thus the integral $\mathfrak{I}$ defined above and extract an imaginary piece from it. As a warm up consider first the simpler case of a semi-infinite interval, where $a_2 \to \infty$ and $a_1=0$ which will serve to exemplify the general case. We have then
\begin{equation}\label{eq:Lhalf31}
\mathfrak{I}_{_\text{half-line}} = \lim_{a_2\to \infty}  \; a_2^2\;\mathfrak{I}(0,a_2) = \int_{t<0}\, \frac{d\tx^+\, d\tx^-}{\tx^+\, \tx^-}
\end{equation}	
which the reader will recognize bears a close resemblance to the integral we computed in \cref{sec:Lren2d}. We will proceed similarly here using an $i\varepsilon$ prescription to pick out the projection onto the vacuum state of the CFT. It will be convenient to introduce an IR cut-off  $L$ which will enter the answer for the semi-infinite interval. We integrate up on $\bulkket$ up to a UV cut-off  restricting $\abs{\tx^+} >\delta$ and obtain

\begin{equation}
\begin{split}
\mathfrak{I}_{_\text{half-line}}
&=
	\lim_{\delta \to 0} \lim_{L \to \infty}\left[ \mathfrak{I}_{_\text{left}} + \mathfrak{I}_{_\text{strip}} +\mathfrak{I}_{_\text{right}}  \right] \qquad
\begin{tikzpicture}[scale=0.5,baseline={([yshift=0.5ex]current bounding box.center)}]
\draw[black, thin,->] (-2,0) -- ++ (4,0)  node[below] {$\scriptscriptstyle{x}$};
\draw[black, thin,->] (0,-2) -- ++ (0,3)  node[left] {$\scriptscriptstyle{t}$};
\draw[black, thin,->] (0,0) -- (1,1)  node[above] {$\scriptstyle{\tx^+}$};
\draw[black, thin,->] (0,0) -- (1.5,-1.5)  node[below] {$\scriptstyle{\tx^-}$};
\draw[blue,thick] (-0.5,0) -- ++(1.5,-1.5);
\draw[blue,thick] (0.5,0) -- ++(1.5,-1.5);
\draw[blue,thick,fill=blue!10] (0.5,0) -- ++(-0.5,-0.5) -- ++(-0.5,0.5)--cycle;
\draw[red,fill=red] (0,0) circle [radius=0.5ex];
\node at (2,-1) [right]{$\scriptstyle{\mathfrak{I}_{_\text{right}}}$};
\node at (-2,-1) [right]{$\scriptstyle{\mathfrak{I}_{_\text{left}}}$};
\end{tikzpicture}  \\
 \mathfrak{I}_{_\text{left}}
  &=
  	 \int_{-L}^{-\delta} \, \frac{d\tx^+}{\tx^+} \; \int_{\tx^+}^L \, \frac{d\tx^-}{\tx^-} =
  	\left[-\log\frac{L}{\delta}\right] \left[ -i\pi +\mathcal{P} \int_{L}^{\tx^+} \, \frac{d\tx-}{\tx^-}  \right]\\
 \mathfrak{I}_{_\text{right}}
  &=
  	\int_{\delta}^{L} \, \frac{d\tx^+}{\tx^+} \; \int_{\tx^+}^L \, \frac{d\tx^-}{\tx^-} =
  	\int_{\delta}^{L} \, \frac{d\tx^+}{\tx^+} \; \log\left(\frac{\tx^+}{L}\right)\\
 \mathfrak{I}_{_\text{strip}}
  &= \int_{\delta}^{L} \, \frac{d\tx^-}{\tx^-} \; \int_{-\delta}^{\delta} \, \frac{d\tx^+}{\tx^+} =
		\left[\log\frac{L}{\delta}\right] \left[ i\pi + \mathcal{P} \int_{-\delta}^{\delta}  \, \frac{d\tx+}{\tx^+}  \right]\\
\end{split}
\end{equation}
where we have used \eqref{eq:prinvalue} and as before analytically continued $\tx^- \to \tx^- + i\epsilon$ while
$\tx^+ \to \tx^+ - i\epsilon$.  We see then that the imaginary parts as before add from the first and third integrals which leads to the final result
\begin{equation}
I_n\big|_{_\text{half-line}} = 2\, n\,  \Im(S^k_\text{gr,fund}) = \frac{c}{48\pi} \left(1-\frac{1}{n^2}\right)  4\pi\, n \log\frac{L}{\delta}
= \frac{c}{12} \left(n-\frac{1}{n}\right)\, \log\frac{L}{\delta}
\end{equation}	
which one can check leads to the correct R\'enyi entropy \eqref{eq:ren2d1int}.\footnote{Note that the result  appears to be missing a factor of $2$, but this is consistent since in the limit of a semi-infinite interval we only pick up the contribution from one branch point. We evaluate the integral a different way in \cref{sec:rindreg} to double check this factor.}

Armed with this understanding it is now clear how to evaluate the integral $\mathfrak{I}(a_1, a_2)$. We again introduce UV and IR regulators $\delta$ and $L$, respectively, and break up the integration range $t<0$ into five domains
\begin{equation}
\begin{split}
\mathfrak{D} &=\mathfrak{D}_1 \cup  \mathfrak{D}_2\cup  \mathfrak{D}_3 \cup \mathfrak{D}_{_\text{strips}}
\,, \qquad
\begin{tikzpicture}[scale=0.6,baseline={([yshift=0.5ex]current bounding box.center)}]
\draw[black, thin,->] (-2.5,0) -- ++ (5.5,0)  node[below] {$\scriptscriptstyle{x}$};
\draw[black, thin,->] (0,-2) -- ++ (0,3)  node[left] {$\scriptscriptstyle{t}$};
\draw[black, thin,->] (0,0) -- (1,1)  node[above] {$\scriptstyle{\tx^+}$};
\draw[black, thin,->] (0,0) -- (1.5,-1.5)  node[below] {$\scriptstyle{\tx^-}$};
\foreach \x in {-1.5,1.5}
{
\draw[blue,thick] (\x-0.5,0) -- ++(1.5,-1.5) ;
\draw[blue,thick] (\x+0.5,0) -- ++(1.5,-1.5);
\draw[blue,thick,fill=blue!10] (\x+0.5,0) -- ++(-0.5,-0.5) -- ++(-0.5,0.5)--cycle;
\draw[red,fill=red] (\x,0) circle [radius=0.5ex];
}
\node at (-1.5,-1) [left] {$\scriptscriptstyle{\mathfrak{D}_1}$};
\node at (1,-1) [left] {$\scriptscriptstyle{\mathfrak{D}_2}$};
\node at (4,-1) [left] {$\scriptscriptstyle{\mathfrak{D}_3}$};
\end{tikzpicture}
\end{split}
\end{equation}	
Three of the domains are analogous to the regions to the left and right of the fixed point in the half-line case considered above. They are demarcated by  constant $\tx^+$ lines:
$\mathfrak{D}_1: \tx^+ \in(-L,a_1-\delta)$, $\mathfrak{D_2}: \tx^+ \in (a_1+\delta, a_2 -\delta)$ and
$\mathfrak{D}_3: \tx^+ \in (a_2 +\delta, L)$ and $\tx^-$ runs up from $\tx^+$   to some IR cut-off value $L$.
 We also now have two strips $\mathfrak{D}_{_\text{strips}} $ once we excise the triangular domains around the fixed points at $a_1$ and $a_2$. We will as before consider the contributions from each region separately. Writing $\mathfrak{I} = \mathfrak{I}_1 +  \mathfrak{I}_2 +  \mathfrak{I}_3 +  \mathfrak{I}_{_\text{strips}} $ we have
\begin{equation}\label{eq:mfI1eval}
\begin{split}
\mathfrak{I}_1
&=
 \int_{-L}^{a_1-\delta} \, \frac{d\tx^+}{(\tx^+-a_1) (\tx^+-a_2)} \,
	\int_{\tx^+}^{L} \, \frac{d\tx^-}{(\tx^- -a_1) (\tx^- - a_2)} \\
&=
	\frac{1}{\abs{a_2-a_1}}\, \int_{-L}^{a_1-\delta} \, \frac{d\tx^+}{(\tx^+-a_1) (\tx^+-a_2)} \
	\left[-\mathcal{P} \int_{\tx^+ - a_1}^{L-a_1} \, \frac{d \tx^-_1}{\tx^-_1} + \mathcal{P} \int_{\tx^+ - a_2}^{L-a_2} \, \frac{d\tx^-_2}{\tx^-_2}
	\right]
\end{split}
\end{equation}	
where we have taken partial fractions   introducing $\tx^-_i = \tx^-- a_i$ and used the principal value prescription.  This term has no imaginary part as should be clear from the fact that we are in the left homology wedge in $\mathfrak{D}_1$. Similarly we can evaluate
the contribution from $\mathfrak{D}_3$ to be purely real, for
\begin{equation}\label{eq:mfI3eval}
\begin{split}
\mathfrak{I}_3
&=
 \int_{a_2+ \delta}^{L} \, \frac{d\tx^+}{(\tx^+-a_1) (\tx^+-a_2)} \,
	\int_{\tx^+}^L \, \frac{d\tx^-}{(\tx^- -a_1) (\tx^- - a_2)} \\
&=
	\frac{1}{\abs{a_2-a_1}} \, \int_{a_2+\delta}^{L} \, \frac{d\tx^+}{(\tx^+-a_1) (\tx^+-a_2)} \,
 	\log\left(\frac{\tx^+ -a_1}{\tx^+-a_2}\right)
\end{split}
\end{equation}	
where we have dropped terms that vanish as $L \to \infty$. We do pick up imaginary parts from the region $\mathfrak{D}_2$ and the strips. The region $\mathfrak{D}_2$ picks out the contribution from the right branch point at $a_2$ as
\begin{equation}\label{eq:mfI2eval}
\begin{split}
\mathfrak{I}_2
&=
 \int_{a_1+\delta}^{a_2-\delta} \, \frac{d\tx^+}{(\tx^+-a_1) (\tx^+-a_2)} \,
	\int_{\tx^+}^L \, \frac{d\tx^-}{(\tx^- -a_1) (\tx^- - a_2)} \\
&=
	\frac{2\pi i}{\abs{a_2-a_1}^2}\, \log \frac{\abs{a_2-a_1}}{\delta} \\
&\qquad
	+ \frac{1}{\abs{a_2-a_1}} \,
	\int_{a_1+\delta}^{a_2-\delta} \, \frac{d\tx^+}{(\tx^+-a_1) (\tx^+-a_2)} \left[\log\left(\frac{\tx^+-a_1}{L-a_1}\right) + \mathcal{P}  \int_{\tx^+-a_2}^{L-a_2} \, \frac{d\tx^-_2}{\tx^-_2}\right]
\end{split}
\end{equation}	
The final contribution comes from the strips which lie a distance $a_i+\delta$ around $\tx^- =0$. These do give non-vanishing imaginary contributions as one of the strips captures the left branch point. To wit,
\begin{equation}\label{eq:mfIstrips}
\begin{split}
\mathfrak{I}_{_\text{strips}}
&=
	 \int_{a_1 - \delta}^{a_1+\delta} \, \frac{d\tx^+}{(\tx^+-a_1) (\tx^+-a_2)} \,
	\int_{a_1+\delta}^{L} \, \frac{d\tx^-}{(\tx^- -a_1) (\tx^-- a_2)}  \\
& \qquad
	 +\int_{a_2-\delta}^{a_2+\delta} \, \frac{d\tx^+}{(\tx^+-a_1) (\tx^+-a_2)} \,
	 \int_{a_2+\delta}^{L} \, \frac{d\tx^-}{(\tx^- -a_1) (\tx^-- a_2)}  \\
&=
	\frac{2\pi i}{\abs{a_2-a_1}^2}\,  \log\frac{\abs{a_2-a_1}}{\delta} \\
&	\qquad  - 	
	 \frac{1}{\abs{a_2-a_1}} \,	
	\mathcal{P} \int_{-\delta}^\delta \, \frac{d\tx^+}{\tx^+}
	\left[
		\mathcal{P} \int_{a_1}^{a_2} \, \frac{d\tx^-}{\tx^- -a_2} - \log\frac{\abs{a_2-a_1}}{\delta}	
	\right]
\end{split}
\end{equation}	

Putting together all the contributions we find
\begin{equation}\label{eq:ImSk11}
\Im(S^k_\text{gr,fund}) = 4\pi\, \log\frac{\abs{a_2-a_1}}{\delta} ,
\end{equation}	
which as one can readily verify leads to the expected result for the R\'enyi entropy \eqref{eq:ren2d1int}. At various points above we have taken the interval to lie on the $t=0$ slice in $\mathbb{R}^{1,1}$ for illustrative purposes. This is however unnecessary, and the result holds for any boosted slice, owing to the boost invariance of the vacuum state of the CFT.

\section{R\'enyi entropies in 2d CFTs: Disjoint intervals}
\label{sec:ads32}

The examples we have discussed thus far comprise of situations where the entropies are computed at a moment of time symmetry. While we see that even in these examples the real-time computations require a careful analysis, we  now turn  to an example where time reflection symmetry is explicitly broken  (in a controllable manner). We explore the R\'enyi entropy for a 2d CFT in its vacuum state on the plane, with the region $\regA$ taken to be the disjoint union of $N$ intervals.

In the Euclidean set-up the computation of the $n^\text{th}$ R\'enyi entropy requires us to compute the CFT partition function on a $n$-sheeted branch cover  of the plane with $2N$ branch points. This is a genus $(n-1)(N-1)$ surface,  albeit one at a special point in moduli space since the moduli are specified by  $2N-3$ parameters (using conformal invariance to fix $3$ points). Unfortunately, one does not have readily available partition functions for generic 2d CFTs on higher genus Riemann surfaces.

Nevertheless one can make progress in certain circumstances. For instance, the problem was  first analyzed using replica methods in CFT in \cite{Calabrese:2009ez} for free 2d CFTs for which the higher genus partition functions are available. One can likewise study large $c$ holographic CFTs.  In fact, the first non-trivial computations of holographic R\'enyi entropies were undertaken in \cite{Headrick:2010zt}, who  analyzed the $N=2$ example for large $c$ CFTs and explicitly demonstrated the holographic entanglement entropy phase transition. Subsequently, \cite{Faulkner:2013yia} analyzed the problem in detail in the gravitational context, constructing the dual gravitational solutions as handlebody geometries, and evaluated the on-shell action to extract the answer. A complementary CFT analysis using properties of Virasoro vacuum blocks was also concurrently given in \cite{Hartman:2013mia}.  We will adapt the discussion of \cite{Faulkner:2013yia} to the real-time setting after reviewing the ingredients of Schottky uniformization that enter the computation in  Euclidean signature. We will keep our discussion general in the main text, though for ease of presentation we will use the $2^\text{nd}$ R\'enyi entropy $n=2$ for $N=2$ intervals to illustrate the general arguments.\footnote{ Details of the geometry  for $N=n=2$ are given in \cref{sec:22ren}.  In \cref{sec:direct2ren} we explicitly evaluate the on-shell action in Euclidean signature for this case. In the bulk of our discussion we will sidestep the evaluation of the R\'enyi entropies, concentrating on obtaining its variation with respect to one of the endpoints.}

\subsection{R\'enyi from Schottky uniformization}
\label{sec:schottky}


We give a quick overview of the Schottky uniformization exploited in \cite{Faulkner:2013yia} to compute the holographic  R\'enyi entropies for disjoint intervals.    For the vacuum entanglement entropy of  $N$ intervals $\regA = \cup_{i=1}^N \, (a_{2i-1},a_{2i})$, we must compute the partition function on the $n$-fold cover $\bdy_{n,N}$ branched over the $N$ intervals.  The manifold $\bdy_{n,N}$ is a compact Riemann surface of genus $(N-1)(n-1)$ with complex structure
\begin{equation}\label{eq:zNn}
z^n = \prod_{i=1}^N \,\left( \frac{v-a_{2i-1}}{v-a_{2i}}\right) .
\end{equation}
Following \cite{Faulkner:2013yia} we will assume that the dominant bulk saddles are replica $\mathbb{Z}_n$ symmetric handlebodies.

A Riemann surface of genus $g$ can be constructed by starting with the Riemann sphere $\mathbb{C}$ and quotienting it by a Schottky group $\Gamma \subset \mathrm{PSL}(2, \mathbb{C})$, which is a discrete subgroup freely generated by $g$ loxodromic elements, constrained such that the closure of the set of fixed points $\Delta$ of its action is not the entirety of $\mathbb{C}$.  The Riemann surface is $\widetilde{\mathbb{C}}/\Gamma$ with $\widetilde{\mathbb{C}} = \mathbb{C}-\Delta$. Operationally, one picks   $2g$ non-intersecting circles $\{\mathfrak{C}_i, \tilde{\mathfrak{C}}_i \}$, lets the generators $\gamma_i$ of $\Gamma$ act by mapping the interior of  the disc bounded by $\mathfrak{C}_i$ to the exterior of the disc bounded by $ \tilde{\mathfrak{C}}_i$, along  with
$\gamma_i( \mathfrak{C}_i ) = \tilde{\mathfrak{C}}_i $. The quotient operation then cuts out the $2 g$ discs to the interior of these circles and identifies the circles themselves, thus creating the handles.

This construction on the Riemann sphere extends to the bulk of Euclidean \AdS{3} where the $\mathrm{PSL}(2, \mathbb{C})$ map acts as on the coordinates $(\xi, y, \bar{y})$ as
\begin{equation}\label{eq:sl2act}
y \rightarrow
	\frac{(a \, y+b)(\bar{c}\, \bar{y}+\ {d})+a\bar{c}\,\xi^2}{|c \, y+d|^2+|c|^2\, \xi^2}, \qquad
	\xi \rightarrow \frac{\xi}{|c \, y+d|^2+|c|^2\, \xi^2}, \qquad
	\left(\begin{matrix} a & b \\ c & d \\ \end{matrix}\right) \in \mathrm{PSL}(2,\mathbb{C}) \,.
\end{equation}	
The quotient acts smoothly in the bulk (because $\Gamma$ has loxodromic elements). However, given a choice of $\Gamma$ which determines the Schottky uniformization of $\bdy_{n,N}$ there may be multiple bulk geometries. These are handlebodies where $g$ commuting cycles of $\bdy_{n,N}$ smoothly pinch off in the bulk.

To determine all the bulk handlebodies that respect the replica $\mathbb{Z}_n$ symmetry, we need to decide which cycles are contractible. Around any single branch point, which is a localized source of stress-energy (see e.g., \eqref{eq:eTd31}), we know the inverse map $y = \pi^{-1}(v)$ of the quotient  $\pi: \mathbb{C} \mapsto \widetilde{\mathbb{C}}/\Gamma $, has local solutions
$(v-a_i)^{\frac{1}{2} \pm \frac{1}{2n}}$, where we coordinatize $\mathbb{C}$ with $\{v,\bv\}$ as before. However, around a loop $\mathfrak{C}$ that contains two or more branch points one picks up a monodromy $M(\mathfrak{C}) \in \mathrm{PSL}(2, \mathbb{C})$.

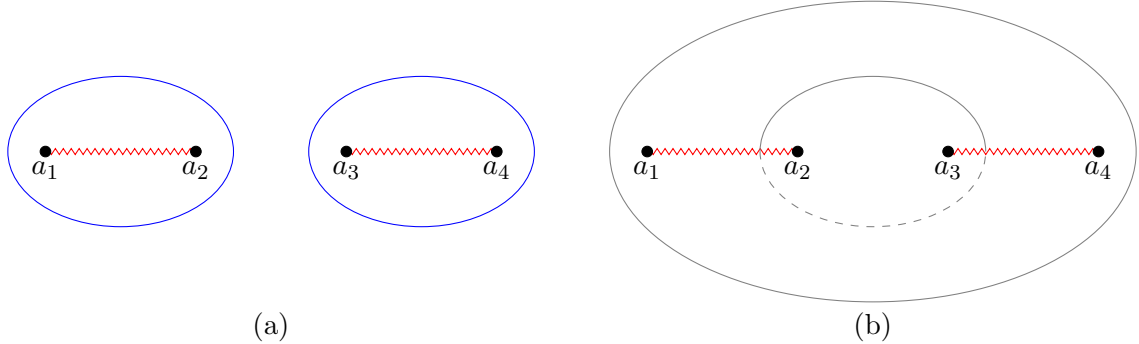
\begin{figure}[h]
\centering
\begin{tikzpicture}

\foreach \x in {- 7, -3,1,5}
{
\draw[snake=zigzag,color=red,segment length=3pt,segment amplitude=0.4mm] (\x,0) -- ++(2,0);
\draw[black,fill=black] (\x,0) circle (2pt);
\draw[black,fill=black] (\x+2,0) circle (2pt);
}
\draw[blue] (-6,0) ellipse (1.5cm and 1cm);
\draw[blue] (-2,0) ellipse (1.5cm and 1cm);
\draw[gray] (4,0) ellipse (3.5cm and 2cm);
\draw[gray] (2.5,0) arc(180:0:1.5cm and 1cm);
\draw[gray,dashed] (2.5,0) arc(180:360:1.5cm and 1cm);
\foreach \x in {-7,1}
{
\node at (\x,0) [below] {$a_1$};
\node at (\x+2,0) [below] {$a_2$};
\node at (\x+4,0) [below] {$a_3$};
\node at (\x+6,0) [below] {$a_4$};
}
\node at ( -4,-2) [below] {(a)};
\node at ( 4,-2) [below] {(b)};
\end{tikzpicture}
\caption{The   choices of cycles around which we can impose trivial monodromy to construct the dual handlebody. (a) trivial monodromy around the two cycles that circle the branch cut, denoted by $\mathfrak{C}_d$, which corresponds to the disconnected RT surface in the limit $n \to 1$; (b) trivial monodromy around the cycle that circles around both branch cuts and the cycle that passes through the branch cuts and laces through all the $n$ sheets, denoted by $\mathfrak{C}_{c}$, which corresponds to the connected RT surface in the limit $n \to 1$.}
\label{fig:twointmonodromies}
\end{figure}

For example with $N=2$, the region $\regA = \regA_1 \cup \regA_2 \equiv  (a_1,a_2) \cup (a_3,a_4)$, the boundary manifold for $n=2$ is a genus $1$ Riemann surface, a torus. There are two distinct bulk geometries that should be considered as the dual handlebody -- we either let the $a$-cycle of the torus shrink smoothly, or let the $b$-cycle shrink. The two choices can equivalently be characterized by the choice of cycles around which we impose trivial monodromy as  depicted in \cref{fig:twointmonodromies}.

To solve the monodromy problem, we realize that the map $y(v)$ satisfies
\begin{equation}\label{eq:imapTdef}
\{y(v)), v\} = T(v) \,, \qquad T_{vv}(v) = \sum_{i=1}^{2N} \, \left[ \frac{\Delta_n}{(v-a_i)^2} + \frac{p_i}{v-a_i} \right] ,
\end{equation}	
where $T(v)$ is the source of the stress-energy on a single sheet arising from the branch structure and $\Delta_n$ is the conformal weight of the defect
\begin{equation}\label{eq:Deltan}
\Delta_n \equiv \frac{1}{2} \left(1-\frac{1}{n^2}\right)
\end{equation}	
This stress-energy is  yet to be fully determined, parameterized as it is by a set of accessory parameters,  $p_i$, which carry information about the covering space topology. Once we solve for these parameters we should have the necessary information to determine the geometry.

One proceeds by solving an auxiliary homogeneous linear differential equation for a function $\psi(v)$,  from whose linearly independent solutions, $\psi_{1,2}(v)$, one can recover $y(v)$, viz.,
\begin{equation}\label{eq:psieq}
\psi''(v) + \frac{1}{2}\, T_{vv}(v)\, \psi(v) = 0 \,, \qquad y(v) = \frac{\psi_1(v)}{\psi_2(v)} \,.
\end{equation}	
We have $2N$ accessory parameters $p_i$. To fix them, consider one sheet of the Riemann surface which is a copy of a sphere with $2N$ punctures. Let $\mathfrak{C}= \{\mathcal{C}_a, a=1,\ldots N\}$ be the set of cycles which contain an even number of punctures. The accessory parameters are fixed by demanding that the solution has trivial monodromy around   $v=\infty$ and around the remaining $N-1$ independent cycles $\mathcal{C}_a$. The absence of monodromy around $v=\infty$ gives three relations:
\begin{equation}\label{eq:accpinfrel}
\sum_{i=1}^{2N} p_{i} = 0	\,, \qquad
\sum_{i=1}^{2N} p_{i}a_{i} = -2N\,\Delta_n \,, \qquad
\sum_{i=1}^{2N} p_{i}a_{i}^2 = -2\Delta_n\, \sum_{i=1}^{2N} a_{i}.
\end{equation}	
By replica symmetry one has actually specified the $(n-1)(N-1)$ cycles on $\bdy_{n,N}$ which have trivial monodromy. Demanding these cycles be contractible in the bulk we have completed the specification of a smooth handlebody.

Note that once we have specified the set of monodromies we fix the accessory parameters, since this suffices to characterize the covering space Riemann surface topology. This implies that $T(v)$ in \eqref{eq:imapTdef} is now completely determined. This will  be sufficient for us to understand the computation of the dual geometry, and in particular the on-shell action.

While the accessory parameters were introduced here to solve the uniformization problem, physically they specify the stress-energy source on a single sheet necessary to build up the Riemann surface. As a result, it should be no surprise to learn that they directly determine the on-shell action of gravity, and thus the R\'enyi entropies. For a given collection of cycles $\mathfrak{C}$  which are contractible one has the result \eqref{eq:renaccesory} obtained in \cite{Faulkner:2013yia} (using results of \cite{Zograf:1988uwp})
\begin{equation}\label{eq:renaccesory}
S^{(n)} = \min_{\mathfrak{C}} \{ S^{(n)}_{\mathfrak{C}} \} \,, \qquad   \pdv{a_i}  S^{(n)}_{\mathfrak{C}} = -\frac{c\, n}{6 \, (n-1)} \, p_i^{\mathfrak{C}} \,.
\end{equation}	
where $\mathfrak{C}$ represents the different sets of choices of cycles which can be made contractible. We present  the details for $N=n=2$ in \cref{sec:22ren} where the branched cover is a torus.

We will broadly content ourselves with obtaining the variation of the R\'enyi entropy with respect to the endpoint, viz., the second expression in \eqref{eq:renaccesory}. There is one special case where $S^{(n)}_{\mathfrak{C}}$ itself is directly computable, which is the second R\'enyi entropy for two disjoint intervals $N=n=2$. In \cref{sec:direct2ren}  we  evaluate the on-shell action of gravity (for the connected phase) to obtain $S^{(2)}$ directly, cf., \eqref{eq:22ren}. We will return to this issue in \cref{sec:generalintervals}.

\subsection{The action from a single fundamental domain}
\label{sec:fdnN}

Let us assume that one has solved the monodromy problem and thus determined the accessory parameters by picking a set of contractible cycles.  Furthermore,  recall that we can use the Fefferman-Graham expansion quite effectively to compute the bulk geometry, cf.,  \eqref{eq:efgmetads3} and \eqref{eq:lfgmetads3} for the Euclidean and Lorentzian signature metrics, respectively. We also know that the computation of the on-shell action in these coordinates is straightforward and one obtains the final results quoted in \eqref{eq:er31all} and \eqref{eq:Lket31}, respectively.

Inspired by their simplicity we can address the problem as follows. Focus for the present on the Euclidean geometries where in the $v$-plane corresponding to a single sheet of the Riemann surface, we have a set of branch points, which are a source of stress-energy. The stress tensor is parameterized  in terms of the accessory parameters $p_i$. Once we solve the monodromy problem and fix these $p_i$ we have determined on a single sheet the local sources of energy-momentum that we need to turn on to construct the Riemann surface. With this knowledge we can immediately compute the on-shell action using  \eqref{eq:er31all} in Euclidean signature.

As a quick check, let us look back at the single-interval case discussed in the main text. We have two branch points, and a single choice of cycle $\mathfrak{C}_1$ which encircles both branch points. It is trivial to check that $p_1= -p_2 = 2\,\frac{\Delta_n}{a_2-a_1}$ are fixed uniquely, and thus we recover $T(v)$ quoted in \eqref{eq:eTd31} which we used to compute the on-shell action in \eqref{eq:eren31intbyparts}. In fact we will borrow extensively from the one-interval analysis for general $n,N$ below.

\subsubsection{The Euclidean computation}
\label{sec:erennN}

We start with the assumption that we have been given the stress tensor on a single fundamental domain \eqref{eq:imapTdef}. This stress tensor is localized on the branch points and excising the sources at these loci, we have to evaluate  \eqref{eq:eren31fin}, i.e.,
\begin{equation}\label{eq:InE}
\hat{I}_n
 = -\frac{c}{24\pi}\, \int_{\mathcal{R}_\epsilon} dv d\bv \, \sqrt{T_{vv}\, T_{\bv \bv}} \, ,
 \end{equation}	
where $\mathcal{R}_\epsilon$ is a domain of the complex $v$-plane with infinitesimal discs $\mathscr{D}^\epsilon_i$ of size $\epsilon$ around each of the branch points $v=a_i$ excised.  We will attempt to evaluate not this integral, but rather its derivative  with respect to the branch point location, viz.,
\begin{equation}
\pdv{a_i} \hat{I}_n =	 -\frac{c}{48\pi}\, \int_{\mathcal{R}_\epsilon} dv d\bv \,    \left[\sqrt{\frac{T_{\bv \bv} }{T_{vv}}} \,\pdv{ T_{vv} }{a_i} +
\sqrt{\frac{T_{vv}} {T_{\bv \bv}}} \,\pdv{T_{\bv\bv}}{a_i}\right] + \text{boundary term} \,,
\end{equation}	
where the boundary term arises from the variation of the discs $\mathscr{D}_i^\epsilon$ about $v=a_i$.

To evaluate the variation of $\hat{I}_n$ with respect to the location of the branch points we are going to employ a trick which will reduce the calculation as in the single-interval case to the evaluation of contour integrals on the boundaries of the discs about each branch point, $\mathscr{C}_i^\epsilon = \partial \mathscr{D}_i^\epsilon$. To facilitate this
analysis let us first introduce  a function $\xf$ which satisfies:
\begin{equation}\label{eq:Fders}
\partial_v \xf(v,\bv) = \sqrt{T_{vv}} \,, \qquad \partial_{\bv} \xf(v,\bv) = \sqrt{T_{\bv \bv}} \,.
\end{equation}	
We can formally write it as a contour integral
\begin{equation}\label{eq:Fintdef}
\xf (v,\bv) \equiv \int_{\mathcal{C}} \, \sqrt{T_{vv}} \, dv +  \int_{\mathcal{C}} \, \sqrt{T_{\bv \bv}} \, d\bv = \xfh(v) + \overline{\xfh}(\bv) \,.
\end{equation}	
To define $\xf$ completely we need to specify the integration contour $\mathcal{C}$. It will however transpire that we will only care about the fact that this contour gets close to the branch points at $a_i$.

By a local analysis in the neighbourhood of each branch point we may deduce that
\begin{equation}
\partial_v \xf =  s_i \left[ \frac{\sqrt{\Delta_n}}{v-a_i} + \frac{p_i}{2\,\sqrt{\Delta_n}} + \order{v-a_i} \right] ,
\end{equation}	
and similarly for $\partial_{\bv} \xf$. Here $s_i =\pm 1$ is a sign, $s_i^2 =1$, which will drop out in our final answer.
Integrating these up we have  the local behaviour near $v =a_i$
\begin{equation}\label{eq:Flocal}
\xf(v,\bv) = s_i \left[ \sqrt{\Delta_n} \log\abs{v-a_i}^2 - C_i(\{a_j\}) + \frac{p_i}{2\sqrt{\Delta_n}} (v+\bv -2\,a_i)
 + \order{\abs{v-a_i}^2}
\right]
\end{equation}	
with undetermined constants $C_i(\{a_j\})$.

While the local analysis thus gives an estimate for the function $\xf$, the function is as yet undetermined owing to the information hidden in the constants $C_i$ which as indicated above depend on  the locations of the branch points. It is this dependence that makes the explicit evaluation of $\hat{I}_n$ quite tricky to obtain (though see \cref{sec:direct2ren} for the $N=n=2$ case). We will see that these constants will drop out in our evaluation of the derivatives $\pdv{a_i} \hat{I}_n $. Given the estimate \eqref{eq:Flocal}, we may  immediately compute the derivatives with respect to the branch point locations $a_j$ obtaining
\begin{equation}\label{eq:Faderlocal}
\begin{split}
\pdv{\xf}{a_j}
	= -s_i\left[\sqrt{\Delta_n} \left(\frac{1}{v-a_i}  + \frac{1}{\bv-a_i} \right) +\frac{p_i}{\sqrt{\Delta_n}}\right] \, \delta_{ij}  - s_i\, \ \pdv{C_i}{a_j} + \order{\abs{v-a_i}} .
\end{split}
\end{equation}	
Note that the derivative of the accessory parameter with respect to the branch point has been ignored as it is of order $v-a_i$.

We will now argue that these local estimates will suffice to compute the variation of the on-shell action with respect to the branch points. One  has under the variation of a branch point, a bulk and a boundary contribution that we will study independently, and write (cf., \cref{sec:complexint})
\begin{equation}
\begin{split}
\pdv{\hat{I}_n}{a_i}
&=
	-\frac{c}{24\pi} \left[ \mathcal{I}_i^\text{bulk} + \mathcal{I}_i^\text{bdy}\right] \\
 \mathcal{I}_i^\text{bulk}
 &\equiv
 	\int_{\mathcal{R}_\epsilon} dv\, d\bv \pdv{a_i}(\pdv{\xf}{v}\, \pdv{\xf}{\bv} ) \\
 \mathcal{I}_i^\text{bdy}
 &\equiv
 	- 2i\,\oint_{\mathscr{C}_i^\epsilon} dv \pdv{\xf}{v}\, \pdv{\xf}{\bv} \,.
\end{split}
\end{equation}	

Consider first the bulk integral.  Using the fact that by definition $\xf$ is a sum of a holomorphic and an anti-holomorphic piece \eqref{eq:Fintdef},
we may rearrange the derivatives in the bulk integral, write it as an integral of a total divergence, and convert it to a boundary integral over the circles $\mathscr{C}_i^\epsilon$:
\begin{equation}\label{eq:bulkF1}
\begin{split}
 \mathcal{I}_i^\text{bulk}
 &=
 	\int_{\mathcal{R}_\epsilon} dv\, d\bv
 	\left[\pdv{v}(\pdv{\xf}{a_i}\, \pdv{\xf}{\bv} ) + \pdv{\bv}(\pdv{\xf}{a_i}\, \pdv{\xf}{v} ) \right]\\
&=
	i\, \sum_{j=1}^{2N}  \, \oint_{\mathscr{C}^\epsilon_j}\, \left[\pdv{\xf}{a_i}\,\pdv{\xf}{\bv}\, d\bv -
	\pdv{\xf}{a_i}\, \pdv{\xf}{v} \,dv \right] 	
\end{split}
\end{equation}	
We may now deduce using  \eqref{eq:Flocal} and \eqref{eq:Faderlocal} that
\begin{equation}
\begin{split}
 \oint_{\mathscr{C}^\epsilon_j}\,\pdv{\xf}{a_i}\, \pdv{\xf}{v} \,dv
 &=
	 -s_j^2   \oint_{\mathscr{C}^\epsilon_j}
	 \left[ \left(\frac{\sqrt{\Delta_n}}{v-a_i}  + \frac{\sqrt{\Delta_n}}{\bv-a_i}
	 			+\frac{p_i}{\sqrt{\Delta_n}}\right)\delta_{ij} + \ \pdv{C_j}{a_i} +\cdots\right]
	 \left[ \frac{\sqrt{\Delta_n}}{v-a_j} + \frac{p_j}{2\,\sqrt{\Delta_n}} + \cdots \right]
	  \\
&=
 	- 2\pi i\, \left(\frac{3}{2}\, p_i \, \delta_{ij}+ \sqrt{\Delta_n}\, \pdv{C_j}{a_i} \right) .
\end{split}
\end{equation}	
Putting together the complex conjugate contribution yields
\begin{equation}
 \mathcal{I}_i^\text{bulk}
  =- 4\pi \left(\frac{3}{2}\, p_i + \sqrt{\Delta_n}\,\sum_{j=1}^{2N}\,  \pdv{C_j}{a_i}\right).
\end{equation}	
The boundary term may be evaluated directly to give
\begin{equation}
 \mathcal{I}_i^\text{bdy} = - 2i \oint_{\mathscr{C}_i^\epsilon} \, dv
 	\left[ \frac{\sqrt{\Delta_n}}{v-a_i} + \frac{p_j}{2\,\sqrt{\Delta_n}} + \cdots \right]
 	\left[ \frac{\sqrt{\Delta_n}}{\bv-a_i} + \frac{p_j}{2\,\sqrt{\Delta_n}} + \cdots \right]
 =2\pi\, p_i.
\end{equation}	
Hence we have
\begin{equation}
\pdv{\hat{I}_n}{a_i}  = \frac{c}{6} \left( p_i  + \sqrt{\Delta_n} \,\sum_{j=1}^{2N}\,  \pdv{C_j}{a_i}\right).
\end{equation}	

To complete the argument we need to deduce the value of $\sum_{j=1}^{2N}\,  \pdv{C_j}{a_i}$, which we may do by judiciously combining $\xfh$ and $\xf$.  We use the fact that the product $\pdv{\xfh}{a_i}\, \partial_v \xf$ dies off  as $v^{-2}$ at large $v$ to deduce
\begin{equation}\label{eq:Cidetermine}
\begin{split}
0
&=	
	\oint_{\abs{v} = \Lambda}\, \pdv{\xfh}{a_i}\, \partial_v \xf  \\
&=
	-\sum_j \, 	\oint_{\mathscr{C}_j^\epsilon}
		\left[\left(\frac{\sqrt{\Delta_n}}{v-a_j} + \frac{p_j}{2\, \sqrt{\Delta_n}}\right)\, \delta_{ij}
			+ \frac{1}{2} \,  \pdv{C_j}{a_i} \right]
		\left[
		\frac{\sqrt{\Delta_n}}{v-a_j}  + \frac{p_j}{2\, \sqrt{\Delta_n}}
		\right]	\\
&=
	-2\pi i \left(p_i+\frac{\sqrt{\Delta_n}}{2}\, \sum_{j=1}^{2N}\,  \pdv{C_j}{a_i}\right) \\
&
\;\; \Longrightarrow \;\;  \sqrt{\Delta_n}\, \sum_{j=1}^{2N}\,  \pdv{C_j}{a_i} = -2\,p_i \,.
\end{split}
\end{equation}

The asymptotic behaviour thus  constrains the derivatives of the constants $C_j$ allowing us to evaluate the quantity we want without detailed knowledge of these constants themselves. Consequently, we have as our final result:
\begin{equation}\label{eq:eIhatnN}
\pdv{\hat{I}_n}{a_i}  = - \frac{c}{6}\, p_i
\end{equation}	
This indeed reproduces the result quoted  in \eqref{eq:renaccesory} for
\begin{equation}\label{eq:NnrenE}
 \pdv{a_i}\, S^{(n)} =  \frac{n}{n-1} \, \pdv{a_i}[\hat{I}_n- I_1]  =  -\frac{n}{6(n-1)}\, c\, p_i \,.
\end{equation}	
%

\subsubsection{The Lorentzian computation}
\label{sec:lrennN}

One reason for our going over the Euclidean computation in some detail was to simplify the ingredients to obtain the result directly in Lorentz signature. We will continue with the computation in a single fundamental domain, and exploit the Fefferman-Graham form of the metric \eqref{eq:lfgmetads3} and distill the computation of the action as in the one-interval case to evaluating an integral of the form \eqref{eq:Lket31}.  In making these observations we are assuming that the form of the boundary stress tensor on a single fundamental domain is known, i.e., one has solved the corresponding monodromy problem. Note that the latter is strictly a non-gravitational computation and thus can be carried out in Euclidean signature, and the result used to set-up the boundary conditions for our Lorentzian gravitational analysis.

In the process of deriving \eqref{eq:Lket31} we have integrated over the bulk radial coordinate and thus have a purely boundary integral to evaluate.  As explained earlier in \cref{sec:lren31} this method is conceptually different from the way we set-up the computation of the action in \cite{Colin-Ellerin:2020mva} where we adapted coordinates to the cosmic-brane in the bulk. While that analysis makes it  easier to see where the imaginary part of the Lorentzian action arises from (viz., from the normal bundle to the splitting surface), we found the  chart adapted to the cosmic brane hard to relate to the coordinates induced by the Schottky construction.  All told the final result for the stress tensor is a function of the location of the end-points of our regions $a_i$ and the stress tensor is parameterized by both $a_i $ and the accessory parameters $p_j(a_i)$. The contribution from the trace of the stress tensor in \eqref{eq:Lket31} is delta-function localized at the entangling surfaces (i.e., at $\tx^+=\tx^- = a_i$ if the intervals are all at $t=0$) and should be dropped in the computation of the cosmic-brane excised action. We are then left with evaluating
\begin{equation}\label{eq:LSkn3Nn}
S^k_\text{gr,fund}=
	\frac{c}{24\pi}\, \int_{\mathscr{R}}\, d\tx^-\, d\tx^+\,  \sqrt{T_{++}\, T_{--}}
\end{equation}
with
\begin{equation}
\begin{split}
T_{--}(\tx^-) = \sum_{i=1}^{2N} \, \left[ \frac{\Delta_n}{(\tx^--a_i)^2} + \frac{p_i(a_j)}{\tx^- -a_i} \right]  , \\
T_{++}(\tx^+) = \sum_{i=1}^{2N} \, \left[ \frac{\Delta_n}{(\tx^+ -a_i)^2} + \frac{p_i(a_j)}{\tx^+ -a_i} \right] .
\end{split}
\end{equation}	
Once again we refrain from evaluating \eqref{eq:LSkn3Nn}  but will take inspiration from the Euclidean computation and evaluate its variation of its imaginary part with respect to $a_i$, i.e.,
\begin{equation}\label{eq:LSkn3Nnai}
\pdv{a_i} \Im(S^k_\text{gr,fund}) =
	\frac{c}{24\pi}\,	
	\Im\left(\pdv{a_i}  \int_{\mathscr{R}}\, d\tx^-\, d\tx^+\,  \sqrt{T_{++}\, T_{--}} \right)
\end{equation}	
The region $\mathscr{R}$ is a part of the space with $t<0$ with neighbourhoods $\mathscr{U}_i^\epsilon$
around each $a_i$ excised.

\begin{figure}
\centering
\begin{tikzpicture}
\draw[black, thin,->] (-5,0) -- ++ (10,0)  node[below] {$\scriptscriptstyle{x}$};
\draw[black, thin,->] (0,-2) -- ++ (0,3)  node[left] {$\scriptscriptstyle{t}$};
\draw[black, thin,->] (0,0) -- (1,1)  node[above] {$\scriptstyle{\tx^+}$};
\draw[black, thin,->] (0,0) -- (1.5,-1.5)  node[below] {$\scriptstyle{\tx^-}$};
\foreach \x in {-4,-2,1,4}
{
\draw[blue,thick,fill=blue!10,->-] (\x-0.5,0) arc (180:360:0.5);
\draw[ultra thin,red] (\x-0.5,-0.5) -- ++(0.5,0.5) -- ++(0.5,-0.5) ;
\draw[red,fill=red] (\x,0) circle [radius=0.5ex];
}
\node at (-4,0) [above] {$\scriptscriptstyle{a_1}$};
\node at (-2,0) [above] {$\scriptscriptstyle{a_2}$};
\node at (1,0) [above] {$\scriptscriptstyle{a_3}$};
\node at (4,0) [above] {$\scriptscriptstyle{a_4}$};
\end{tikzpicture}
\caption{The domain of integration $\mathscr{R}$ for \eqref{eq:LSkn3Nnai}  is the lower half space $t<0$ with half-discs $\mathscr{U}_i^\delta$ around each $a_i$ removed.  The imaginary contributions to $\pdv{a_i} \Im(S^k_\text{gr,fund})$ arise from the causal past of $a_i$.}
\label{fig:discsnN}
\end{figure}
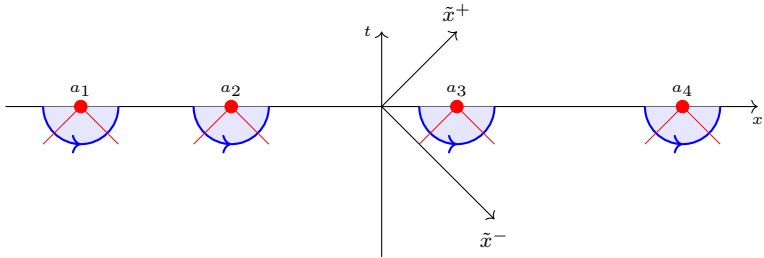

Even before we set out to compute \eqref{eq:LSkn3Nnai} let us convince ourselves that the general arguments of \cite{Colin-Ellerin:2020mva} suffice to give us the desired result. To infer this let us look back to the Euclidean computation described in \cref{sec:erennN} and note that the final result \eqref{eq:eIhatnN} indicates that the variation of the stress tensor integral,  $\pdv{a_i}\int dv d\bv \sqrt{T_{vv}\, T_{\bv\bv}} $, evaluates simply to $4\pi\, p_i$.  We view this result as saying that the local contribution arises from the Euler character which changes because of the source of energy-momentum tensor at the branch points.

To motivate this interpretation we recall again that we have carried out the integral over the radial coordinate and are left with an integral along the boundary directions to evaluate in \eqref{eq:InE}. On the contrary, \cite{Colin-Ellerin:2020mva} used a Gaussian normal chart adapted to the splitting surface to argue for the use of the complex Gauss-Bonnet theorem for the bulk Einstein-Hilbert action (supplemented by boundary terms). Continuing to carry out the integration as we have done, when we consider the variation of the bulk action with respect to the parameter $a_i$  we isolate the section of the splitting surface that is anchored at $a_i$ on the boundary.

This can be understood as follows: variation with respect to $a_i$ is a pure boundary term from the bulk perspective since one is evaluating the change of the on-shell action with respect to modified boundary conditions.  Even if we had carried out the computation using the Gauss-Bonnet theorem adapting coordinates to the splitting surface, we would have only picked up the contribution from the vicinity of the boundary -- there would have been no bulk integral to compute. The essential upshot of the Euclidean calculation is that the net variation is localized in the vicinity of the branch point at $a_i$. The simplicity of the result suggests a natural interpretation based on the above: there is a local contribution to the Euler character set by $p_i$.

Given this interpretation, we can deduce that one indeed obtains the expected result for  $ S^k_\text{gr,fund}$, viz.,
\begin{equation}\label{eq:LrennNaians}
\frac{24\pi}{c} \,  \pdv{a_i} \Im(S^k_\text{gr,fund}) = - 2\pi\, p_i \,,
\end{equation}	
by invoking the complex version of Gauss-Bonnet theorem.   Let us see this in a bit more detail. An imaginary contribution to \eqref{eq:LSkn3Nn} can arise because of the singularities at $\tx^+ = \tx^- = a_i$ which  extend into the bulk along the splitting surface. The precise value  of this imaginary part depends on the terms in the metric involving the accessory parameters $p_i$. These, by themselves, are hard to isolate in the on-shell action directly (see below). However, they can be straightforwardly extracted by considering the variation with respect to an endpoint $a_i$. In the process of taking the variation we effectively localize the computation to the neighbourhood of the branch point. In fact, in the Fefferman-Graham parameterization of the bulk geometry, the terms of interest are completely localized onto a neighbourhood of the branch point at the boundary of the spacetime.

With this picture in mind, one can trace the imaginary part to the contribution from the cut-off surfaces around the $a_i$ at the asymptotic boundary of the spacetime.  Suppose, for example, we take the cut-offs to be half-discs $\mathscr{U}_i^\delta$ as illustrated in \cref{fig:discsnN}. This choice (or indeed any other cut-off choice), will intersect the past light-cone from $a_i$. Indeed, the  local structure is dictated completely by these light-cone crossings.  The complex Gauss-Bonnet theorem  would suggest that we pick up a factor of $-2\pi i$ from such crossings.   For $\pdv{S^k_\text{gr,fund}}{a_i}$ using the Fefferman-Graham coordinate chart  we can deduce that there  is no bulk radial integral to perform along the splitting surface. However, from the earlier Euclidean analysis one learns that the contribution to the Euler characteristic is  augmented by the local source of stress energy , which is  captured by $p_i$. Putting these pieces together one is thus led to the final result quoted in \eqref{eq:LrennNaians}.

One can understand the localized nature of the contribution by referring back to the one-interval computation in \cref{sec:lren31} (which was also reduced to computing an integral along the boundary). There we had carried out the integral over the domain $\mathscr{R}$ directly after having used the fact that the accessory parameters $p_1$ and $p_2$ are fixed to be $p_1 = -p_2 = 2\, \frac{\Delta_n}{a_2-a_1}$. In that case we obtained imaginary contributions from light-cone crossings (using the principal value prescription) leading to \eqref{eq:ImSk11}. One can readily check that this result agrees with \eqref{eq:LrennNaians}.  In the evaluation of the gravity action itself we see parts where the imaginary parts cancel -- for example in the domain
$\mathfrak{D}_1$ in  \eqref{eq:mfI1eval} which is crossed by the past directed light-rays from both
$a_1$ and $a_2$. Such partial cancellations do not occur in the variation   $\pdv{a_i} \Im(S^k_\text{gr,fund}) $ which is another reason to consider it.

We emphasize the use of the complex Gauss-Bonnet theorem in the evaluation of the \eqref{eq:LSkn3Nnai} as it illustrates quite generally the moral of the discussion in \cite{Colin-Ellerin:2020mva}. One can of course check that these statements hold by choosing an explicit regulator. For instance, in \cref{sec:lrenNnApp} we employ  the light-cone regulators following the one-interval discussion. At the end of the day we find indeed
\begin{equation}\label{eq:NnrenL}
\pdv{a_i} \Im(S^k_\text{gr,fund}) = -\frac{c}{12} \, p_i \;\; \Longrightarrow \;\; \pdv{a_i} S^{(n)}
=- \frac{n}{6(n-1)} \, c\, p_i \,.
\end{equation}	
%

\subsubsection{Generalizations}
\label{sec:generalintervals}

We can use  the mnemonic that the variation of the action with respect to the end-points gives an imaginary contribution to the Lorentz signature on-shell action as in \eqref{eq:NnrenL} for more general configurations.  For instance, while we have explicitly carried out the integrals when all the intervals are taken to lie at $t=0$, we can more generally take the regions to be spacelike regions on an arbitrary boundary Cauchy slice. In this case the accessory parameters $p_i$ are complex even in Euclidean signature. We expect that they should analytically continue to real accessory parameters in Lorentz signature and lead to real stress-energy sources, and real values for the entropies.

To see this in a particular example, consider the case of two intervals $N=2$, one relatively boosted with respect to the other.  In Euclidean signature, working with the invariant cross-ratio $\chi$, the boost corresponds to rotating the finite interval $(0,\chi)$, allowing $\chi$ to have a non-zero imaginary part. For example, for $n=2$,  the branched cover geometry for the computation  of the second R\'enyi entropy is a torus with a general complex structure and the dual geometry is the rotating BTZ black hole in a suitable conformal frame.\footnote{This can be seen directly from the analysis in \cref{sec:22ren}: the cross-ratio $\chi$ is complex if one of the end-points is displaced in real-time, and the complex structure $\tau(\chi) $ then is no longer purely imaginary. }  This rotation has no effect on the monodromy differential equation which did not require any assumption of the reality, nor does it affect our conditions to determine the accessory parameter by demanding trivial monodromy around certain cycles. The main difference is that with $\chi$ complex, the accessory parameter $p_\chi$ is likewise manifestly complex.
The entropies are nevertheless real; this implies that we should integrate up \eqref{eq:NnrenL} along a suitable contour choice to obtain the physically relevant  real answers.

In the Lorentzian context, our analytic continuation $v \to \tx^-$ should be accompanied by $\chi \to \chi^-$ for $\chi$ being rotated in the Euclidean time direction (as usual we treat $\chi$ and $\bar{\chi}$ as independent in the analytic continuation). The restriction to spacelike intervals demands that $\chi_x > \chi_t > 0$. With this choice the accessory parameter $p_\chi$ is real in Lorentz signature, as is therefore the source of energy-momentum necessary to construct the branched cover geometry. The computation of the on-shell action proceeds as before, and the result for the variation of the entropies with respect to the accessory parameters  is manifestly real. Integrating with respect to $\chi^-$ leads to the expected real answers for the entropies.

There is one limiting case to consider of our example, viz.,  the limit $\chi \to 1$ whence $\chi^- = x_\chi -t_\chi \to 1$. The interval $\regA_1$ has left endpoint at $(t,x) = (0,0)$ and right endpoint at $\left( \frac{\chi^+ -\chi^-}{2},\frac{\chi^++\chi^-}{2}\right)$, while $\regA_2$ runs from $(0,1)$ to infinity.  Now as $\chi^- \to -1$, the two intervals start to approach null separation. In the limit there is no spacelike surface containing both intervals and we should see this in the result, cf., \cite{Kusuki:2017jxh}. Indeed, focusing on the  $SL(2,\mathbb{C})$ invariant mutual R\'enyi information (MRI), cf., \eqref{eq:MIren} which is a function of $\chi$ alone, purity of the global state demands that
\begin{equation}\label{eq:mriN2rel}
I^{(n)}(\chi) = I^{(n)}(1-\chi) + \frac{c}{6}\, \left(1+\frac{1}{n}\right)  \log\left(\frac{\chi}{1-\chi}\right)
\end{equation}	
In Euclidean signature  \eqref{eq:mriN2rel} implies that $I^{(n)}$ diverges as $\chi \to 1$, for using $I^{(n)}(0) =0$ we have
\begin{equation}
I^{(n)}(\chi) \sim - \frac{c}{6}\, \left(1+\frac{1}{n}\right)  \log(1-\chi) \,, \qquad 1-\chi \ll 1
\end{equation}	
Equivalently, this divergence can also be seen in the accessory parameter -- from
\eqref{eq:2inpcd} we find that $p_\chi \sim \frac{1}{2(\chi-1)}$ as $\chi \to 1$ in the connected phase (which dominates in this regime). This holds under the analytic continuation $\chi \to \chi^-$ and is the signature of the intervals failing to be on a common Cauchy slice. We expect that the result  of the two interval case generalizes to arbitrary intervals, with divergences encountered when the intervals enter into each other's causal domains.

One interesting generalization to consider is to directly evaluate the on-shell action $ \Im(S^k_\text{gr,fund})$ itself.  As mentioned earlier, we have been able to carry out the evaluation of the bulk Euclidean action, $S^E_\text{gr}$,  for the case $n=N=2$. The reader can find a detailed account of the computation in \cref{sec:direct2ren}. We work in the Fefferman-Graham gauge (in a suitable boundary conformal frame), evaluate the bulk action with a suitable cut-off of the radial coordinate (see \cref{fig:bulkfunddom}), and exploit some useful incomplete elliptic function integral identities. The mechanics of this computation being highly adapted to the Euclidean setting, we were unable to translate it directly to the Lorentzian context, in particular, were unable to extract the desired imaginary pieces from the light-cone crossings. It should be possible to do better by working in a bulk coordinate chart adapted to the splitting surface as envisaged in  \cite{Colin-Ellerin:2020mva}.

Alternately, one could at least see how to integrate up \eqref{eq:renaccesory} (the latter is blind to the spacetime signature, compare the Lorentzian \eqref{eq:NnrenL} and Euclidean results \eqref{eq:NnrenE}) to obtain the R\'enyi entropy $S^{(n)}$. As mentioned above for generic intervals with relative boosts this will require understanding an appropriate contour prescription. For two disjoint intervals with the intervals on a time symmetric slice (real cross-ratio $\chi$) this was carried out numerically  in \cite{Faulkner:2013yia}, see Figure 5 of that paper.  We note that the expressions for the accessory parameters themselves are quite simple when the intervals are far separated (for instance, for $N=n=2$ from \eqref{eq:2inpcd} we have $p_\chi \sim -\frac{3}{64}\, \chi$ for $\chi \ll 1$), but since the R\'enyi entropies are not invariant under change of conformal frame, one should pass again to working with the MRI $I^{(n)}$ which likewise has a simple variation,  $\pdv{\chi} I^{(2)}(\chi)  \sim \frac{c}{64}\, \chi$ for small $\chi$. If we consider relatively boosted intervals then $\chi$ becomes complex. However,  as we noted above, the accessory parameters are expected to be real in Lorentz signature and one should be able to obtain $ \Im(S^k_\text{gr,fund})$ without too much trouble. Moreover, this observation suggests that the contour prescription for computing the on-shell action with complex $\chi$ in Euclidean signature should be inherited  from the Lorentzian geometry.

\section{Discussion}
\label{sec:discuss}

We have exemplified the general discussion of \cite{Colin-Ellerin:2020mva} with some explicit low-dimensional examples, demonstrating a first-principles evaluation of stationary points of the real-time gravitational path integral. In particular, the on-shell action for these configurations was evaluated directly in Lorentz signature and shown to agree with the result obtained by analytically continuing the Euclidean saddle-point computations to real-time.

While our investigations were confined to analysis of the R\'enyi or swap entropies in simple states (thermofield double in JT-gravity and the vacuum state in \AdS{3}), it is clear that the  principles outlined in \cite{Colin-Ellerin:2020mva} hold more generally. The essential point is that the contributions to the gravitational path integral are localized and isolated by suitable use of the complex Gauss-Bonnet theorem. In particular, entropies can be extracted by performing the analysis in Euclidean  signature and thence analytically continuing the parameters to the real-time domain (say by moving the entangling surfaces appropriately). While this has been the modus operandi for computations of von Neumann and R\'enyi entropies both in field theory and gravity thus far, our results demonstrate the rationale behind the agreement.  In particular, they lend support to the recent investigations in the gravitational context for the evolution of the fine-grained von Neumann entropy in the context of the black hole information problem.

There are several  directions that would be interesting to pursue in the future. It would for instance be helpful to understand the evolution of entropies following a quantum quench directly in Lorentz signature. These were first investigated in two dimensional CFTs in \cite{Calabrese:2005in,Calabrese:2007rg} and studied in holography using properties of Virasoro conformal blocks in \cite{Asplund:2014coa}. Reanalyzing the results of the latter discussion directly in real-time would pave the way for more general gravitational analysis such as the fine grained entropies in black hole collapse (which has been discussed in \cite{Anous:2016kss}).

Of direct relevance to the black hole information  problem would be to construct  the real-time replica wormholes relevant to obtaining the Page curve from an evaporating black hole (even in a simple model). This investigation will be aided by computation of the bulk quantum corrections to the entropies which we have not attempted to do here.

Ideally, it would be useful to extend the gravitational computations to higher dimensional scenarios with dynamical gravitational degrees of freedom. The non-trivial aspect here would be to deal with the gravitational backreaction. Developing numerical techniques to determine complex geometries for the class of real-time boundary value problems would greatly facilitate such explorations.

\acknowledgments
It is a pleasure to thank  Veronika Hubeny, R.~Loganayagam, and Henry Maxfield for discussions. We would also like to thank Jesse Held for catching an error in  the JT example (\cref{sec:2dgrav}) in the original version (v1). 
SCE was  supported by U.S.\ Department of Energy grant {DE-SC0019480} under the HEP-QIS QuantISED program.
XD was supported in part by the National Science Foundation under Grant No.\ PHY-1820908 and by funds from the University of California.  DM and ZW were supported by NSF grant PHY1801805 and funds from the University of California. MR  was supported by  U.S. Department of Energy grant DE-SC0009999 and by funds from the University of California.

\appendix
\section{A Rindler regulator for on-shell action of  the semi-infinite interval}
\label{sec:rindreg}

In this appendix we provide an alternate calculation to that given in \cref{sec:lren31} for the R\'enyi entropy of a semi-infinite interval from the Lorentzian on-shell action in a single fundamental domain. The calculation in the main text used a small polygonal cut-off around the branch point with an $i\varepsilon$ prescription. The imaginary part of the action then came from the principal value prescription. Here we will instead evaluate the Lorentzian action in Rindler coordinates with cut-off surfaces of constant Rindler radius. The imaginary part of the action now comes from the excursion into the Euclidean time direction as we pass between the different wedges.

\begin{figure}[h]
\begin{center}
\begin{tikzpicture}[scale=1]


\tikzset{>=stealth}
\draw[->,black] (-4,-1) -- (-4,3);
\draw[->,black] (-7,0) -- (-1,0);
\draw[black] (-1.45,2.78) -- (-1.06,2.78);
\draw[black] (-1.45,2.78) -- (-1.45,3.2);
\node at (-1.2,3) {${\sf t}_{_L}$};
\draw[ thick,orange] (-6.5,0) -- (-4,0);
\draw[ thick,orange] (-6.5,1) -- (-1.5,1);
\draw[ thick,orange] (-4,2) -- (-1.5,2);
\draw[ thick,dashed,orange] (-6.5,0) -- (-6.5,1);
\draw[ thick,dashed,orange] (-1.5,1) -- (-1.5,2);
\draw[->, thick,orange] (-5.27,0) -- (-5.28,0);
\draw[->, thick,orange] (-6.5,0.6) -- (-6.5,0.6);
\draw[->, thick,orange] (-4,1) -- (-3.9,1);
\draw[->, thick,orange] (-1.5,1.6) -- (-1.5,1.6);
\draw[->, thick,orange] (-2.77,2) -- (-2.78,2);
\node at (-6.55,-0.4) {$-T$};
\node at (-1.45,-0.4) {$T$};
\node at (-3.75,1.3) {$\frac{\pi}{2}i$};
\node at (-3.75,2.3) {$\pi i$};


\begin{axis}
[xmin=-5.5,xmax=5.5, ymin=-5,ymax=5,
axis line style={draw=none},
tick style={draw=none},
xticklabels={,,},yticklabels={,,},
xshift=1.6cm,yshift=-0.75cm]
\addplot [blue,thick,domain=-1.5:0] ({2*cosh(x)}, {2*sinh(x)});
\addplot [blue,thick,domain=-1.5:0] ({-2*cosh(x)}, {2*sinh(x)});
\addplot [blue,thick,domain=-1.5:1.5] ({-2*sinh(x)}, {-2*cosh(x)});
\addplot [orange, thick,domain=-1:0] ({3*cosh(x)}, {3*sinh(x)});
\addplot [orange, thick,domain=-1:0] ({-3*cosh(x)}, {3*sinh(x)});
\addplot [orange, thick,domain=-1:1] ({-3*sinh(x)}, {-3*cosh(x)});
\addplot[black,dashed,domain=-5:5] expression {-abs(x)};
\end{axis}
\draw[black,->] (5.04,1.7) -- (5.04,2);
\draw[black,->] (5.04,1.3) -- (5.04,1.0);
\node at (5.04,1.5) {$\delta$};
\draw[black,fill=black] (5.04,2.1) circle (2pt);
\draw[snake=zigzag,color=green!60!black,segment length=3pt,segment amplitude=0.4mm] (5.04,2.1) -- (8,2.1);
\node at (6.6,2.4) {$A$};
\draw[->, thick,orange] (2.863,1.1) -- (2.75,0.9);
\draw[->, thick,orange] (2.56,-0.253) -- (2.565,-0.258);
\draw[->, thick,orange] (5.12,0.386) -- (5.13,0.386);
\draw[->, thick,orange] (7.583,-0.1638) -- (7.588,-0.159);
\draw[->, thick,orange] (7.313,0.9) -- (7.20,1.1);
\draw[dashed, thick,orange] (2.86,-0.525) -- (2.18,0.095);
\draw[dashed, thick,orange] (7.18,-0.53) -- (7.87,0.095);
\draw (1.8,2.1) -- (8,2.1);
\node at (8.5,2.1) {$t=0$};
\end{tikzpicture}
\caption{Left: integration contour in the complex ${\sf t}_{_L}$-plane. Right: boundary spacetime $\mathbb{R}^{1,1}$ for $t<0$ with cut-offs (blue) at Rindler radius $r=\delta$. The ${\sf t}_{_L}$ contour (orange) has an excursion into the Euclidean time domain (dashed) as it passes between wedges.}
\label{fig:Rindlerregulator}
\end{center}
\end{figure}

We start with the Lorentzian action after integration over the bulk coordinates which is given by \eqref{eq: LSkn31} and \eqref{eq:Lhalf31}. In particular, we want to evaluate $\mathfrak{I}_{_\text{half-line}}$ which we rewrite here for the reader's convenience:
\begin{equation}\label{eq:Lhalf31_2}
\mathfrak{I}_{_\text{half-line}} = \int_{t<0}\, \frac{d\tx^+\, d\tx^-}{\tx^+\, \tx^-}.
\end{equation}
We transform to Rindler coordinates $({\sf t},\rho)$ and impose cut-offs at some very small Rindler radius $\rho = \delta$. Recall that to pass between wedges we shift ${\sf t}$ in the imaginary direction by $i \frac{\pi}{2}$. We require that the time contour be continuous so we must include the integration along this imaginary direction from $0$ to $i\frac{\pi}{2}$. Therefore, the time contour for ${\sf t}_{_L}$ is given by
\begin{equation}\label{eq:Rindlertimecontour}
C_{T} = [0,-T] \cup \left[-T,-T+i\frac{\pi}{2}\right] \cup \left[-T+i\frac{\pi}{2},T+i\frac{\pi}{2}\right] \cup \left[T+i\frac{\pi}{2},T+i\pi\right] \cup \left[T+i\pi,i\pi\right],
\end{equation}
where we have put in some large time cut-off $T$. The contour is depicted in \cref{fig:Rindlerregulator}.

The integral giving $\mathfrak{I}_{_\text{half-line}}$ is now trivial:
\begin{equation}\label{eq:Lhalf31_Rindler}
\mathfrak{I}_{_\text{half-line}} = 2 \lim_{T \to \infty}\int_{C_{T}} \, d {\sf t}_{_L}  \, \int_{\delta}^{L} \frac{dr}{r} = 2\pi i \log \left(\frac{L}{\delta}\right).
\end{equation}
This agrees with the result from \cref{sec:eren31}. In particular, it verifies that there is a missing factor of $2$ if one only considers the branch point at the origin. This way of doing the calculation makes it manifest how the imaginary part of the Lorentzian action gives the Euclidean action because the imaginary part comes from an explicit integration over Euclidean time.

\section{Lorentzian action for disjoint interval R\'enyi entropies}
\label{sec:lrenNnApp}

%
\begin{figure}
\centering
\begin{tikzpicture}
\draw[black, thin,->] (-5,0) -- ++ (10,0)  node[below] {$\scriptscriptstyle{x}$};
\draw[black, thin,->] (0,-2) -- ++ (0,3)  node[left] {$\scriptscriptstyle{t}$};
\draw[black, thin,->] (0,0) -- (1,1)  node[above] {$\scriptstyle{\tx^+}$};
\draw[black, thin,->] (0,0) -- (1.5,-1.5)  node[below] {$\scriptstyle{\tx^-}$};
\foreach \x in {-4,-2,1,4}
{
\draw[blue,thick,fill=blue!10,-<-] (\x+0.5,0) -- ++(-0.5,-0.5) -- ++(-0.5,0.5)--cycle;
\draw[red,fill=red] (\x,0) circle [radius=0.5ex];
}
\node at (-4,0) [above] {$\scriptscriptstyle{a_1}$};
\node at (-2,0) [above] {$\scriptscriptstyle{a_2}$};
\node at (1,0) [above] {$\scriptscriptstyle{a_3}$};
\node at (4,0) [above] {$\scriptscriptstyle{a_4}$};
\end{tikzpicture}
\caption{The domain of integration $\mathscr{R}$ for \eqref{eq:LSkn3Nnai}  is the lower half space $t<0$ with triangular regions $\mathscr{U}_i^\epsilon$ around each $a_i$ removed. This choice is particularly convenient for the light-cone like coordinates $\tx^\pm$ that we work with, since the boundaries of the region $\mathscr{R}$ lie at
constant $\tx^\pm = a_i \mp \delta$. }
\label{fig:stripsnN}
\end{figure}
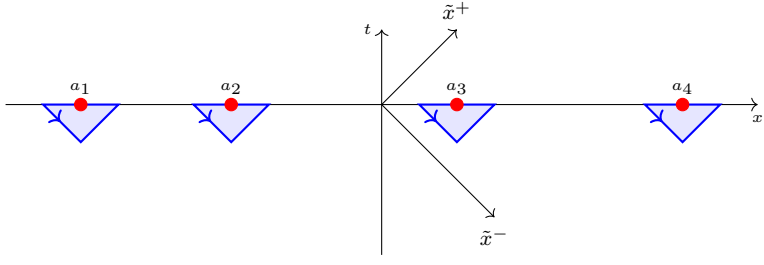

To evaluate the integral in \eqref{eq:LSkn3Nnai} directly we consider the lower half of the $(x,t)$ plane ($t<0$)
and use the past directed light-rays from $x= a_i\pm \delta$ to carve out little triangular regions which we excise, see \cref{fig:stripsnN}. Thus,
\begin{equation}
\begin{split}
\mathscr{R}
&=
	\bigg(\mathbb{R}^{1,1} \cap \{(x,t) | t<0\} \bigg)
	\backslash
		\bigcup_{i=1}^{2N} \mathscr{U}_i^\delta \\
\mathscr{U}_i^\delta
&=
		\bigg\{ (\tx^+,\tx^-) \big| \; \tx^+ \in (a_i-\delta, a_i+\delta)  \;\; \& \;\;
		\tx^- <a_i+\delta \;\; \& \;\;
		\tx^+ -\tx^- < 0
		\bigg\} .
\end{split}
\end{equation}	
For future use let us also define the boundaries of $\mathscr{U}_i^\delta$ as
\begin{equation}
\begin{split}
\partial \mathscr{U}_{i}^+ &=
	\{ \tx^+ = a_i-\delta \,, \  \tx^- \in [a_i-\delta, a_i+\delta] \} \\
\partial \mathscr{U}_{i}^- &=
	\{ \tx^- = a_i+\delta \,, \  \tx^+ \in [a_i+\delta, a_i-\delta] \}	
\end{split}
\end{equation}	
where we have specified the ranges consistent with the orientation of the boundaries.

To compute the integral we will introduce a function $\mathfrak{T}(\tx^+,\tx^-) $ whose light-cone derivatives  give the two terms in the integrand
\begin{equation}
\partial_- \mathfrak{T}(\tx^+,\tx^-)  = \sqrt{T_{--}} \,, \qquad \partial_+ \mathfrak{T}(\tx^+,\tx^-)  = \sqrt{T_{++}} \,.
\end{equation}	
We will content ourselves with local behaviour near the sources $a_i$ which are given by the Lorentzian analog of \eqref{eq:Flocal}
\begin{equation}\label{eq:FlocalL}
\begin{split}
\mathfrak{T}(\tx^+,\tx^-) &=
 	s_i \left[ \sqrt{\Delta_n} \log \left[(\tx^- -a_i)(\tx^+-a_i)\right] - C_i + \frac{p_i}{2\sqrt{\Delta_n}} (\tx^-+\tx^+ -2\,a_i)
	 + \cdots\right] \\
\partial_\pm\mathfrak{T}(\tx^+,\tx^-) &=
	 s_i \left[ \frac{\sqrt{\Delta_n}}{\tx^\pm -a_i} + \frac{p_i}{2\sqrt{\Delta_n}}  + \cdots\right] \\
 \pdv{a_j}\mathfrak{T}(\tx^+,\tx^-) &=
	-s_i\left[\sqrt{\Delta_n} \left(\frac{1}{\tx^--a_i}  + \frac{1}{\tx^+-a_i} \right) +\frac{p_i}{\sqrt{\Delta_n}}\right] \, \delta_{ij}  - s_i\, \ \pdv{C_i}{a_j} + \cdots \,,
\end{split}
\end{equation}	
where the ellipses denote higher order terms in the local expansion about $\tx^+= \tx^- = a_i$.

We therefore have to evaluate
\begin{equation}
\begin{split}
 \pdv{a_i} S^k_\text{gr,fund}
 &=
  \frac{c}{24\pi} \left[ \mathfrak{I}_\text{bulk} + \mathfrak{I}_\text{bdy}\right]\\
\mathfrak{I}_\text{bulk}
&=
	 \int_{\mathscr{R}} \, d\tx^- \, d\tx^+ \,
	\pdv{a_i} (\partial_-\mathfrak{T} \; \partial_+\mathfrak{T} ) \\
\mathfrak{I}_\text{bdy}
&=
  \int_{\partial\mathscr{U}_i^+} \, d\tx^-   \partial_-\mathfrak{T} \;\partial_+\mathfrak{T} \bigg|_{\tx^+=a_i-\delta}
  +
    \int_{\partial\mathscr{U}_i^-} \, d\tx^+   \partial_-\mathfrak{T} \;\partial_+\mathfrak{T} \bigg|_{\tx^-=a_i+\delta}  	
\end{split}
\end{equation}	

Let us first evaluate the boundary integral which is straightforward as we have to compute contributions of the form
\begin{equation}
\begin{split}
\mathfrak{I}_\text{bdy}
&=
	\int_{a_i-\delta}^{a_+\delta} \, d\tx^-   \left[\frac{\sqrt{\Delta_n}}{\tx^- -a_j} + \frac{p_j}{2\,\sqrt{\Delta_n}} + \cdots\right]
		\left[\frac{\sqrt{\Delta_n}}{-\delta} + \frac{p_i}{2\,\sqrt{\Delta_n}} + \cdots\right]\\
&\qquad
	+\int_{a_i+\delta}^{a_i-\delta} \, d\tx^+   \left[\frac{\sqrt{\Delta_n}}{\delta} + \frac{p_i}{2\,\sqrt{\Delta_n}} +\cdots
		\right]
		\left[\frac{\sqrt{\Delta_n}}{\tx^+-a_j} + \frac{p_j}{2\,\sqrt{\Delta_n}} + \cdots \right] .
\end{split}
\end{equation}	
We see that the only part that contributes is the one where the terms align, i.e., only from $i=j$, since this is the only situation when the integral has a non-vanishing imaginary part from the  principal value prescription. Therefore,  keeping track of the orientation of the boundary we find the two terms add to give
\begin{equation}
\Im(\mathfrak{I}_\text{bdy}) =
	- \pi \left[\frac{\Delta_n}{-\delta} + \frac{p_i}{2}\right] - \pi 	\left[\frac{\Delta_n}{\delta} + \frac{p_i}{2}\right]
	= -\pi\, p_i\,.
\end{equation}	

The bulk terms can be evaluated along similar lines. We first exchange the order of integration and use the fact that $\partial_+\partial_- \mathfrak{T}$ has no support in the region of integration: it is localized at the branch points following the same chain of logic that led to the first line of \eqref{eq:lT31}. Hence,
\begin{equation}
\begin{split}
\mathfrak{I}_\text{bulk}
&=
	 \int_\mathscr{R} \, d\tx^-\,d\tx^+
			\partial_- \left(\partial_{a_i}\mathfrak{T} \; \partial_+\mathfrak{T} \right)
			+
			\partial_+\left(\partial_{a_i}\mathfrak{T} \; \partial_-\mathfrak{T} \right) \\
&=
	\sum_{j=1}^{2N}
		\left[\int_{\partial \mathscr{U}_j^-} \, d\tx^+\, \partial_{a_i}\mathfrak{T} \; \partial_+\mathfrak{T}
		+
		\int_{\partial \mathscr{U}_j^+}  \, d\tx^-\, \partial_{a_i}\mathfrak{T} \; \partial_-\mathfrak{T} \right]
\end{split}
\end{equation}	
It is now straightforward to use \eqref{eq:FlocalL} and compute each of the terms in the above. For instance we have to compute integrals of the form:
\begin{equation}
\begin{split}
	- \int_{a_j+\delta}^{a_j-\delta} \, d\tx^+ &
	\left[
		\left( \frac{\sqrt{\Delta_n}}{\delta}  + \frac{\sqrt{\Delta_n}}{\tx^+-a_j}
		 +p_j\right) \, \delta_{ij} +  \pdv{C_j}{a_i} + \cdots\right]  \left[\frac{\sqrt{\Delta_n}}{\tx^+-a_j} + \frac{p_j}{2\,\sqrt{\Delta_n}} + \cdots \right]\\
&
	= i\pi \left[\frac{3p_i}{2}\, \delta_{ij}  + \pdv{C_j}{a_i} + \frac{\Delta_n}{\delta} \, \delta_{ij}\right] + \cdots \\
- \int_{a_j-\delta}^{a_j+\delta} \, d\tx^- &
	\left[
		\left( -\frac{\sqrt{\Delta_n}}{\delta}  + \frac{\sqrt{\Delta_n}}{\tx^- - a_j}
		 +p_j\right) \, \delta_{ij} +  \pdv{C_j}{a_i} + \cdots\right]  \left[\frac{\sqrt{\Delta_n}}{\tx^- - a_j} + \frac{p_j}{2\,\sqrt{\Delta_n}} + \cdots \right]  \\
&
	= i\pi \left[\frac{3p_i}{2}\, \delta_{ij}  + \pdv{C_j}{a_i} - \frac{\Delta_n}{\delta} \, \delta_{ij}\right] + \cdots
\end{split}
\end{equation}
where we have only indicated explicitly the imaginary parts that arise from the principal value prescription.
The terms combine nicely together to give
\begin{equation}
\Im(\mathfrak{I}_\text{bulk})  = 2\pi  \left( \frac{3p_i}{2}\,  + \sqrt{\Delta_n}\, \sum_{j=1}^{2N}\pdv{C_j}{a_i}\right)  = - \pi \, p_i
\end{equation}	
where we finally used \eqref{eq:Cidetermine}.

Putting it all together we have the expected result from the Lorentzian replica computation, viz., \eqref{eq:NnrenL}.
As noted earlier this was to be expected owing to the contributions arising from the regions where the metric becomes complex.

\section{The second R\'enyi entropy for two intervals: Geometry}
\label{sec:22ren}
In this appendix we give explicit details for the 2-interval second R\'enyi entopy. We focus on the Schottky construction on the boundary and the determination of the bulk handlebody geometries. We will use these results in  \cref{sec:direct2ren}  to compute the on-shell action of the gravitational dual.

\subsection{The boundary geometry}
\label{sec:22renbdy}

The boundary manifold has a complex structure
\begin{equation}\label{eq:torusc}
z^2 = \frac{(v-a_1)(v-a_3)}{(v-a_2)(v-a_4)} \;\; \Longrightarrow \;\; z^2 = \frac{v(v-1)}{(v-\chi)} \,,
\end{equation}	
where we have used a M\"obius transformation to set $a_1=0$, $a_2 = \chi$, $a_3 =1$, and $a_4 \to \infty$, respectively.  In particular, we have
\begin{equation}\label{eq:chiadef}
\chi = \frac{(a_1-a_2)(a_3-a_4)}{(a_1-a_3)(a_2-a_4)}\,.
\end{equation}	
The modulus of the torus is given in terms of the elliptic integral, cf., \eqref{eq:Kelliptic},
\begin{equation}\label{eq:tauchi}
\tau(\chi) = i\,  \frac{K(1-\chi)}{K(\chi)}\,,
\end{equation}	
which implies that a modular transform $\tau \leftrightarrow -\frac{1}{\tau}$ corresponds to the exchange $\chi \leftrightarrow 1-\chi$.

Note however, that the R\'enyi entropy is not invariant under $\mathrm{SL}(2,\mathbb{C})$ transformations which we used to gauge fix $a_i$.  On the contrary the mutual R\'enyi information  defined by
\begin{equation}\label{eq:MIren}
I^{(2)}_{\regA_1 \cup \regA_2} = S^{(2)}_{\regA_1} + S^{(2)}_{ \regA_2} - S^{(2)}_{\regA_1 \cup \regA_2}\,,
\end{equation}	
is invariant under $\mathrm{SL}(2,\mathbb{C})$. Using the purity of the vacuum state one can relate $I^{(n)}(1-\chi)$ to $I^{(n)}(\chi)$. The swap $\chi \leftrightarrow 1-\chi$ which is achieved by $a_2 \leftrightarrow a_4$  exchanges the two choices of cycles, $\mathfrak{C}_d \leftrightarrow \mathfrak{C}_c$.  One can use Schottky uniformization\footnote{We explain the elements underlying the Schottky uniformization calculation in  \cref{sec:schottky} and derive the result by explicitly evaluating the on-shell gravitational action in Euclidean signature in \cref{sec:direct2ren}.}  to directly determine  \cite{Faulkner:2013yia}
\begin{equation}\label{eq:rmi2int}
\begin{split}
I^{(2)}(\chi)
&=
	\max\left\{I^{(2)}_{\mathfrak{C}_d}(\chi), I^{(2)}_{\mathfrak{C}_c}(\chi)\right\} \\
&=
	\max \left\{ -\frac{c}{12}\, \log\left(2^8 \, \frac{1-\chi}{\chi^2}\right) - i\, \frac{\pi}{6}\, c\, \tau(\chi) ,
	-\frac{c}{12}\, \log\left(2^8 \, \frac{1-\chi}{\chi^2}\right) + i\, \frac{\pi}{6}\, c\,  \frac{1}{\tau(\chi)}
	\right\} .
\end{split}
\end{equation}	
This result was first derived in \cite{Headrick:2010zt} and leads to the aforesaid phase transition since $I^{(2)}_{\mathfrak{C}_d}(\chi)$ dominates for $\chi <\frac{1}{2}$.

As explained above, one could construct directly the covering space handlebody, and recover from it the on-shell action for the geometry.
A direct evaluation of on-shell action turns out to be formidable even for the case of the $2^\text{nd}$ R\'enyi entropy for two-intervals. We were however able to derive \eqref{eq:rmi2int} directly by computing the gravitational action in Euclidean signature. As this computation has not been reported in the literature we present it in \cref{sec:direct2ren}. However, we found it somewhat  cumbersome to manipulate for the real-time analysis, so we resorted to a different approach in the main text.

\subsection{The Euclidean handlebodies}
\label{sec:ehandle}

For $N=2$ and $n=2$ the monodromy problem relies on the following differential equation
\begin{equation}\label{eq:montorus}
\psi''(v) +
	\frac{1}{2}\Bigg[\Delta_2\, \bigg(\frac{1}{v^2}+\frac{1}{(v-1)^2}+\frac{1}{(v-\chi)^2}-\frac{2}{v(v-\chi)}\bigg)
	-\frac{p_{\chi}\, \chi\, (\chi-1)}{v(v-1)(v-\chi)}\Bigg]\psi(v)=0 \,.
\end{equation}	
In writing this expression we have gauge fixed the  branch points using \eqref{eq:chiadef} and set $p_\chi = - p_2$ and set $n=2$ after using the relations in \eqref{eq:accpinfrel}. While the natural map on the cover is  \eqref{eq:torusc}, for solving \eqref{eq:montorus} it will be useful to introduce a new elliptic coordinate $\eta(v)$; see \eqref{eq:etavint} and rewrite \eqref{eq:montorus} using the conformal transformation properties of $\psi(v)$ and $T(v)$. Under $v\to \mathfrak{f}(v)$ one has
\begin{equation}
\psi(v) = \left(\pdv{\mathfrak{f}}{v}\right)^{-\frac{1}{2}} \, \psi(\mathfrak{f}(v)) \,, \qquad T_{vv} =  \left(\pdv{\mathfrak{f}}{v}\right)^2\, T_{\mathfrak{f}\mathfrak{f}} + \{ \mathfrak{f},v\} \,.
\end{equation}	
This implies that the monodromy equation can be brought to the form of a standard differential equation
\begin{equation}\label{eq:montoruseta}
\dv[2]{\psi(\eta)}{\eta} - \frac{2K(\chi)^2}{\pi^2} \left(\frac{\chi-2}{4} + p_\chi \, \chi (\chi-1) \right) \psi(\eta)  =0 \,.
\end{equation}	
This has  solutions in terms of simple exponentials if we also reparameterize the accessory parameter as
\begin{equation}\label{eq:pchibeta}
p_\chi = \frac{1}{\chi(\chi-1)} \left[ \frac{2-\chi}{4} + \frac{\pi^2 }{2\, K(\chi)^2} \, \mathfrak{p}^2 \right]  \,.
\end{equation}	
Altogether we find that the desired solution to the inverse map $\tilde{y}(v)$ is given by
\begin{equation}\label{eq:schottkycoord}
\tilde{y}(v) = e^{2\, \mathfrak{p} \, \eta(v)} \,.
\end{equation}	
To complete the solution we need to fix $p_\chi$ by computing the monodromies around the two possible choices of cycles: the disconnected one $\mathfrak{C}_d$ and the connected one $\mathfrak{C}_c$ in \cref{fig:twointmonodromies}.

For two intervals the second R\'enyi entropy computation leads to the following stress energy on a single sheet (using $\Delta_2 = \frac{3}{8}$):
\begin{equation}\label{eq:Tvv2int}
T_{vv}(v) = \frac{3}{8}\, \bigg(\frac{1}{v^2}+\frac{1}{(v-1)^2}+\frac{1}{(v-\chi)^2}-\frac{2}{v(v-\chi)}\bigg)
	-\frac{p_{\chi}\, \chi\, (\chi-1)}{v(v-1)(v-\chi)}.
\end{equation}	
To complete its specification we fix $p_\chi$ by computing the monodromies around the two possible choices of cycles $\mathfrak{C}_d$ and $\mathfrak{C}_c$ in \cref{fig:twointmonodromies}.  One finds:
\begin{equation}\label{eq:2inpcd}
\begin{split}
\mathfrak{p}_{d} = -\frac{i}{2}
	& \;\; \Longrightarrow \;\; p_{\chi}\big|_{\mathfrak{C}_d} = \frac{1}{4\chi(\chi-1)} \left[ 2-\chi - \frac{\pi^2}{2\,K(\chi)^2}\right] \,,\\
\mathfrak{p}_{c} = \frac{i}{2\,\tau(\chi)}
	& \;\; \Longrightarrow \;\; p_{\chi}\big|_{\mathfrak{C}_c} = \frac{1}{4\chi(\chi-1)} \left[ 2-\chi  + \frac{\pi^2}{2\,K(1-\chi)^2}\right] \,,
\end{split}
\end{equation}
where $\tau(\chi)$ is the modulus of the torus and is defined in \eqref{eq:tauchi}.

These are given in \eqref{eq:2inpcd} as a function of the cross-ratio $\chi$.

\paragraph{The torus elliptic map:}

The elliptic map from the complex $v$ plane to the torus is
\begin{equation}\label{eq:etavint}
\begin{split}
\eta(v) = \frac{\pi}{2 K(\chi)} \, \int_0^v \, \frac{d\zeta}{\sqrt{\zeta(\chi-\zeta) (1-\zeta)}} \,.
\end{split}
\end{equation}
We can either invoke Legendre integral definition of the incomplete elliptic function\footnote{ We define $K(x)$ to be the incomplete elliptic integral of the first kind as in \eqref{eq:Kelliptic}. The definition  differs from some traditional forms, which define the integral in \eqref{eq:Kelliptic} as $ F(\frac{\pi}{2}, \sqrt{x})$; see for example \cite[Eq.~19.2.4]{NIST-DLMF}.} or the inverse Jacobi elliptic sine (denoted $\sn(z,m)$) amplitude, and write
\begin{equation}\label{eq:etavel}
\begin{split}
\eta(v) = \frac{\pi}{K(\chi)} \,  F\left( \arcsin(\sqrt{\frac{v}{\chi} }),  \chi\right)
= \frac{\pi}{K(\chi)} \, \sn^{-1}\left( \arcsin(\sqrt{\frac{v}{\chi} }),  \chi\right)
\end{split}
\end{equation}
which fixes the function in the principal domain $v \in [0,\chi]$. For the other domains we  analytically continue past the cuts  which are at $(0,\chi)$ and $(1,\infty)$.  The normalization factor is the complete elliptic integral of the first kind
\begin{equation}\label{eq:Kelliptic}
K(x) = \int_0^\frac{\pi}{2} \, \frac{d\theta}{\sqrt{1-x\, \sin^2\theta}} \equiv F\left(\frac{\pi}{2} ,x\right) .
\end{equation}	
%

\section{The second R\'enyi entropy for two intervals: Euclidean on-shell action}
\label{sec:direct2ren}

In this appendix we evaluate the on-shell gravitational action in Euclidean signature for the second R\'enyi entropy for two intervals. We will compute the action of the covering space $\bulk_2$ using the Schottky construction outlined in \cref{sec:schottky}. The action was evaluated numerically for higher R\'enyi entropies ($n>2$) in \cite{Faulkner:2013yia} and here we will evaluate it analytically for $n=2$. From this we can extract the second mutual R\'enyi information and thus derive \eqref{eq:rmi2int}.

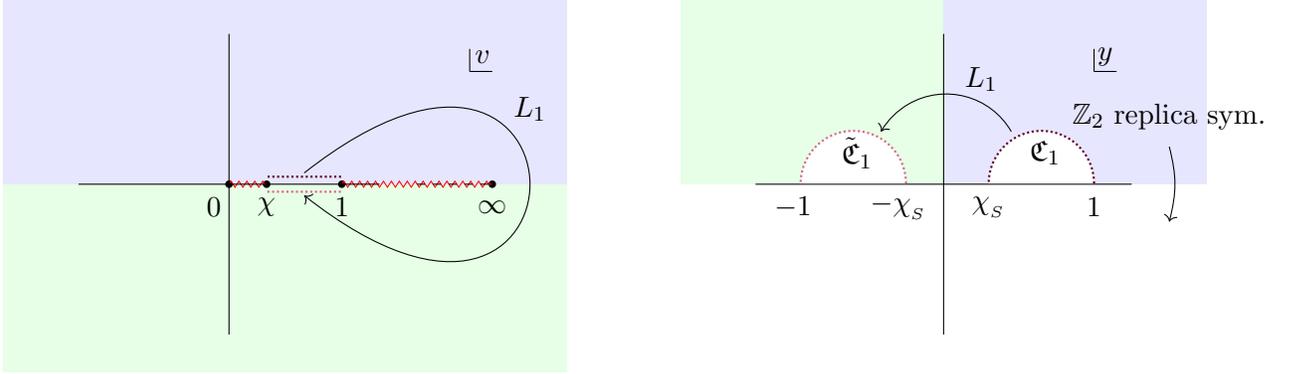
\begin{figure}[h]
\begin{center}
\begin{tikzpicture}[scale=1]


\fill[blue!10!white] (-10,0) rectangle (-2.5,2.5);
\fill[green!10!white] (-10,0) rectangle (-2.5,-2.5);
\draw[->] (-6,0.15) .. controls (-2,3.3) and (-2,-3.3) .. (-6,-0.15);
\node at (-7.2,-0.3) {$0$};
\node at (-6.5,-0.3) {$\chi$};
\node at (-5.5,-0.3) {$1$};
\node at (-3.5,-0.3) {$\infty$};
\fill (-7,0) circle (0.05cm);
\fill (-6.5,0) circle (0.05cm);
\fill (-5.5,0) circle (0.05cm);
\fill (-3.5,0) circle (0.05cm);
\draw[snake=zigzag,color=red,segment length=3pt,segment amplitude=0.4mm] (-7,0) -- (-6.5,0);
\draw[snake=zigzag,color=red,segment length=3pt,segment amplitude=0.4mm] (-5.5,0) -- (-3.5,0);
\draw[black] (-9,0) -- (-5,0);
\draw[black] (-7,-2) -- (-7,2);
\draw[black,dashed] (-4.5,0) -- (-3.5,0);
\draw[black] (-3.8,1.5) -- (-3.5,1.5);
\draw[black] (-3.8,1.5) -- (-3.8,1.8);
\node at (-3.63,1.7) {$v$};
\draw[thick,densely dotted,purple!60!white] (-6.49,-0.1) -- (-5.5,-0.1);
\draw[thick,densely dotted,purple!50!black] (-5.51,0.1) -- (-6.5,0.1);
\node at (-3,1) {$L_{1}$};


\fill[green!10!white] (-1,0) rectangle (4,2.5);
\fill[blue!10!white] (2.5,0) rectangle (6,2.5);
\draw[black] (0,0) -- (5,0);
\draw[black] (2.5,-2) -- (2.5,2);
\draw[black] (4.8,1.5) -- (4.5,1.5);
\draw[black] (4.5,1.5) -- (4.5,1.8);
\node at (4.65,1.7) {$\yt$};
\draw[thick,densely dotted,draw=purple!50!black,fill=white] (4.5,0.01) arc (0:180:0.7cm);
\filldraw[thick,densely dotted,draw=purple!60!white,fill=white] (2,0.01) arc (0:180:0.7cm);
\node at (3.85,0.42) {$\mathfrak{C}_{1}$};
\node at (1.35,0.42) {$\tilde{\mathfrak{C}}_{1}$};
\node at (3.1,-0.3) {$\xs$};
\node at (4.5,-0.3) {$1$};
\node at (1.9,-0.3) {$-\xs$};
\node at (0.5,-0.3) {$-1$};
\draw[->] (3.4,0.7) arc (30:150:1cm);
\node at (3,1.4) {$L_{1}$};
\draw[->] (5.5,0.5) .. controls (5.6,0.1) and (5.6,-0.1) .. (5.5, -0.5);
\node at (5.5,0.9) {$\mathbb{Z}_{2}$ replica sym.};
\end{tikzpicture}
\caption{Left: one sheet of the boundary geometry $\bdy_{2,2}$ with the generator $L_{1}$ of the Schottky group corresponding to non-trivial monodromy around the cycle containing one of the branch cuts. Right: the image of $\bdy_{2,2}$ in the $\yt$-plane with the two circles $\mathfrak{C}_1, \tilde{\mathfrak{C}}_1$ identified by the action of the Schottky group and their interiors removed to give the fundamental domain. The upper and lower $\yt$-plane are related by the $\mathbb{Z}_{2}$ replica symmetry with each corresponding to a sheet of $\bdy_{2,2}$.}
\label{fig:tildeyplane}
\end{center}
\end{figure}

For definiteness we will focus on the choice of cycles $\mathfrak{C}_c$, but the other choice of cycles follows similarly. We previously obtained the coordinate $\tilde{y}(v)$ for the Schottky domain of the boundary torus \eqref{eq:schottkycoord}. For the purposes of evaluating the action, it is nicer to use a different coordinate for the Schottky domain which is related to $\tilde{y}(v)$  in \eqref{eq:schottkycoord} by a $\mathrm{PSL}(2,\mathbb{C})$ transformation.\footnote{This is not strictly a $\mathrm{PSL}(2,\mathbb{C})$ transformation because the determinant is not equal to $1$, but the Schottky construction is only defined up to an overall scaling, which we have chosen such that $\yt(1)=1$.} We choose our new coordinate $\yt(v)$ to diagonalize the monodromy around $a_{1}$ which gives
\begin{equation}\label{eqn:newycoord}
\yt(v) = \tanh(\pi \mathfrak{p}_{c} )\tanh\left(\mathfrak{p}_{c} \eta(v)\right) .
\end{equation}
To keep future expressions legible we also introduce a parameter $\xs$ encoding the complex structure via
\begin{equation}\label{eq:xsdef}
\xs \equiv \yt(\chi) = \tanh^2(\pi\mathfrak{p}_c),
\end{equation}	
where we have used $\eta(\chi) = \pi$.
The fundamental domain $\mathcal{D}_{\mathrm{bdy}}$ of the Schottky quotient is the exterior of the two discs bounded by the circles $\mathfrak{C}_1, \tilde{\mathfrak{C}}_1$ which we have illustrated in \cref{fig:tildeyplane}. The generator of the Schottky group identifies these two circles as discussed previously. The replica symmetry acts simply on the fundamental domain described by the $\yt(v)$ coordinate: $\yt(v) \to -\yt(v)$.

\begin{figure}
\begin{center}
\begin{tikzpicture}[scale=1]

\fill[color=green!30!white] (-7,-0.35) .. controls (-3,-0.6) and (-2,0.1) .. (3,-0.4) .. controls (3.9, 1) and (4.5,0.5) .. (5,1.8) .. controls (1,1.4) and (-1,2.1) .. (-5,1.8) .. controls (-5.3,0.8) and (-6.2,0.9) .. (-7,-0.35);
\node at (-7,-0.7) {$\xi = \xi_c(\yt,\bar{\yt})$};
\fill[ball color=blue!70!white, opacity=0.50]  (-4.55,0.9) .. controls (-4.35,0.23) and (-1.55,0.374) .. (-1.54,1.05) -- (-1.545,1) arc (362:183:1.501cm) -- cycle;
\fill[ball color=blue!70!white, opacity=0.50]  (-0.45,0.93) .. controls (-0.25,0.26) and (2.55,0.404) .. (2.56,1.08) -- (2.555,1.03) arc (362:183:1.501cm) -- cycle;
\filldraw[draw=black,fill=red!50!white, fill opacity=0.7] (-5,2) -- (5,2) -- (3,0) -- (-7,0) -- (-5,2) -- cycle;
\filldraw[draw=blue,fill=white,rotate=2.5] (-3,1.1) ellipse (1.5 and 0.5);
\filldraw[draw=blue,fill=white,rotate=2.5] (1.1,0.95) ellipse (1.5 and 0.5);
\node at (1.15,0.8) {$\mathfrak{C}_{1}$};
\node at (-3.1,0.8) {$\tilde{\mathfrak{C}}_{1}$};
\draw[<-] (-2.4,-0.6) arc (210:330:1.6cm);
\node at (-1,-1.8) {$L_{1}$};
\draw[->] (5.2,1.7) -- (5.2,-1);
\node at (5.6,-0.8) {$\xi$};
\draw (4.1,1.6) -- (4.6,1.6);
\draw(4.1,1.6) -- (4.5,2.0);
\node at (4.55,1.79) {$\yt$};
\end{tikzpicture}
\caption{The bulk fundamental domain of the Schottky construction consisting of two hemispheres excised from \AdS{3} with their boundaries identified by the action of the Schottky group. The bulk coordinate $\xi$ is cut off by the surface $\xi = \xi_c(\yt,\bar{\yt})$ (green).}
\label{fig:bulkfunddom}
\end{center}
\end{figure}
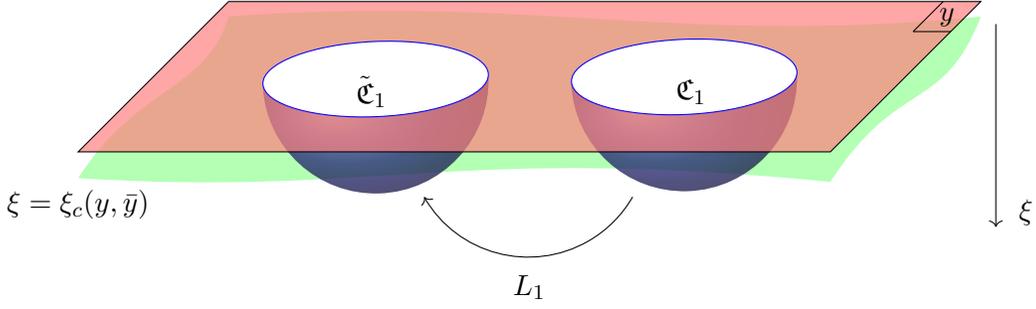

We construct the bulk geometry by filling in the cycles $\mathfrak{C}_c$. The bulk geometry has the standard Poincar\'e metric
\begin{equation}\label{eqn:metricschottkyquot}
ds^{2} = \frac{d\xi^{2}+d\yt \, d\bar{\yt}}{\xi^{2}}
\end{equation}
with the fundamental domain in the bulk obtained by extending the boundary circles, whose radius is
$\ell = \frac{1-\xs}{2}$, into hemispheres and identifying these hemispheres by the action of the Schottky group, as illustrated in \cref{fig:bulkfunddom}.

\subsection{On-shell gravitational action}
\label{sec:22renaction}

We now proceed to evaluate the Euclidean gravitational action for the metric \eqref{eqn:metricschottkyquot} on the bulk fundamental domain $\mathcal{D}_{\mathrm{bulk}}$. We need to evaluate the action
\begin{equation}\label{eq:22renactioncovering}
S^E_\text{gr}[\bulk_2]= -\frac{1}{16\pi G_N} \left[ \int_{\bulk_2} \, d^3x\, \sqrt{g}\, (R+2)  + 2\, \int_{\bdy_c} \sqrt{\gamma}\, K - 2\int_{\bdy_c} \, \sqrt{\gamma} \right] .
\end{equation}	
The boundary curvature counterterm in \eqref{eq:er31exp} is absent here since the torus is flat.

We will use Fefferman-Graham coordinates $(\rho, v, \bar{v)}$ to define the cut-off surface because these give a simple way to extract the contribution from the branch points. The contribution comes from the conformal factor between the $(\xi,\yt,\bar{\yt})$ coordinates and the Fefferman-Graham coordinates, cf., \eqref{eq:zmet1}. The transformation between the coordinates is given by
\begin{equation}\label{eq:FGmapyt}
\xi  = \frac{\sqrt{\rho} \, e^{-\varphi}}{1+ \rho\, e^{-2\varphi} \, \abs{\partial_z \varphi}^2} \,, \qquad
\yt = w +\frac{\rho \, e^{-2\varphi} \, \partial_{\bar{z}} \varphi}{1+ \rho\, e^{-2\varphi} \, \abs{\partial_z \varphi}^2} \,,
\end{equation}
where we set $\Omega \equiv e^{-\varphi}$ in \eqref{eq:FGglMap} and have
\begin{equation}\label{eq:varphi}
\varphi = -\log\left[\frac{\pi\sqrt{\xs}\,\mathfrak{p}_{c}}{2K(\chi)}\frac{1}{\sqrt{|v(v-1)(v-\chi)|}}\,\sech(\mathfrak{p}_{c}\eta(v))\sech(\mathfrak{p}_{c}\bar{\eta}(\bv))\right].
\end{equation}

We define the cut-off surface by $\rho = \rho_c$ which describes a non-trivial cut-off surface $\bdy_c$ in Poincar\'e coordinates described by $\xi = \xi_{c}(\yt,\bar{\yt})$ restricted to $\mathcal{D}_{\mathrm{bulk}}$.

The three contributions to the action can be evaluated directly. We find
\begin{equation}\label{eq:EHGHct22}
\begin{split}
S_\text{EH}[\bulk_2]
&=
	 \int_{\bulk_2} \, d^3x\, \sqrt{g}\, (R+2) = -2\int_{\bulk_2} d\yt\,d\bar{\yt}\,\frac{d\xi}{\xi^3} \,, \\
S_\text{GH}[\bulk_2]
&=
	2\, \int_{\bdy_c} \sqrt{\gamma}\, K = 2\int_{\bdy_c}d\yt\,d\bar{\yt}\,\left(\frac{e^{2\varphi}}{\rho_c}+2\abs{\partial_{\yt}\varphi}^2-2\partial_{\yt}\partial_{\bar{\yt}}\varphi\right)\,,\\	
S_\text{ct}[\bulk_2]
&=
	2\int_{\bdy_c} \, \sqrt{\gamma} = \int_{\bdy_c}d\yt\,d\bar{\yt}\,\left(\frac{e^{2\varphi}}{\rho_c}+4\abs{\partial_{\yt}\varphi}^2\right) .	
\end{split}
\end{equation}
In deriving the boundary quantities we used
\begin{equation}\label{eq:Kgam22ren}
\begin{split}
\gamma_{\yt \yt} &=
	 \frac{1}{\xi_c^{2}}\left(\partial_{\yt}\xi_c\right)^{2}\,, \qquad
	 \gamma_{\bar{\yt}\bar{\yt}} = \frac{1}{\xi_c^{2}}\left(\partial_{\bar{\yt}}\xi_c\right)^{2}\,,
	 \qquad \gamma_{\yt\bar{\yt}} = \frac{1}{\xi_c^{2}}\left(\abs{\partial_{\yt}\xi_c^{2}}+\frac{1}{2}\right) \,,\\
K &= 2\left(2\xi_c\partial_{\yt}\partial_{\bar{\yt}}\xi_c-4\abs{\partial_{\yt}\xi_c}^{2} + 1\right)
\end{split}
\end{equation}

The boundary integrals above are straightforward, but the bulk integral in $S_\text{EH}$ has two distinct contributions: one contribution comes from the region of the bulk below the cut-off surface and the other comes from the region below the hemispheres. Picking $\theta$ to be the azimuthal coordinate around the hemisphere (whose radius we recall is $\frac{1-\xs}{2}$) we can evaluate the two contributions and obtain
\begin{equation}\label{eq:EHaction2contributions}
\begin{split}
S_\text{EH}[\bulk_2] &= -2\int_{\bulk_2} d\yt\,d\bar{\yt}\,\frac{d\xi}{\xi^3}
	= -\left[\int_{\bdy_c} d\yt\,d\bar{\yt}\,\frac{1}{\xi_c(\yt,\bar{\yt})^2} + 2\int_{\mathrm{hemi}} d\yt\,d\bar{\yt}\,\frac{1}{\xi_{\mathrm{hemi}}(\yt,\bar{\yt})^2}\right] \\
	&=  -\left[\int_{\bdy_c} d\yt\,d\bar{\yt}\left(\frac{e^{2\varphi}}{\rho_c}+2\abs{\partial_{\yt}\varphi}^2\right) +
			2\int_{0}^{2\pi} d\theta\,\int_{0}^{r_{\mathrm{cut-off}}(\theta)}dr\,\frac{r}{\ell^2-r^2} \right]\\
	&= -\left[\int_{\bdy_c} d\yt\,d\bar{\yt}\left(\frac{e^{2\varphi}}{\rho_c}+2\abs{\partial_{\yt}\varphi}^2\right) + 2\int_{0}^{2\pi} d\theta\,\left(2\varphi-\log\left(\frac{\rho_c}{\ell^{2}}\right)\right)\right],
\end{split}
\end{equation}

Putting all of the pieces in \eqref{eq:EHGHct22} together we see that the leading divergences cancel as they must and the Euclidean action \eqref{eq:22renactioncovering} becomes
\begin{equation}\label{eq:Eucaction22ren_explicit}
S^E_\text{gr}[\bulk_2]= \frac{1}{8\pi G_N}\left[\int_{\bdy_c}d\yt\,d\bar{\yt}\,\left(\abs{\partial_{\yt}\varphi}^2+2\partial_{\yt}\partial_{\bar{\yt}}\varphi\right)+2\int_{0}^{2\pi} d\theta\,\varphi +4\pi\log(\ell)-2\pi\log(\rho_c)\right].
\end{equation}
We will evaluate each of these terms in turn.

The first term in \eqref{eq:Eucaction22ren_explicit} can be computed very similarly to the one interval case \eqref{eq:eren31intbyparts} where we integrate by parts to reduce the integral to the contributions from the boundaries of the domain. There are boundary terms from the circles $\{\mathfrak{C}_1, \tilde{\mathfrak{C}}_1 \}$ in the $\yt$-plane and boundary terms from the discs $D_{i}^{\delta}$ of radius $\delta$ that we cut out around each branch point $a_{i}$ in the $v$-plane. Finally, there are boundary terms from the IR cut-offs in the $\yt$- and $v$-planes. We thus find:
\begin{equation}\label{eq:varphikineticterm}
\begin{split}
\int_{\bdy_c}d\yt\,d\bar{\yt}\,\abs{\partial_{\yt}\varphi}^2
	&= \delta\sum_{i=1}^{3}\oint_{a_i} \, \varphi \partial_{\abs{v}} \varphi + \int_{0}^{2\pi} d\theta\,\ell\varphi\partial_{r}\varphi|_{r=\ell} + S_\mathrm{IR}\\
	&= -\pi\log\left[\left(\frac{\pi\, \mathfrak{p}_{c} \, \sqrt{\xs}}{2\, K(\chi) \sqrt{\delta}}\right)^3 \,
	\frac{\sech^{2}(\pi \,\mathfrak{p}_{c})\csch^{2}(\pi \,\mathfrak{p}_{c})}{\chi(1-\chi)} \right] - \int_{0}^{2\pi} d\theta\,\varphi + S_\mathrm{IR}\,,
\end{split}
\end{equation}
where the contribution from the boundaries in the $v$-plane has a factor of $2$ owing to the two sheets of $\bdy_{2,2}$ and in the last line we have used the fact that $\partial_{r}\varphi|_{r=\ell}=-1/\ell$. The term labeled $S_\text{IR}$ is the contribution from large radius region in the $\yt$ or $v$-planes, and in particular includes the contribution from the branch point $a_{4}$. We evaluate these separately in \cref{sec:IRdiv} as they are involved, but quote here the final result:
\begin{equation}\label{eq:IRdiv}
S_{\mathrm{IR}} = \pi \left[
	\log\left(\frac{K(\chi) \, \sqrt{\delta\, \xs}}{2^{5} \pi\, \mathfrak{p}_{c}}\right) + 2 \, \log(\rho_c) + 3
	\log(a_{4}) - 8\, \log(R_{v}) \right] ,
\end{equation}
where we are meant to take the limit $R_v, a_4 \to \infty$.

The second term in \eqref{eq:Eucaction22ren_explicit} reduces to a sum of localized delta functions as in the one-interval case and thus vanishes,
\begin{equation}\label{eq:deltafunction22ren}
\int_{\bdy_c} d\bar{\yt}\,d\yt\,\partial_{\yt}\partial_{\bar{\yt}}\varphi = 2\int_{\widehat{\mathbb{C}}\backslash\cup_{i}D_{i}^{\delta}}\,d\bv\,dv\,\partial_{v}\partial_{\bv}\varphi = \frac{\pi}{4}\int_{\widehat{\mathbb{C}}\backslash\cup_{i}D_{i}^{\delta}}\,d\bv\,dv\,\delta\left(|v|\right)+\delta\left(|v-1|\right)+\delta\left(|v-\chi|\right) = 0.
\end{equation}
The third term in \eqref{eq:Eucaction22ren_explicit} turns out to be formidable. We use various elliptic function identities to evaluate it in \cref{sec:hemisphereint} and  find when all the dust settles the result
\begin{equation}\label{eq:varphiint}
\int_{0}^{2\pi} d\theta\,\varphi = 4\pi\log\left(\frac{2^{\frac{3}{2}}\, \pi\, \mathfrak{p}_{c}}{K(\chi)}\cosh(\pi\,\mathfrak{p}_{c})\right) - 4\pi^{2}\mathfrak{p}_{c}.
\end{equation}

Plugging all of these pieces into \eqref{eq:Eucaction22ren_explicit}, we arrive at our final answer for the Euclidean action:
\begin{equation}\label{eq:Eucaction22ren_final}
S^E_\text{gr}[\bulk_2] = -\frac{\pi\,c}{3}\mathfrak{p}_{c} + \frac{c}{12}\log\big(\delta^{2}\chi(1-\chi)\big) + \frac{c}{4}\log(a_{4}) - \frac{2c}{3}\log(R_{v}).
\end{equation}
To obtain the R\'enyi entropy, we need to normalize by the gravitational action of the sphere $S^E_\text{gr}[\bulk_1]$. However, one needs to be careful because we have chosen a particular IR regularization scheme to deal with the fact that we placed one of the branch points at infinity (this is the same as the regularization scheme used in \cite{Lunin:2000yv}). As a result, the action on the sphere is no longer unity like in the single interval case, instead one finds
\begin{equation}\label{eq:Eucactionsphere}
S^E_\text{gr}[\bulk_1] = \frac{c}{6}\log\left(\frac{\rho_c}{R_{v}^{2}}\right) - \frac{c}{3}.
\end{equation}
The second R\'enyi entropy for two intervals is thus  (using $\mathfrak{p}_c$ from \eqref{eq:2inpcd})
\begin{equation}\label{eq:22ren}
\begin{split}
S_{\regA_{1} \cup \regA_{2}}^{(2)}
	&= S^E_\text{gr}[\bulk_2]-2S^E_\text{gr}[\bulk_1] \\
	&= -i\frac{\pi c}{6\tau(\chi)}+\frac{c}{12}\log\left(\delta^{2}\chi(1-\chi)\right) + \frac{2c}{3} + \frac{c}{4}\log(a_{4}) -\frac{c}{3}\log(\rho_c).
\end{split}
\end{equation}
We emphasize that this only gives the second R\'enyi entropy for the connected phase $1/2 \leq \chi < 1$. For the second mutual R\'enyi information $I^{(2)}$ we need the second R\'enyi entropy for the single interval using our choice of regularization and thus it will differ from \eqref{eq:1nren}. It is given by
\begin{equation}\label{eq:1nren_LMreg}
S_{\regA_{i}}^{(2)} = \frac{c}{4}\log\left(\delta^{\frac{1}{3}}|a_{2i}-a_{2i-1}|\right)-\frac{c}{3}\log(2)-\frac{c}{6}\log(\rho_c) + \frac{c}{3}.
\end{equation}
We thus arrive at the second mutual R\'enyi information (with the regulators $\delta, a_4, \rho_c$ canceling)
\begin{equation}\label{eq:22mri}
I^{(2)}(\chi) = -\frac{c}{12}\, \log\left(2^8 \, \frac{1-\chi}{\chi^2}\right) + i\, \frac{\pi}{6}\, c\,  \frac{1}{\tau(\chi)} .
\end{equation}
This is in complete agreement with the result \eqref{eq:rmi2int} obtained from the accessory parameter (in the connected phase). The disconnected case proceeds along similar lines with $\mathfrak{p}_c \to \mathfrak{p}_d$.

\subsection{Hemisphere integral}
\label{sec:hemisphereint}

We now turn to the calculation of the integral of $\varphi = -\log \Omega$ along the azimuthal angle of the hemisphere appearing in \eqref{eq:Eucaction22ren_explicit}. To do this, we first need to rewrite the coordinate $\yt$ along the semi-circles given by the intersection of the circles $\mathfrak{C}_1, \tilde{\mathfrak{C}}_1$ with the upper-half $\yt$-plane. These semi-circles are the images of the intervals $[\chi+i\epsilon,1+i\epsilon]$ and $[\chi-i\epsilon,1-\epsilon]$ in the $v$-plane, respectively; see \cref{fig:tildeyplane}. In the interval $v \in [\chi,1]$, the torus elliptic map\footnote{ We find it useful to employ Jacobian notation cf., \cite[Sec.~22.1]{NIST-DLMF}, to avail of various identities. A useful reference for elliptic function properties is \cite{Byrd:2013hbk}. } is given by continuing \eqref{eq:etavel} outside the principal domain,
\begin{equation}\label{eq:toruselliptic_chito1}
\eta(v\pm i\epsilon) = \pm \pi + i\frac{\pi}{K(\chi)} \sn^{-1}\left(\Theta(v),1-\chi\right)\,,
\end{equation}
with
\begin{equation}\label{eq:Theta}
\sin\Theta = \sqrt{\frac{(v-\chi)}{(1-\chi)v}}\,,
\qquad v = \frac{\chi}{1-(1-\chi)\sin^{2}\Theta}.
\end{equation}
Using this the  map $\yt(v)$ in the interval $v \in [\chi,1]$ then takes the form
\begin{equation}\label{eq:tildeysemicirc}
\begin{split}
\yt(v \pm i \epsilon)
	&= \tanh(\pi\,\mathfrak{p}_{c} )\tanh\left(\mathfrak{p}_{c} \eta(v)\right)
	= \frac{\pm \xs + i\zeta(v)}{1 \pm i\zeta(v)},
\end{split}
\end{equation}
where we have defined a new map $\zeta(v)$ using the addition formula, viz.,
\begin{equation}\label{eq:zeta}
\begin{split}
\zeta(v)
	&= \tanh(\pi\,\mathfrak{p}_{c} )\tan\left(i\mathfrak{p}_{c}(\pi-\eta(v))\right)
	= \tanh(\pi\,\mathfrak{p}_{c} )\tan\left(\frac{\pi\,\mathfrak{p}_{c}}{K(\chi)}\sn^{-1}\left(\Theta(v),1-\chi\right)\right).
\end{split}
\end{equation}
This gives the desired description of the semi-circle. Note that we can invert $\Theta(\zeta)$ and write
\begin{equation}\label{eqn:sinTheta}
\sin\Theta = \sn(w,1-\chi) \,,
\qquad
w = \frac{2K(1-\chi)}{\pi} \coth(\pi\,\mathfrak{p}_c)\, \zeta(v) \,.
\end{equation}

Armed with these definitions we can evaluate $\varphi(v)$  in the interval $v \in [\chi,1]$ to be
\begin{equation}\label{eq:varphi_semicirc}
\varphi(v) = -\frac{1}{2}\log\left(\frac{d\yt}{dv}\frac{d\bar{\yt}}{d\bv}\bigg|_{v=\bv}\right) = -\log\left(\frac{(1-\xs)\zeta'(v)}{\zeta(v)^{2}+1}\right) .
\end{equation}
Likewise, the azimuthal angle as a function of $v$ is
\begin{equation}\label{eq:theta}
\theta(v) = \tan^{-1}\bigg(\frac{2\zeta(v)}{\zeta(v)^{2}-1}\bigg).
\end{equation}
The desired integral thus becomes
\begin{equation}\label{eq:trickyint}
\int_{0}^{2\pi} d\theta\,\varphi = 2\int_{0}^{\pi} d\theta\,\varphi = -4\int_{\chi}^{1}dv\,\left(\frac{\zeta'(v)}{\zeta(v)^{2}+1}\right)\log\left(\frac{(1-\xs)\,\zeta'(v)}{\zeta(v)^{2}+1}\right).
\end{equation}
Evaluating the argument of the logarithm we find it convenient to split integral into two pieces, one of which can be integrated directly, leading to
\begin{equation}\label{eq:trickyint2}
\begin{split}
\int_{0}^{2\pi} d\theta\,\varphi
	= 2\pi\log\left(\frac{8K(\chi)\cosh^{3}\left(\pi\,\mathfrak{p}_{c} \right)\sinh\left(\pi\,\mathfrak{p}_{c} \right)}{\pi\,\mathfrak{p}_{c}}\right)-4\pi^{2}\,\mathfrak{p}_{c} + \mathcal{I}(\chi),
\end{split}
\end{equation}
with
\begin{equation}\label{eq:Ichi}
\mathcal{I}(\chi) = 4\int_{\chi}^1 \, dv\, \frac{\zeta'}{1+\zeta^2}\, \log\left(\sqrt{v(1-v)(v-\chi)}\right) .
\end{equation}	

To evaluate $\mathcal{I}(\chi)$ we evaluate the integrand in terms of Jacobi elliptic functions:
\begin{equation}\label{eqn:trickyintegrand}
\sqrt{v(1-v)(v-\chi)}
	= \frac{\chi(1-\chi)\cos\Theta\sin\Theta}{\big(1-(1-\chi)\sin^{2}\Theta\big)^{\frac{3}{2}}}
	= \frac{\chi(1-\chi)\,\sn(w,1-\chi)\,\cn(w,1-\chi)}{\dn^{3}(w,1-\chi)},
\end{equation}
where we have used the relations $\sn^{2}(z,m)+\cn^{2}(z,m) = 1$ and $m\,\sn^{2}(z,m)+\dn^{2}(z,m)=1$,
and $w$ is defined above in \eqref{eqn:sinTheta}. The integral changing variables to $w$, with $\tilde{w} = \frac{\pi w}{2K(\cx)}$, is
\begin{equation}\label{eq:trickyint3}
\begin{split}
\mathcal{I}(\chi)
&=
	2\pi\log\left(\chi(1-\chi)\right) + \mathcal{J}(1-\chi) \\
\mathcal{J}(x)
&=
	\frac{2\pi\,\coth(\pi\,\mathfrak{p}_c)}{K(\cx)}
	\int_{0}^{K(\cx)}dw\,
	\frac{\sec^{2}\tilde{w} }{\coth^2(\pi\,\mathfrak{p}_{c})+ \tan^{2}\tilde{w}}
	\log\left(\frac{\sn(w,\cx)\, \cn(w,\cx)}{\dn^{3}(w,\cx)}\right).
\end{split}
\end{equation}
We can now exploit the fact that Jacobian elliptic functions have an infinite product representation:
\begin{equation}\label{eq:Jacobianellipticfn_product}
\begin{split}
\sn(w,\cx)
	&= 2\bigg(\frac{q_{\cx}}{ \cx }\bigg)^{\frac{1}{4}} \, \sin\tilde{w}  \,
	\prod_{n=1}^{\infty}\frac{1-2\,q_{\cx}^{2n}\,\cos(2\tilde{w}) +q_{\cx}^{4n}}{1-2\,q_{\cx}^{2n-1}\,\cos(2\tilde{w})+q_{\cx}^{4n-2}} \\
\cn(w,\cx)
&=
	2\bigg(\frac{(1-x) \,q_{\cx}}{\cx}\bigg)^{\frac{1}{4}} \, \cos\tilde{w} \,
	\prod_{n=1}^{\infty}\frac{1+2\,q_{\cx}^{2n}\,\cos(2\tilde{w})+q_{\cx}^{4n}}{1-2\,q_{\cx}^{2n-1}\cos(2\tilde{w})
	+q_{\cx}^{4n-2}} \\
\dn(w,\cx)
&=
	(1-x)^{\frac{1}{4}}\,
	\prod_{n=1}^{\infty} \frac{1+2\,q_{\cx}^{2n-1}\, \cos(2\tilde{w})+q_{\cx}^{4n-2}}{1-2\,q_{\cx}^{2n-1}\,
	\cos(2\tilde{w})+q_{\cx}^{4n-2}}.
\end{split}
\end{equation}
where $q_{\cx} = e^{\pi i \tau(\cx)}$ is the elliptic nome. These products inside the logarithm become an infinite sum of logarithms and a change of variables to $\tan(\tilde{w})$ allows for a straightforward evaluation of the resulting integrals. Once the dust settles, we arrive at
\begin{equation}\label{eq:trickyint4}
\begin{split}
I(\chi)
&=
	 \pi\log\left(
	 	\frac{16\,q_{1-\chi}\, \chi(1-\chi) \, \coth^{2}(\pi\,\mathfrak{p}_{c})}{(\coth(\pi\,\mathfrak{p}_{c})+1)^{4}}
	 	\right)
	  + 4\pi \sum_{n=1}^\infty\, \log\left(
	  	\frac{1+q_{1-\chi}^{2n+1}}{1-q_{1-\chi}^{2n}} \, \frac{1-q_{1-\chi}^{2n+1}}{1-q_{1-\chi}^{2n}}\,  \frac{1+q_{1-\chi}^{2n}}{1-q_{1-\chi}^{2n}}
	  \right)
\end{split}
\end{equation}
which can be simplified using \eqref{eq:Jacobianellipticfn_product} evaluated at special values of $w$ to give
\begin{equation}\label{eq:trickyint4result}
\mathcal{I}(\chi) = 2\pi\log\left(
	\frac{\pi^3\, \mathfrak{p}_c^3\, \sech(\pi\,\mathfrak{p}_{c})\csch(\pi\,\mathfrak{p}_{c})}{K(\chi)^{3}}\right).
\end{equation}
Inserting this result into \eqref{eq:trickyint2} gives the result \eqref{eq:varphiint} quoted earlier.

\subsection{IR divergences}
\label{sec:IRdiv}

The final ingredient in our computation is the evaluation of the long-distance contributions encoded in $S_\mathrm{IR}$, which originate from several different places and we will discuss each of them in turn.

\begin{itemize}[wide,left=0pt]
\item Firstly, the integration by parts  of the `kinetic term' for $\varphi$ in \eqref{eq:varphikineticterm} contributes. Imposing large radius cut-offs $R_{\yt}$ and $R_{v}$ in the $\yt$- and $v$-planes, respectively, we obtain the following boundary contributions to \eqref{eq:varphikineticterm}
\begin{equation}\label{eq:kineticdiv}
\begin{split}
&
	\lim_{R_{v},R_{\yt} \to \infty}\int_{0}^{2\pi} d\theta\,R_{v}\varphi\partial_{r}\varphi|_{r=R_{v}} + \frac{1}{2}\int_{0}^{2\pi} d\theta\,R_{\yt}\varphi\partial_{r}\varphi|_{r=R_{\yt}} 	 \\
&=
	 \lim_{R_{v},R_{\yt} \to \infty} 2\int_{0}^{2\pi} d\theta\,\varphi(R_{v},\theta) - \int_{0}^{2\pi} d\theta\,\varphi(R_{\yt},\theta) = \lim_{R_{v},R_{\yt} \to \infty} 4\pi\varphi(R_{v}) - 2\pi\varphi(R_{\yt})\,,
\end{split}
\end{equation}
using the fact that $\partial_{r}\varphi|_{r=R_{v}} = 2/R_{v}$ and $\partial_{r}\varphi|_{r=R_{\yt}} = -2/R_{\yt}$. Furthermore, $\varphi$ becomes angle independent in the infinite radius limit (as we shall see later).

\item The second contribution comes from working in Poincar\'e coordinates which misses an extra term coming from the curvature of the $\yt$ sphere (which is pushed off to infinity in these coordinates). To find this extra term, we pass to global coordinates with metric
\begin{equation}\label{eq:globaltildeymetric}
ds^{2} = \frac{d\xi^2}{\xi^{2}}+R_{\yt}^{2}\left(\frac{R_{\yt}}{\xi} - \frac{\xi}{R_{\yt}}\right)^{2}\frac{d\yt\,  d\bar{\yt}}{(R_{\yt}^{2}+|\yt|^{2})^{2}},
\end{equation}
which recovers the Poincar\'e metric for $R_{\yt} \to \infty$. One finds the extra contribution by computing the Einstein-Hilbert action with this metric in global coordinates and comparing to Einstein-Hilbert action in Poincar\'e coordinates \eqref{eq:EHGHct22}. One thus finds the missing term to be
\begin{equation}\label{eq:spherecurvaturecontr}
\lim_{R_{\yt} \to \infty} \frac{1}{8\pi G_{N}}\int d\yt\,d\bar{\yt}\,\frac{R_{\yt}^{2}}{(R_{\yt}^{2}+|\yt|^{2})^{2}}\left(\log\bigg(\frac{\rho_c}{R_{\yt}^{2}}\bigg)-2\varphi\right) = \lim_{R_{\yt} \to \infty}\frac{1}{4G_{N}}\log\left(\frac{\rho_c}{R_{\yt}^{2}}\right)-\frac{1}{2G_{N}}\varphi(R_{\yt})
\end{equation}

\item The third and final contribution,  requires careful analysis of the contribution to the action from the branch point $a_{4}$ which we have sent to infinity. This was discussed in Appendix D of \cite{Faulkner:2013yia} for the case $n > 2$. The main challenge with obtaining this contribution is that when we set $a_{4} = \infty$, we have $\yt(a_{4}=\infty) = \infty$ so we cannot distinguish the contribution of the branch point from the contribution of the sphere curvature at $\yt = \infty$ discussed above. We give an analytic estimate for $n=2$ below.
\end{itemize}

To understand the $a_4$ contribution, we deform the map $\yt$ slightly so that $a_{4}$ does not map to infinity, instead $\yt(a_{4}) = y_{4} \gg 1$ with some point $v_{\infty} \approx a_{4}$ on one sheet such that $\yt(v_{\infty}) = \infty$. This will allow us to separately find the contribution from $a_{4}$ and from the sphere curvature \eqref{eq:spherecurvaturecontr}. We will then take the limits  $y_{4} \to \infty$ followed by $a_{4} \to \infty$.

Now that $a_{4}$ is finite, we can use that $\yt(v)$ is a power series in $(v-a_{4})^{\frac{1}{2}}$ near $a_{4}$ by the Schottky construction to write
\begin{equation}\label{eq:tildeyneara4}
\yt(v) = y_{4} + \mu_{4}\, (v-a_{4})^{\frac{1}{2}} + \order{v-a_{4}}\,.
\end{equation}
Therefore, the contribution of $a_{4}$ to the integral of the `kinetic term' for $\varphi$ evaluated in \eqref{eq:varphikineticterm} is given by
\begin{equation}\label{eq:a4contribution}
-\oint_{a_4} \, \varphi \partial_{\abs{v}} \varphi = \pi\log\left(\frac{|\mu_{4}|}{2\epsilon^{\frac{1}{2}}}\right).
\end{equation}
We next find the behavior of $\yt$ and $\varphi$ at $R_{\yt}$ and $R_{v}$. Since we only put $v_{\infty}$ on one sheet, $\yt(v)$ must have an order one pole at this point so near $v_{\infty}$ (with residue $\nu_\infty$), we have
\begin{equation}\label{eq:tildeyvinf}
\yt(v) \approx \frac{\nu_{\infty}}{v-v_{\infty}} \implies \lim_{R_{\yt} \to \infty}\varphi(R_{\yt}) = \lim_{R_{\yt} \to \infty}\log\left(\frac{|\nu_{\infty}|}{R_{\yt}^{2}}\right).
\end{equation}
Recall that the accessory parameters are chosen such that $\yt$ is regular at $v=\infty$. It is not branched at this point since $a_{4}$ is finite, thus near $v=\infty$
\begin{equation}\label{eq:tildeyvtoinf}
\yt(v) = y_{\infty} + \frac{\mu_{\infty}}{v} \implies \lim_{R_{v} \to \infty}\varphi(R_{v}) = \lim_{R_{v} \to \infty}\log\left(\frac{R_{v}^{2}}{|\mu_{\infty}|}\right).
\end{equation}
Having extracted all the necessary contributions from $a_{4},R_{v},$ and $R_{\yt}$, we can now take the desired limits $y_{4} \to \infty$ and then $a_{4} \to \infty$. From \eqref{eq:a4contribution} and plugging \eqref{eq:tildeyvinf} and \eqref{eq:tildeyvtoinf} into \eqref{eq:kineticdiv} and \eqref{eq:spherecurvaturecontr}, one finds that the contribution to the action from the newly defined parameters is
\begin{equation}\label{eq:newparams_action}
\mathfrak{S} = -\frac{1}{4G_{N}}\log\left(\frac{|\nu_{\infty}|\, |\mu_{4}|^{\frac{1}{2}}}{|\mu_{\infty}|^{2}}\right).
\end{equation}
We need to understand the behavior of these three parameters when we take the desired limits. To take the limit $y_{4} \to \infty$ (or equivalently $a_{4} \to v_{\infty}$), we use that $\yt$ is 2-branched at $a_{4}$ to write the inverse function $v(\yt)$ near $y_{4}$ as
\begin{equation}\label{eq:tildeyinv}
v(\yt) \approx a_{4} + \frac{y_{4}^{4}}{\mu_{4}^{2}}\, \left(\frac{1}{\yt}-\frac{1}{y_{4}}\right)^{2},
\end{equation}
where we fixed the coefficient of the quadratic term by comparison with  \eqref{eq:tildeyneara4}.
One can then extract from \eqref{eq:tildeyinv} the relation between $\nu_{\infty}$ and $y_{4},\; \mu_{4}$ by taking $\yt$ large and then $y_{4} \to \infty$ with the result
\begin{equation}\label{eq:nuinf}
v-v_{\infty} = \lim_{y_{4} \to \infty} (v-a_{4})
\;\; \implies \;\;  \nu_{\infty} = -2\frac{y_{4}^3}{\mu_{4}^2}.
\end{equation}
Furthermore, observe that in the limit $y_{4} \to \infty$ the behavior of $\yt$ near $a_{4}$ is given by
\begin{equation}\label{eq:tildeyneara4y4toinf}
\lim_{y_{4} \to \infty} \yt(v) \approx - \frac{\abs{y_4}^2}{\abs{\mu_4}\, (v-a_{4})^{\frac{1}{2}}}.
\end{equation}

Taking the limit $a_{4} \to \infty$ we can find the relationship between $\frac{y_4^2}{\mu_4}$ and $\mu_{\infty}$ as follows. We compute the residue at $v=\infty$ of $\yt(v)^2$ using \eqref{eq:tildeyvtoinf} and equate it to the limit $a_{4} \to \infty$ of the residue at $v=a_{4}$ of $\yt(v)^2$ using \eqref{eq:tildeyneara4y4toinf}. We repeat the same procedure for the function $\yt(v)^2/v$ and then plug the latter equation into the former. The final result is
\begin{equation}\label{eq:muinf}
\lim_{a_{4} \to \infty}|\mu_{\infty}| = \lim_{a_{4} \to \infty}\frac{\sqrt{a_{4}}\, \abs{y_4}^2}{2\, \abs{\mu_4}}.
\end{equation}
This completes the analysis of the parameters in $\mathfrak{S}$ in the desired sequence of limits, in particular we can write the argument of the logarithm in terms of $a_{4},y_{4},$ and $\mu_{4}$. We want to compute the latter two parameters from the analytic solution for $\yt(v)$ \eqref{eqn:newycoord}. However, this was obtained by solving the monodromy problem which assumed that $a_{4} = \infty$. The function $\yt(v)$ is regular at $v=\infty$ for finite $a_{4}$, but this is no longer the case when $a_{4}=\infty$ so that the map is 2-branched at $v=\infty$ with the following behavior
\begin{equation}\label{eq:tildeyinf}
\yt(v) \approx \hat{\mu}_{4}\, v^{\frac{1}{2}}.
\end{equation}
Taking the limit $a_{4} \to \infty$ and then $v \to \infty$ in \eqref{eq:tildeyneara4y4toinf} and comparing to \eqref{eq:tildeyinf} gives
\begin{equation}\label{eq:muhat4}
\hat{\mu}_{4} = \lim_{a_{4} \to \infty}\frac{\abs{y_4}^2}{a_{4}\, \abs{\mu_4}} \,.
\end{equation}

Finally, it remains to find an explicit expression for $\hat{\mu}_{4}$ from \eqref{eqn:newycoord}. Using \eqref{eq:etavint}, one finds for small $\delta$
\begin{equation}
\eta\left(\frac{1}{\delta}\right)
	= \pi\,\tau-\frac{\pi}{K(\chi)} \, \sqrt{\delta} + \order{\delta^{\frac{3}{2}}}
	 \implies
	 \lim_{\delta \to 0}\, 	 \yt\left(\frac{1}{\delta}\right)
	 = -\frac{K(\chi) \,\sqrt{\xs}}{\pi\,\mathfrak{p}_{c}\,\sqrt{\delta}}
	 \implies \hat{\mu}_{4} = -\frac{K(\chi) \,\sqrt{\xs}}{\pi\,\mathfrak{p}_{c}}.
\end{equation}
Therefore,
\begin{equation}\label{eq:newparams_action_final}
\mathfrak{S} =
	 -\frac{1}{4G_{N}}
	 	\log\left(\frac{8}{a_{4}^{\frac{3}{2}}|\widehat{\mu}_{4}|^{\frac{1}{2}}}\right)
	 = \frac{1}{8G_{N}}\,\log\left(\frac{a_{4}^{3} \,K(\chi)\,\sqrt{\xs}}{2^{6}\,\pi\, \mathfrak{p}_{c}}\right).
\end{equation}
Putting all of this together, we find the total contribution from the long-distance pieces:
\begin{equation}\label{eq:IRdiv_final}
S_{\mathrm{IR}} = \pi \left[
	\log\left(\frac{K(\chi) \, \sqrt{\delta\, \xs}}{2^{5} \pi\, \mathfrak{p}_{c}}\right) + 2 \, \log(\rho_c) + 3
	\log(a_{4}) - 8\, \log(R_{v}) \right].
\end{equation}
%

\section{Actions, signs, and all that}
\label{sec:actions}
We collect here some useful facts about actions and signs that the reader might find helpful in checking various details of the paper.

\subsection{Signs of gravitational action}
\label{sec:EHsigns}

The Lorentzian gravitational action $S$ which enters in the path integral measures as $e^{iS_\text{gr}}$ for standard time-ordered scattering computations, or as $e^{i (S^k_\text{gr} - S^b_\text{gr})}$ is given by
\begin{equation}\label{eq:SEHLorentz}
S_\text{gr}  \equiv S^k_\text{gr} = \frac{1}{16\pi G_N} \left[ \int d^{d+1}x\, \sqrt{-g} \left( R + d(d-1) \right) + 2 \, \int d^dx\, \sqrt{-\gamma}\, K + S_\text{ct}\right]
\end{equation}	
The Euclidean path integral on the other hand is defined to be one with a real measure $e^{-S^E_\text{gr}}$ which in turn can be obtained by analytic continuation. When we Wick rotate $t \to - i\, t_{_\text{E}}$ we pick end up picking a factor of $-i$ from the integration measure, which combined with the $i$ in the quantum weighting, gives $+1$. A more straightforward statement is that the Euclidean action should correspond to the Euclidean Hamiltonian and generically be positive definite. This is why one defines:
\begin{equation}\label{eq:SEHEuclid}
S^E_\text{gr} =  -\frac{1}{16\pi G_N} \left[ \int d^{d+1}x\, \sqrt{g }\left( R + d(d-1) \right) + 2 \, \int d^dx\, \sqrt{\gamma}\, K + S_\text{ct}\right]
\end{equation}	

The evaluation of the functional integral is supposed to give a generating function (or a partition function), $Z$ which in turn is  expressed as a free energy (to pick up the connected components).  We usually define therefore
\begin{equation}
Z = e^{-I} = \int_L \, [Dg]\,  e^{iS_\text{gr}} \,, \qquad Z = e^{-I} = \int_E \, [Dg]\,  e^{-S^E_\text{gr}}
\end{equation}	

In thermodynamic systems $I = \beta F$ where $F$ is the free energy, which for sensible thermal systems is negative $F = E - TS$. This is necessary for positivity of entropy and for the usual intuition that systems with lower free energy dominate the canonical ensemble (since $S = - \pdv{F}{T}$ using $dF = - S\, dT$).  This implies $I < 0$ (it is negative of the pressure). A saddle point or stationary phase evaluation of the above path integrals then gives:
\begin{equation}
I = S^E_{\text{gr}}\big|_\text{on-shell} \,, \qquad I = - i\, S_\text{gr} = - i (S^k_\text{gr} - S^b_\text{gr}) = 2 \Im(S^k_\text{gr})
\end{equation}	

These statements can be checked for the  planar-Schwarzschild-\AdS{5} black hole which does define a sensible thermodynamic system for the dual CFT plasma. With a UV cut-off at $r=r_c$ in Schwarzschild coordinates one finds:
\begin{equation}
\begin{split}
\int d^{5}x\, \sqrt{g }\left( R +12  \right)  &=  - 2 (r_c^4 - r_+^4) \\
2 \, \int d^4x\, \sqrt{\gamma}\, K  & = 8\, r_c^4 - 4 r_+^4 \\
S_\text{ct} &= - 6 r_c^4 + 3\, r_+^4
\end{split}
\end{equation}	
giving $S^E_{\text{gr}}\big|_\text{on-shell} = I = -r_+^4$ which is the expression that correctly reproduces the pressure of the dual plasma.

\subsection{Complex integral identities}
\label{sec:complexint}

In our evaluation of the $\hat{I}_n$ in Euclidean signature for $N$-intervals we made use of two identities which we quote here in generality.  First, consider an integral $\mathcal{I}$
\begin{equation}
\mathcal{I} = \int_{\mathscr{R}_\epsilon} \, dv d\bv\, \mathfrak{F}(v,\bv) = i\,
 \int_{\mathscr{R}_\epsilon} \,   \mathfrak{F}(v,\bv)\, dv \wedge d\bv
\end{equation}	
over a domain $\mathscr{R}_\epsilon$ of the complex plane
defined by excising discs  $\mathscr{D}_i^\epsilon$ centered at $a_i$
\begin{equation}
\mathscr{R}_\epsilon = \mathbb{C} \backslash \left( \cup_j \mathscr{D}_j^\epsilon\right)
\end{equation}	
If one wishes to consider  the variations of the integral with respect to the locations $a_i$ then not only should one consider the explicit variation of the integrand but also account for the variation of the domain $\mathcal{R}_\epsilon$. The latter is a boundary integral and the general result we need is
\begin{equation}
\pdv{a_i} \mathcal{I}  = i\,   \int_{\mathscr{R}_\epsilon} \,  \pdv{a_i} \mathfrak{F}(v,\bv)\, dv \wedge d\bv
	+ i\, \oint_{\partial \mathscr{D}_i^\epsilon} \, \mathfrak{F}(v,\bv)\, d\bv  -
	 i\, \oint_{\partial \mathscr{D}_i^\epsilon} \, \mathfrak{F}(v,\bv)\, dv
\end{equation}	
Another relation we have employed is the Stokes' theorem on the Dolbeault complex ($d = \partial + \bar{\partial}$).  For a holomorphic $\mathfrak{f}(v)$ we have
\begin{equation}
i\int_{\mathscr{R}} \left( \partial_v \mathfrak{f}(v)  + \partial_{\bv} \bar{\mathfrak{f}}(\bv)\right)\, dv \wedge d\bv
=
i\int_{\mathscr{R}} \, d\left(\mathfrak{f} \,d\bv - \bar{\mathfrak{f}}\, dv\right) = i \, \int_{\partial \mathscr{R}} \,\left( \mathfrak{f} \,d\bv - \bar{\mathfrak{f}}\, dv\right)
\end{equation}	
%


\providecommand{\href}[2]{#2}\begingroup\raggedright\endgroup

\end{document}